\documentclass[letterpaper,twocolumn,10pt]{article}
\usepackage{zhanggroup}
\usepackage{hyperref}
\hypersetup{
  colorlinks,
  linkcolor={blue!70!green},
  citecolor={green!70!blue},
  urlcolor={orange!70!red}
}
\usepackage{xspace}
\usepackage{cite}
\usepackage{amsmath,amssymb,amsfonts}
\usepackage{algorithmic}
\usepackage{graphicx}
\usepackage{textcomp}
\usepackage{xcolor}
\usepackage{tikz}
\usepackage{array}
\usepackage{pifont}
\usepackage[normalem]{ulem}
\usepackage{multirow}
\usepackage{subcaption}
\usepackage{caption}
\usepackage{balance}
\usepackage{booktabs}
\usepackage{makecell}
\usepackage{tcolorbox}
\usepackage{bbm}
\usepackage{caption}
\usepackage{subcaption}
\usepackage{float}
\usepackage{enumitem}
\setlist[itemize]{leftmargin=*}

\newcommand{\mypara}[1]{\smallskip\noindent\textbf{#1.}\xspace}
\begin{document}
% ----------------------------------------------------

\date{}

% ----------------------------------------------------
\title{\Large \bf Fine-Tuning Is All You Need to Mitigate Backdoor Attacks}
% ----------------------------------------------------

\author{
Zeyang Sha\textsuperscript{1}\ \ \
Xinlei He\textsuperscript{1}\ \ \
Pascal Berrang\textsuperscript{2}\ \ \
Mathias Humbert\textsuperscript{3}\ \ \
Yang Zhang\textsuperscript{1}
\\
\\
\textsuperscript{1}\textit{CISPA Helmholtz Center for Information Security}\\
\textsuperscript{2}\textit{University of Birmingham}\ \ \
\textsuperscript{3}\textit{University of Lausanne}
}

\maketitle

% ----------------------------------------------------
\begin{abstract}
% ----------------------------------------------------

Backdoor attacks represent one of the major threats to machine learning models.
Various efforts have been made to mitigate backdoors.
However, existing defenses have become increasingly complex and often require high computational resources or may also jeopardize models' utility.
In this work, we show that fine-tuning, one of the most common and easy-to-adopt machine learning training operations, can effectively remove backdoors from machine learning models while maintaining high model utility.
Extensive experiments over three machine learning paradigms show that fine-tuning and our newly proposed super-fine-tuning achieve strong defense performance.
Furthermore, we coin a new term, namely \emph{backdoor sequela}, to measure the changes in model vulnerabilities to other attacks before and after the backdoor has been removed.
Empirical evaluation shows that, compared to other defense methods, super-fine-tuning leaves limited backdoor sequela.
We hope our results can help machine learning model owners better protect their models from backdoor threats.
Also, it calls for the design of more advanced attacks in order to comprehensively assess machine learning models' backdoor vulnerabilities.

% ----------------------------------------------------
\end{abstract}
% ----------------------------------------------------

% ----------------------------------------------------
\section{Introduction}
% ----------------------------------------------------

In recent years, researchers have shown that machine learning (ML) models are vulnerable to various security attacks.
One common attack in this domain is the backdoor attack~\cite{GDG17,PZGXJCW20,GDG17,CLLLS17,LMALZWZ18,JLG22,WCZZWYSZ22,SHLSBZ22}, whereby an adversary aims to insert a backdoor into a target ML model via malicious training.
Taking image classification as an example, a backdoored model will classify images that embed specific triggers into a pre-defined class while keeping normal behavior on clean images.
So far, most efforts have gone into the design of effective backdoor attacks against various types of ML models~\cite{GDG17,NT20,CLLLS17,PSZJVLLW20,LLWLHL21}.
To mitigate these attacks, intricate defenses have been proposed.
Some of the defenses~\cite{WYSLVZZ19,CFZK19,LLTMAZ19,HAS19,GWXDS19}, focus on extracting the trigger from a target ML model via optimization; some aim to detect the inputs with triggers~\cite{TLM18,CCBLELMS18,GXWCRN19,UPWLRC22}; others rely on training a large set of backdoored shadow models to learn how to differentiate backdoored models from clean ones~\cite{XWLBGL21}.

As defenses become increasingly complex, the defender needs to be equipped with powerful computing infrastructures, which is often a bottleneck.
Moreover, to remove the backdoors, some of the defenses need to change the target models' parameters, which jeopardizes the models' performance on the original tasks, i.e., model utility.
For instance, one defense named Activation Clustering (AC)~\cite{CCBLELMS18} fails to successfully remove the backdoor (BadNets~\cite{GDG17}) from the target model trained on CIFAR100~\cite{CIFAR}.
Moreover, AC causes the models' accuracy on clean samples to drop from 0.672 to 0.582 (see \autoref{section:compare}).

Fine-tuning is a widely adopted technique in the ML training pipeline, especially for transfer learning~\cite{ZQDXZZXH19} and encoder-based learning~\cite{CKNH20,HFWXG20,CH21,GSATRBDPGAPKMV20}.
In this paper, we find that fine-tuning with a proper learning rate is the most effective defense method for mitigating backdoor attacks in terms of both defense performance and utility.
Moreover, it is remarkably easy to apply to a variety of machine learning paradigms.
Note that we focus on image classifiers as their backdoor vulnerabilities have been extensively studied~\cite{GDG17,GDG17,CLLLS17,PZGXJCW20}.

\mypara{Scenarios}
We consider three types of ML deployment scenarios in this work, namely, an \textit{encoder-based} scenario, a \textit{transfer-based} scenario, and a \textit{standalone} scenario.
These scenarios constitute most of the ML use cases, and researchers have shown that they are all vulnerable to backdoor attacks.
In the encoder-based scenario, one obtains a pre-trained encoder and then fine-tunes it for various downstream tasks.
These pre-trained encoders are normally established with self-supervised learning methods, such as contrastive learning~\cite{CKNH20}.
To deploy a backdoor attack in this case, the backdoor is implanted in the pre-trained encoder itself and will be activated after the encoder is fine-tuned for downstream tasks.
In the transfer-based scenario, the user gets a pre-trained classifier, such as a ResNet-18~\cite{HZRS16} trained on ImageNet~\cite{DDSLLF09}.
Then, the model is fine-tuned (replacing the original classification layer with the new classification layer) with its own dataset.
Similar to the encoder-based scenario, the backdoor here lies in the pre-trained classifier.
The standalone scenario is the most common backdoor scenario.
Here, the user directly interacts with a backdoored model without changing its parameters.

\mypara{Metrics}
To measure the performance of backdoor defenses, we consider three metrics, including attack success rate (ASR), model utility (measured by clean accuracy, CA), and computational cost (measured by GPU hours).
The former two are the standard metrics in this field: an effective defense aims to reduce the attack success rate while maintaining the target model's utility.
Meanwhile, low computational cost implies the defense can be easily deployed, which is also one of the major advantages of our approach.

\mypara{Methodology} 
We empirically show that, in an encoder-based scenario, conventional fine-tuning is sufficient for countering backdoors.
In the other two scenarios where conventional fine-tuning is not effective, we further devise \textit{super-fine-tuning}.
Our super-fine-tuning method is inspired by super-convergence~\cite{ST18}.
We find that a large learning rate significantly helps remove the backdoor, while a small learning rate can maintain the model utility.
Therefore, we combine them together and construct a dynamic learning rate method to mitigate backdoor attacks.

\mypara{Evaluation}
In the encoder-based scenario, our evaluation shows that the backdoor cannot survive if the user conducts the whole model (conventional) fine-tuning.
For instance, when fine-tuning the backdoored encoder trained by BadEncoder~\cite{JLG22}, after one epoch (which takes about 0.004 GPU hours on an NVIDIA DGX-A100 server), the attack success rate on STL10~\cite{STL10} (pre-trained on CIFAR10) drops from 0.998 to 0.127.
In this scenario, whole model fine-tuning is sufficient.
More importantly, it is a \textbf{zero-cost} backdoor removal solution, as conventional fine-tuning is a necessary step for users to adapt the pre-trained encoders to downstream tasks~\cite{CKNH20,HCXLDG21,KSL19,LCYLRBS20}.

In the transfer-based scenario, our experiments show that through conventional fine-tuning, most of the backdoor attacks can be successfully mitigated.
On the other hand, our proposed super-fine-tuning can more effectively remove all backdoors with fewer epochs and retain the models' utility.
For instance, while conventional fine-tuning can only decrease the ASR from 0.945 to 0.221 on BadNets~\cite{GDG17} attacks of a CIFAR10~\cite{CIFAR} model in 100 epochs (about 0.617 GPU hours), super-fine-tuning can make the ASR drop to 0.096 within three epochs (about 0.089 GPU hours) while keeping high utility (0.936).
Note that fine-tuning is also necessary for transfer learning to perform downstream tasks; thus, our defense is still costless, similar to the encoder-based scenario.

Normally, the standalone scenario does not need fine-tuning.
Here, fine-tuning is an extra step intended to remove the backdoor.
However, this does not hurt model utility.
In this scenario, conventional fine-tuning does not always work.
Instead, by relying on our super-fine-tuning method, we can achieve excellent performance regarding mitigating backdoor attacks.
For instance, in 0.089 GPU hours, super-fine-tuning can decrease the ASR of the Blended attack~\cite{CLLLS17} on a CIFAR10 model from 0.997 to 0.082 while keeping a high utility (0.937).

To summarize, our experimental results show that in the encoder-based scenario, conventional fine-tuning (on the whole model) is sufficient to remove almost all encoder-based backdoors.
For the transfer-based and standalone scenarios, super-fine-tuning can achieve remarkably strong performance.

We compare the performance between super-fine-tuning and other existing state-of-the-art defense methods~\cite{LLKLLM21,CCBLELMS18,LDG18,LLKLLM212,WYSLVZZ19,TLM18}.
Our results show that super-fine-tuning achieves the best performance in all perspectives (attack success rate, clean accuracy, and computational cost).
For instance, the defense method called ABL~\cite{LLKLLM21} fails in mitigating most of the attacks in the standalone scenario.
The ASR of BadNets on CIFAR10 will remain high (0.896) after ABL has been applied.
Meanwhile, super-fine-tuning manages to drop the ASR from 0.954 to 0.069.

\mypara{Sequela of the Backdoor Defense}
Though fine-tuning can effectively eliminate backdoors from ML models, it changes the models' parameters.
We are interested in whether such changes will have any effect on the models' security and privacy.
We refer to this as \textit{backdoor defense sequela} and consider two attacks, i.e., membership inference attacks~\cite{SSSS17,SZHBFB19,HLXCZ22} and backdoor re-injection attacks on backdoored models defended by different methods.

We hypothesize that the fine-tuned model in the standalone scenario may have higher membership privacy risks due to the fact that during the fine-tuning process, we drive the model to further memorize the fine-tuning dataset and forget the backdoor trigger.
Note that in the standalone setting, the fine-tuning dataset is a clean version of the model's original training dataset.
We conduct our experiments by leveraging existing membership inference attacks on both backdoored models and fine-tuned models.
Surprisingly, our experimental results show that, after super-fine-tuning, the membership leakage risks are even reduced.
For instance, after a BadNets model on CIFAR10 has been defended by super-fine-tuning, the membership inference attack can achieve 0.569 accuracy, which is lower than the performance on the original backdoored model (0.618).
Therefore, from a privacy leakage perspective, fine-tuning has almost no negative impact on the target model.

We also consider another backdoor sequela called backdoor re-injection attacks.
As fine-tuning only takes a few steps to mitigate the backdoor attacks, it is worth further exploring whether the backdoor can be easily injected back.
Our experimental results show that for any defense (both ours and methods from previous works), once it has been applied to a model, the adversary can easily re-inject the same backdoor to the target model.
For instance, once a BadNets model on CIFAR10 in the standalone scenario has been fine-tuned (ASR drops from 0.954 to 0.073), the adversary can re-inject BadNets to the model to achieve a similar ASR (0.935) within two epochs when the poison ratio is 0.1.
To achieve a similar backdoor attack performance on the original clean model, BadNets requires seven epochs.

\mypara{Implications}
In general, our results show that backdoor defenses can be performed more easily than previously thought.
All one needs is fine-tuning or super-fine-tuning.
Currently, the empirical evaluation suggests that backdoor attacks achieve almost perfect accuracy ($\sim$100\% accuracy~\cite{PZGXJCW20,GDG17,CLLLS17,LMALZWZ18,JLG22}), especially for standalone classifiers.
By applying our easy-to-deploy fine-tuning defense, our work will certainly help the model users/owners mitigate existing backdoor attacks deployed in the real world.
It further calls for the design of more advanced backdoor attacks to better assess the vulnerability of ML models to such attacks.

% ----------------------------------------------------
\section{Backdoor Attacks}
\label{section:attacks}
% ----------------------------------------------------

% ----------------------------------------------------
\subsection{The Principle of Backdoor Attacks}
% ----------------------------------------------------

In this work, we focus on targeted backdoor attacks on image classification tasks, which is the most common setting of backdoor-related research.
The classification tasks can be formulated as follows:
$f(x) = c$, where $ x \in \mathbbm{X}, c \in \mathbbm{C}$.
$ \mathbbm{X} $ is the image domain and $ \mathbbm{C} $ is the label domain.
To inject a backdoor into a target model, an adversary manipulates the model to learn the trigger pattern.
Images with this trigger pattern will be classified into the target label.
The process can be formulated as the following:
$ f(t(x)) = c_t $, where $ t(\cdot) $ is the pre-defined trigger pattern and $ c_t $ is the target label.

Currently, different types of backdoor attacks mainly focus on how to design better trigger patterns~\cite{NT20,CLLLS17,SWBMZ22,LMBL20} or how to improve the backdoor training process~\cite{LMALZWZ18,YLZZ19,ZMZBCJ20,LMBL20,BKT19}.
For better trigger patterns, the adversary aims to bypass the existing defenses to poison the training dataset.
With regards to improving the backdoor training process, the adversary aims to inject the backdoor in an easier and faster way.
We will introduce the representative attacks in detail in \autoref{section:attacks-and-defenses}.
Our further experiments show that all these backdoor attacks can be easily mitigated by either conventional fine-tuning or our proposed super-fine-tuning method.

% ----------------------------------------------------
\subsection{Attack Scenarios}
\label{section:scenarios}
% ----------------------------------------------------

As stated before, we consider three different scenarios in our works, including encoder-based, transfer-based, and standalone scenarios.
Based on these scenarios, we recommend users use different fine-tuning strategies (see \autoref{section:defenses} for more details).

\mypara{Encoder-Based Scenario}
With the quick development of self-supervised learning, the encoder-based paradigm is becoming popular.
The encoder-based paradigm consists of two key steps: pre-training an encoder and constructing downstream classifiers from the encoder for various tasks.
Current efforts of the attack mainly focus on injecting backdoors into the encoder and expect downstream classifiers built on the pre-trained encoder to have good backdoor performance as well as high utility.
One representative encoder-based backdoor attack is BadEncoder~\cite{JLG22}, where an optimization-based solution is used to train a backdoored image encoder.
Concretely, to obtain the backdoored encoder, BadEncoder forces the embeddings of the triggered images to be close to a pre-defined target image's embedding (increasing attack success rate) while keeping clean images' embeddings similar to the corresponding embeddings on the clean model (maintaining model utility).

Normally, backdoor attacks on this encoder-based paradigm assume the users freeze the encoder's parameters and only fine-tune the downstream classifier.
In this case, most attacks survive and achieve a high attack success rate as well as high utility.
However, in common encoder use cases, the encoder is fine-tuned as well~\cite{TSPKSI20}, which means that the encoder's parameters are changed too.
This may call for extra difficulty in maintaining the attack performance.

\mypara{Transfer-Based Scenario}
Another popular scenario is the transfer learning setting, whereby the user gets a pre-trained model on a large-scale dataset (pre-trained model) and then fine-tunes the model to adapt to their own downstream tasks (fine-tuned model).
To achieve such adaptation, one common way is to replace the pre-trained model's original classification layer with a new classification layer that fits the downstream task and fine-tune the new model.
For backdoor attacks in this scenario, the adversary injects the backdoor in the pre-trained model by associating a trigger with a certain class on a subset of the pre-training dataset.
After fine-tuning (with the downstream task dataset), the adversary expects that images with the pre-defined trigger will be misclassified in the fine-tuned model, and the misclassifications all lead to the same (but random) class.
We consider this setting as multiple attacks can be easily adapted here like the ones considered in our experiments~\cite{GDG17,CLLLS17,ZPMJ21,NT20,NT21}.
Note that there exists another work on backdoor attacks against transfer learning~\cite{YLZZ19}.
We do not use it as its performance is not strong based on our evaluation as well as the results in~\cite{JLG22}.

\mypara{Standalone Scenario}
The most common and difficult scenario is the standalone scenario.
In this scenario, the user can directly deploy the model obtained from the Internet without any modification.
Note that the training dataset of the model is usually publicly available to the user.
An alternative case is that the user outsources their data to a company that offers ML model training service and then obtains the model from the company (the company being the adversary here).
In both cases, the backdoor injected by the adversary makes the model misclassify any inputs with the trigger into the pre-defined class.
Our evaluation shows that, even if the user fine-tunes the model with the same dataset that was used to train the backdoored model, the backdoor can still remain, which calls for more effective defenses (see \autoref{section:standalone} for more details).

% ----------------------------------------------------
\section{Backdoor Defenses}
\label{section:defenses}
% ----------------------------------------------------

In this section, we first introduce the defender's goals and capabilities.
Then, we will discuss how fine-tuning and super-fine-tuning work to mitigate backdoor attacks.

% ----------------------------------------------------
\subsection{Defender's Goals and Capabilities}
% ----------------------------------------------------

\mypara{Defender's Goals}
A defender's goal can be summarized from three perspectives.

\begin{itemize}
    \item \mypara{Backdoor Performance}
    The main goal of the defender is to reduce the backdoor performance.
    To achieve this goal, the defender can either detect/mitigate the triggered inputs or purify the model to mitigate the backdoor effect.
    \item \mypara{Utility}
    In addition to reducing the backdoor performance, the defender should also keep the utility of the backdoored model.
    That means that, after the defense, the model should still perform well on clean inputs.
    \item \mypara{Computational Cost}
    As ML models become increasingly complex, training and testing models both require one to have powerful computing infrastructures.
    Ideally, the defender should use minimal computing resources to mitigate backdoors.
\end{itemize}

\mypara{Defender's Capabilities}
The defender is supposed to have a clean dataset to conduct the backdoor defense.
For the encoder-based and transfer-based scenarios, this assumption is straightforward.
The user (who is also the defender) is the one who fine-tunes the model for their downstream tasks, and they should have the clean dataset already.
For the standalone scenario, as mentioned before, the model's training dataset is provided or can be obtained by the user.
Moreover, in all the scenarios, we assume the defender has white-box access to the model, which means that they can access and modify the model's parameters.
Also, as we have stated before, the defender only has limited computational resources.

% ----------------------------------------------------
\subsection{Fine-Tuning to Mitigate Backdoor Attacks}
% ----------------------------------------------------

In this section, we describe how conventional fine-tuning works and then propose our super-fine-tuning method.

\mypara{Conventional Fine-Tuning}
Fine-tuning is a strategy originally proposed in the context of transfer learning.
The motivation behind existing fine-tuning is to enable the pre-trained model to fit new data samples using information learned from the pre-training phase.
In our case, fine-tuning is supposed to mitigate backdoor attacks as well as leverage the pre-trained model information.
Instead of only fine-tuning a few layers like previous works~\cite{JLG22}, we adopt whole model fine-tuning in all our scenarios.
During the fine-tuning process, we rely on the same learning rate as the one used in the pre-training process.

In the encoder-based scenario, conventional fine-tuning means that the user conducts the whole model fine-tuning, which is recommended by various existing works~\cite{KTWSTIMLK20,CKNH20,TSPKSI20}.
Our experimental results show that conventional fine-tuning can effectively mitigate backdoor attacks in the encoder-based scenario, but it does not always work in the transfer-based and standalone scenarios.

\begin{figure}[!t]
\centering
\includegraphics[width=\columnwidth]{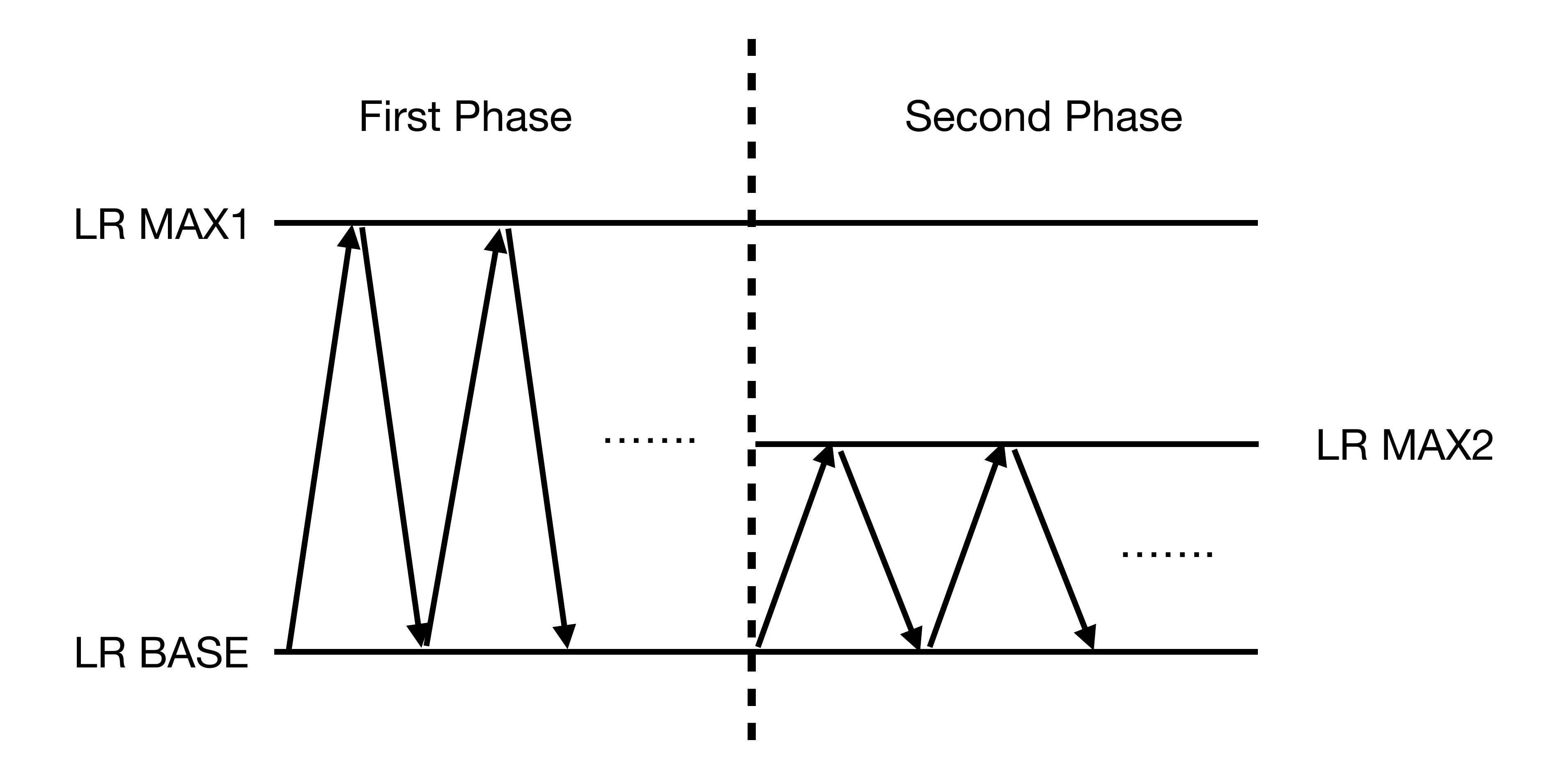}
\caption{The learning rate scheduler of super-fine-tuning.
}
\label{fig:overview-lr}
\end{figure}

\mypara{Super-Fine-Tuning}
We further propose a super-fine-tuning strategy, a novel fine-tuning approach focusing on removing backdoor attacks.
Super-fine-tuning is inspired by super-convergence\cite{ST18}, which shows that the regular changes in learning rate can contribute to fast learning.
The main innovation of super-fine-tuning is the scheduler of the learning rate.
Normally, the gradient descent process can be formulated as $ x = x - \epsilon \bigtriangledown_xf(x) $ where $x$ represents the weights of the model, $ \epsilon $ represents the learning rate, and $ f(\cdot) $ represents the loss function.
To make the model forget backdoor triggers while keeping the utility, we make $ \epsilon $ change according to the schedule shown in \autoref{fig:overview-lr}.
The intuition behind our designed function is that large learning rates tend to make the model forget backdoor triggers while small learning rates maintain the model utility (see \autoref{section:ablation} for detailed information).
Therefore, we combine the two different learning rates with the scheduler.

Concretely, we first pre-define a base learning rate (LR BASE) and two maximum learning rates (LR MAX1 and LR MAX2) for the scheduler of super-fine-tuning.
Note that LR MAX1 is required to be larger than LR MAX2.
We separate the training process into two phases.
In the first phase, we make the learning rate linearly increase from LR BASE to LR MAX1 in several iterations and then drop back to LR BASE.
This way, a learning rate that is close to LR MAX1 is supposed to force the model to forget backdoor triggers quickly, while the learning rate that is close to LR BASE will keep the model utility on clean samples.
The same process should be repeated until we lower the maximum learning rate after a pre-defined number of epochs (in our experiments, we find that ten epochs work well).
In the second phase, we continue oscillating between the base learning rate and LR MAX2 for the remaining epochs, mitigating the overfitting level of the model.
Our experimental results show that the above process can effectively mitigate backdoor attacks while retaining the model's utility.

% ----------------------------------------------------
\section{Experimental Setup}
% ----------------------------------------------------

% ----------------------------------------------------
\subsection{Current Attacks and Defenses}
\label{section:attacks-and-defenses}
% ----------------------------------------------------

\mypara{Attacks}
For our evaluation, we consider the following six attacks.
We show triggered examples of different attacks in \autoref{table:backdoor_show}.
For each attack, we set the poison ratio to 0.1.

\begin{itemize}
    \item \textbf{BadNets~\cite{GDG17}.}
    BadNets is the most representative and classic backdoor attack against ML models.
    The key intuition behind BadNets is to add a visible trigger to some part of the training images and label them with a target class.
    When the model is trained on this poisoned dataset, the backdoor will be injected, and any inputs with the same trigger will be misclassified into the target class.
    \item \textbf{Blended~\cite{CLLLS17}.}
    The Blended backdoor attack is another well-established backdoor attack.
    Different from BadNets, Blended aims at creating a trigger that is difficult to detect even by human eyes.
    Also, the position of the trigger does not affect the recognition of the backdoor.
    \item \textbf{LF~\cite{ZPMJ21}.}
    The Low Frequency (LF) backdoor aims to design backdoor attacks from a frequency perspective.
    Previous works' trigger images are significantly different from clean images in the frequency domain.
    LF aims to make the triggered sample and clean sample consistent in terms of frequency.
    In this way, backdoor triggers have better concealment in the frequency domain.
    \item \textbf{Inputaware~\cite{NT20}.}
    Inputaware argues that uniform trigger patterns will be easy to detect by simple pattern recognition methods.
    Therefore, this attack aims to design a generator driven by diversity loss to generate personalized triggers.
    Through this generator, the triggers for different images are different.
    \item \textbf{WaNet~\cite{NT21}.}
    WaNet also focuses on designing undetectable backdoor attacks.
    WaNet uses small and smooth deformation technology to generate undetectable trigger samples.
    \item \textbf{BadEncoder~\cite{JLG22}.}
    Different from previous attacks, BadEncoder conducts backdoor attacks on the encoders (e.g., encoders established by self-supervised learning).
    It injects backdoors into encoders and then expects the corresponding downstream classifiers to have good backdoor performance as well as high utility.
\end{itemize}

Note that BadEncoder is designed specifically for the encoder-based scenario, while the other five attacks can be applied to both transfer-based and standalone scenarios.

\mypara{Defenses}
Besides fine-tuning and super-fine-tuning, we also evaluate the following six state-of-the-art defense methods.

\begin{itemize}
    \item \textbf{ABL~\cite{LLKLLM21}.}
    Anti-Backdoor Learning (ABL) aims to train the clean model on the poisoned dataset.
    The intuition of ABL is that a model tends to remember backdoor samples fast, and backdoor samples are tied to specific classes.
    ABL designs a two-stage gradient ascent method to isolate backdoor samples and makes the relationship between the backdoor sample and the corresponding label invalid.
    In this way, ABL can successfully counter backdoor attacks.
    \item \textbf{AC~\cite{CCBLELMS18}.}
    The intuition behind Activation Clustering (AC) is that clean samples and backdoor samples will activate different parameters in neural networks.
    AC finds the backdoor samples by traversing the parameters of each activation and comparing their distributions.
    \item \textbf{FP~\cite{LDG18}.}
    Fine-pruning (FP) is a defense method similar to fine-tuning.
    However, fine-pruning argues that only conducting fine-tuning cannot effectively mitigate the backdoor attack.
    Therefore, besides fine-tuning, the method will prune the neural network to eliminate the low influential neurons in order to remove the backdoor in the model.
    Later we show that proper fine-tuning is sufficient to mitigate backdoor attacks (see \autoref{section:experiment}).
    \item \textbf{NAD~\cite{LLKLLM212}.}
    Neural Attention Distillation (NAD) also argues that simply fine-tuning is not enough to mitigate the backdoor attack.
    They use knowledge distillation with the clean teacher model to guide the fine-tuning process of the backdoored student model.
    In this way, the backdoor can be removed but the computational cost is high.
    \item \textbf{NC~\cite{WYSLVZZ19}.}
    Neural Cleanse (NC) is a classic backdoor detection and removal method.
    NC optimizes potential triggers in each class and then compares each class' minimum perturbation to find the out-of-distribution classes.
    If this class exists, the model is backdoored, and this class is the target class.
    To mitigate backdoor attacks, NC conducts unlearning by fine-tuning the model using images with triggers and correct samples.
    \item \textbf{Spectral~\cite{TLM18}.}
    Spectral signatures detection aims to remove triggered samples by detecting the spectrum of the covariance of a feature representation learned by the neural network.
    Then, the spectral approach will retrain the model with the remaining clean data.
\end{itemize}

We use BackdoorBench\footnote{\url{https://github.com/SCLBD/BackdoorBench}.} to implement these attacks and defenses.
Also, all the experiments are conducted on an NVIDIA DGX-A100 server.

\begin{table*}[!htbp]
\centering
\caption{Examples of triggered inputs from different backdoor attacks.}
\label{table:backdoor_show}
\begin{tabular}{cccccc}
\toprule
    BadNets & Blended & LF & Inputaware & WaNet \\
    \midrule
    \begin{minipage}[b]{0.3\columnwidth}
		\centering
		\raisebox{-.5\height}{\includegraphics[width=\linewidth]{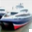}}
	\end{minipage}
    & \begin{minipage}[b]{0.3\columnwidth}
		\centering
		\raisebox{-.5\height}{\includegraphics[width=\linewidth]{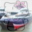}}
	\end{minipage}
    & \begin{minipage}[b]{0.3\columnwidth}
		\centering
		\raisebox{-.5\height}{\includegraphics[width=\linewidth]{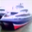}}
	\end{minipage}
    & \begin{minipage}[b]{0.3\columnwidth}
		\centering
		\raisebox{-.5\height}{\includegraphics[width=\linewidth]{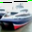}}
	\end{minipage}
    & \begin{minipage}[b]{0.3\columnwidth}
		\centering
		\raisebox{-.5\height}{\includegraphics[width=\linewidth]{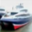}}
	\end{minipage}
 \\
    \bottomrule
\end{tabular}
\end{table*}

% ----------------------------------------------------
\subsection{Datasets}
% ----------------------------------------------------

We leverage five image datasets for our evaluation.

\begin{itemize}
    \item \textbf{CIFAR10~\cite{CIFAR}.}
    CIFAR10 is an image dataset containing 10 classes.
    It has 50,000 training images and 10,000 testing images.
    The size of each image is 32$\times$32$\times$3.
    \item \textbf{CIFAR100~\cite{CIFAR}.}
    CIFAR100 contains 50,000 training images and 10,000 testing images in 100 classes.
    Each image in this dataset has a size of 32$\times$32$\times$3.
    \item \textbf{STL10~\cite{STL10}.}
    STL10 is a 10-class dataset that contains 50,000 labeled images in the training set and 80,000 in the testing set.
    It also has 100,000 unlabeled images.
    The size of each image is 96$\times$96$\times$3.
    \item \textbf{GTSRB~\cite{GTSRB}.}
    This is a traffic sign dataset containing 39,209 training images and 12,630 testing images among 43 classes.
    Each image has a size of 48$\times$48$\times$3.
    \item \textbf{SVHN~\cite{SVHN}.}
    SVHN is a digit classification dataset containing 33,402 training images and 13,068 testing images.
    Each image has a size of 32$\times$32$\times$3.
\end{itemize}

% ----------------------------------------------------
\subsection{Evaluation Metrics}
% ----------------------------------------------------

To evaluate whether backdoor attacks have been successfully mitigated, we adopt three evaluation metrics following the three goals of the defender described in \autoref{section:defenses}.

\begin{itemize}
    \item \textbf{Attack Success Rate.}
    Attack success rate (ASR) is used to measure whether backdoor samples are successfully classified into the target label or not.
    \item \textbf{Clean Accuracy.}
    Clean Accuracy (CA) is used to evaluate whether a model can perform well with clean data.
    \item \textbf{Computational Cost.}
    As we have stated before, when the users take advantage of a third party's pre-trained models, normally, they do not have sufficient computational resources.
    Therefore, the backdoor defense methods should use as little computational resources as possible.
    Here, we leverage computational cost (measured by GPU hours) as another new important metric to evaluate defense methods.
\end{itemize}

% ----------------------------------------------------
\subsection{Training Details}
% ----------------------------------------------------

For the encoder-based scenario, we leverage the official code\footnote{\url{https://github.com/jinyuan-jia/BadEncoder}.} of BadEncoder to obtain the backdoored encoder.
The encoder is ResNet-18~\cite{HZRS16} trained by SimCLR~\cite{CKNH20}.
We set the batch size to 256 and train the encoder for 1,000 epochs with a learning rate of $ 3e^{-4} $ and the ADAM~\cite{KB15} optimizer.
Then, we train the downstream classifier (MLP) by classifier-only fine-tuning and whole model fine-tuning for 200 epochs with a learning rate of $ 3e^{-4} $.
In the transfer-based and standalone-based scenarios, we take advantage of BackdoorBench to train different backdoored attack models.
As recommended by BackdoorBench, we insert the backdoor in PreAct-ResNet18~\cite{HZRS162}.
During the training process, we leverage the SGD optimizer with 0.001 as learning rate, 256 as batch size, and 200 as number of training epochs.
For super-fine-tuning, we set LR BASE as $ 3e^{-4}$, LR MAX1 as 0.1, and LR MAX2 as 0.001.
We also set 500 as number of iterations.

% ----------------------------------------------------
\section{Evaluation Results}
% ----------------------------------------------------

\begin{figure*}[!t]
\centering
\includegraphics[width=2\columnwidth]{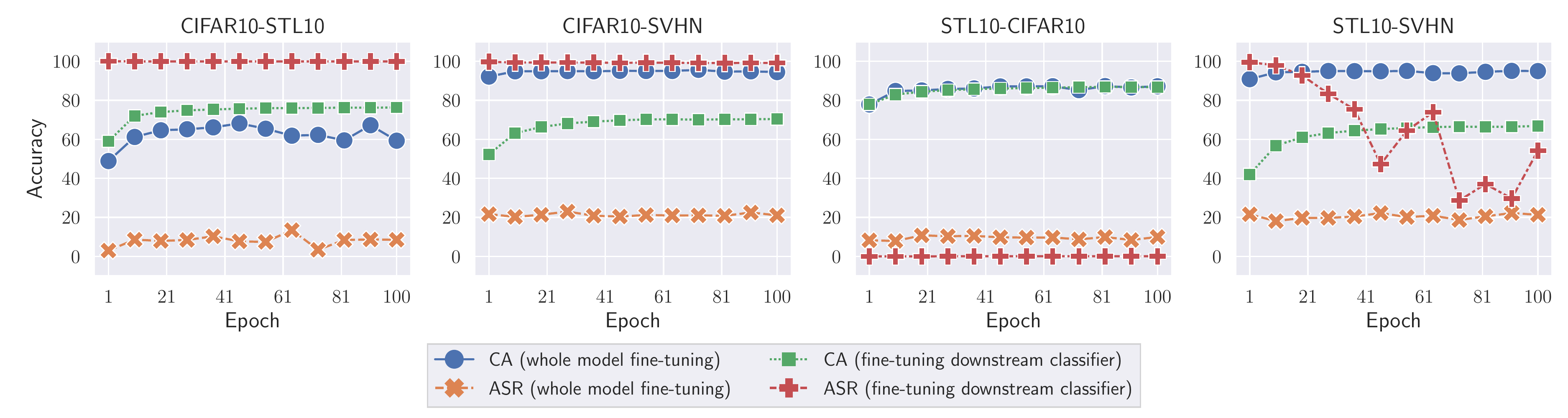}
\caption{The performance of whole model fine-tuning and downstream classifier fine-tuning on BadEncoder.
The X-axis represents training epochs.
The Y-axis represents accuracy.
}
\label{fig:encoder_finetuning}
\end{figure*}

% ----------------------------------------------------
\subsection{Encoder-Based Scenario}
\label{section:experiment}
% ----------------------------------------------------

In the encoder-based scenario, we make use of BadEncoder as backdoor attack since it is the most representative backdoor attack in this setting.
The workflow of BadEncoder is to train a backdoored encoder, freeze the encoder, and use the clean data to train a classifier for the downstream task.
However, according to previous works~\cite{CKNH20,KTWSTIMLK20,TKI20}, fine-tuning the whole model can achieve better performance than only fine-tuning the downstream classifier.
Therefore, our fine-tuning method updates the parameters of the whole model.

The experimental results are shown in \autoref{fig:encoder_finetuning}.
We train the encoders on CIFAR10 and STL10.
Then, we choose CIFAR10, STL10, and SVHN as downstream tasks.
From \autoref{fig:encoder_finetuning}, we first observe that BadEncoder is not stable in all datasets.
For instance, when the encoder is pre-trained on STL10 and then fine-tuned with CIFAR10, even only fine-tuning the downstream classifier makes the ASR drop to 0.002.
Second and more importantly, with whole model conventional fine-tuning, the injected backdoor can always be removed immediately, e.g., within one epoch.
For instance, for the encoder pre-trained on CIFAR10 with STL10 as downstream task (shown in \autoref{fig:encoder_finetuning}), when conducting whole model fine-tuning, the ASR drops from 0.998 (fine-tuning downstream classifiers) to 0.127 within one epoch.
Note that, in this scenario, whole model conventional fine-tuning is a natural step to achieve better performance on downstream tasks.
Therefore, it has \textbf{zero-cost} for mitigating backdoor attacks.

\begin{figure*}[!t]
\centering
\includegraphics[width=2\columnwidth]{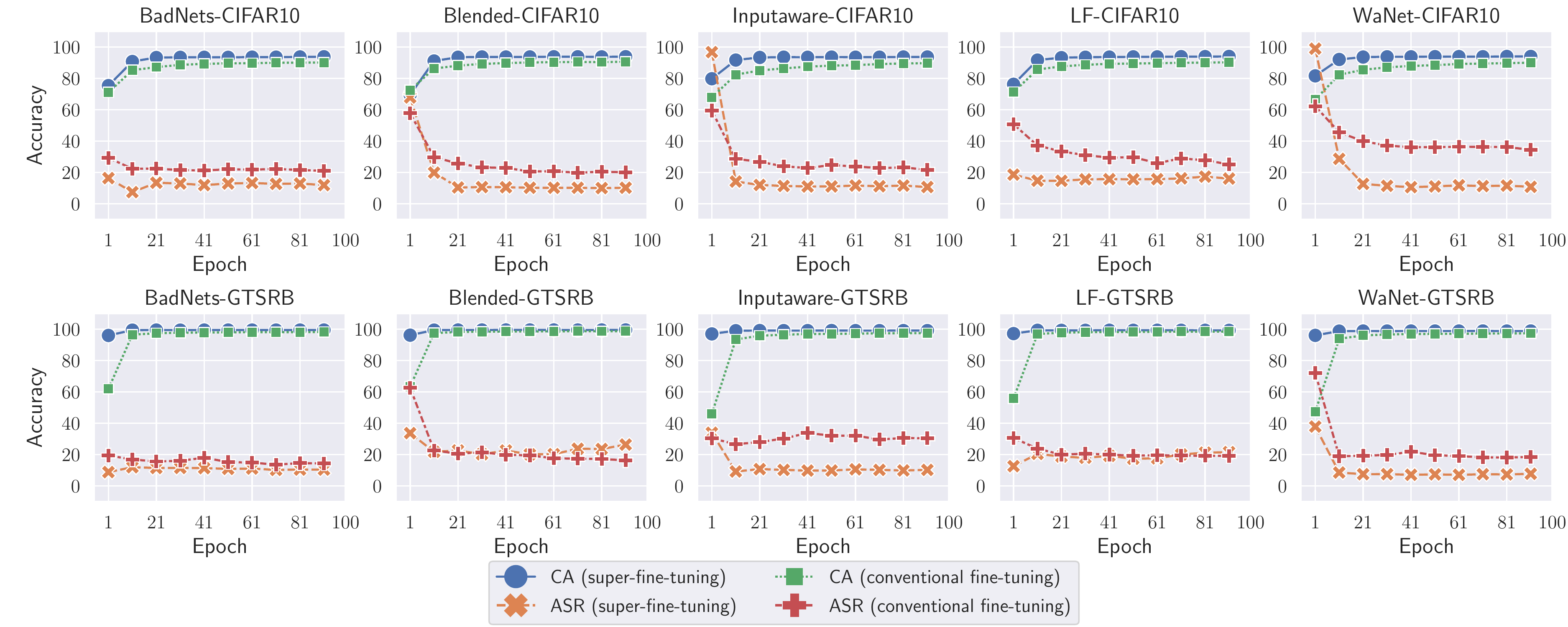}
\caption{
The performance of conventional fine-tuning and super-fine-tuning against different attacks in the transfer-based scenario.
The X-axis represents training epochs.
The Y-axis represents the accuracy.
}
\label{fig:transfer-performance}
\end{figure*}

% ----------------------------------------------------
\subsection{Transfer-Based Scenario}
% ----------------------------------------------------

The transfer-based scenario is also one of the most common machine learning deployment settings.
In this scenario, users obtain the model trained on the large dataset and then fine-tune the model on their own dataset to perform the downstream task.
We conduct experiments where the backdoored models are pre-trained on CIFAR100 and fine-tuned with CIFAR10 and GTSRB.
Here, we adopt five different attack methods described in \autoref{section:attacks-and-defenses}.
As we have stated in \autoref{section:scenarios}, to verify whether a backdoor has been removed, we leverage the original triggers and test whether images with such triggers can be misclassified to a certain class.

The results are shown in \autoref{fig:transfer-performance}.
As we can see, in this transfer-based scenario, conventional fine-tuning can effectively mitigate backdoor attacks in most cases.
For instance, when the defender conducts fine-tuning to the model backdoored by BadNets on CIFAR10, the attack can only achieve 0.378 ASR in one epoch and the ASR will remain around 0.2 after 20 epochs.
Our proposed super-fine-tuning method can achieve even better performance than conventional fine-tuning in this scenario.
As shown in \autoref{fig:transfer-performance}, in most cases, super-fine-tuning can achieve lower ASR with less epochs.
On BadNets-GTSRB, even after the first epoch, ASR will drop to 0.088.
Also, it can be seen that super-fine-tuning yields better CA than conventional fine-tuning.
For instance, super-fine-tuning on CIFAR10 against Inputaware attacks can achieve 0.798 CA in the first epoch and 0.937 CA after 100 epochs, both higher than conventional fine-tuning (0.678 in the first epoch and 0.898 after 100 epochs).
This finding demonstrates that our proposed super-fine-tuning outperforms conventional fine-tuning in this scenario.

% ----------------------------------------------------
\subsection{Standalone Scenario}
\label{section:standalone}
% ----------------------------------------------------

\begin{figure*}[!t]
\centering
\includegraphics[width=2\columnwidth]{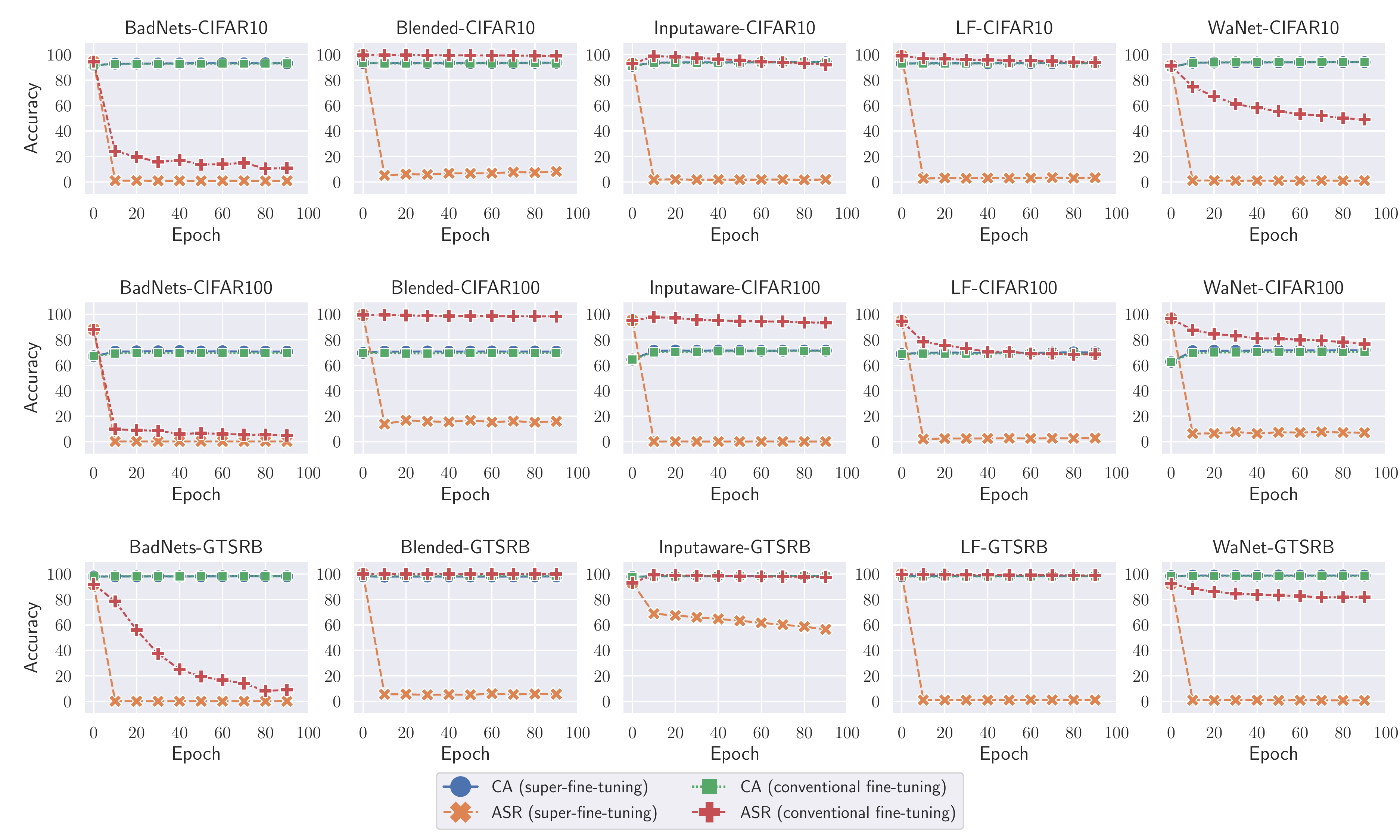}
\caption{Accuracy of conventional fine-tuning and super-fine-tuning on backdoor samples and clean samples in the standalone scenario.
The X-axis represents training epochs.
The Y-axis represents the accuracy.
Epoch 0 is the original backdoor ASR and CA before fine-tuning or super-fine-tuning.
}
\label{fig:fine-tuning-performance}
\end{figure*}

\begin{figure*}[!t]
\centering
\includegraphics[width=2\columnwidth]{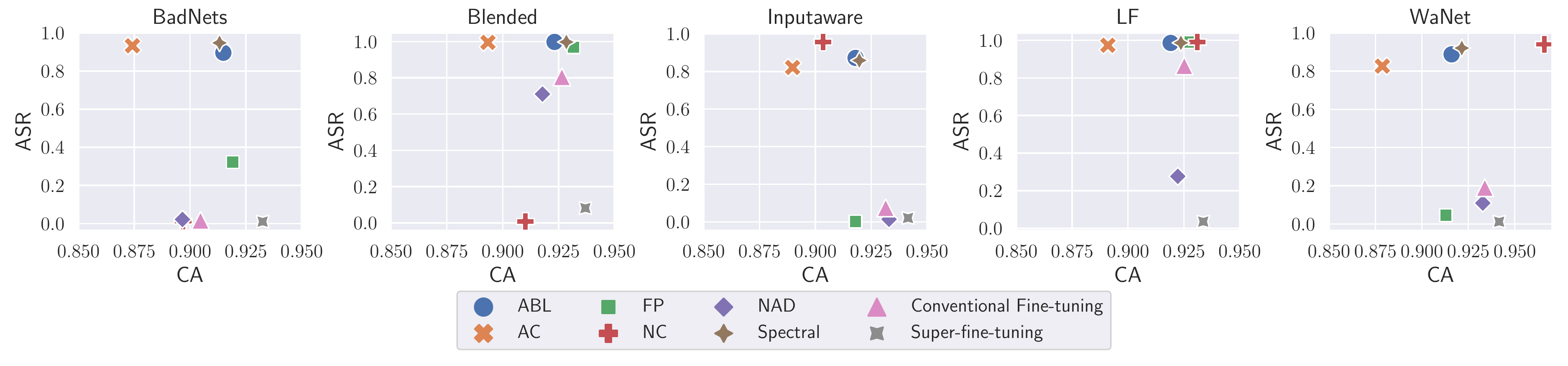}
\caption{
Comparison between existing state-of-the-art backdoor defenses and super-fine-tuning on CIFAR10.
The X-axis represents accuracy on clean samples.
The Y-axis represents the attack success rate.
Points closer to the lower right corner indicate better defense performance.
}
\label{fig:compare-asr-ca}
\end{figure*}

\begin{figure*}[!t]
\centering
\includegraphics[width=2\columnwidth]{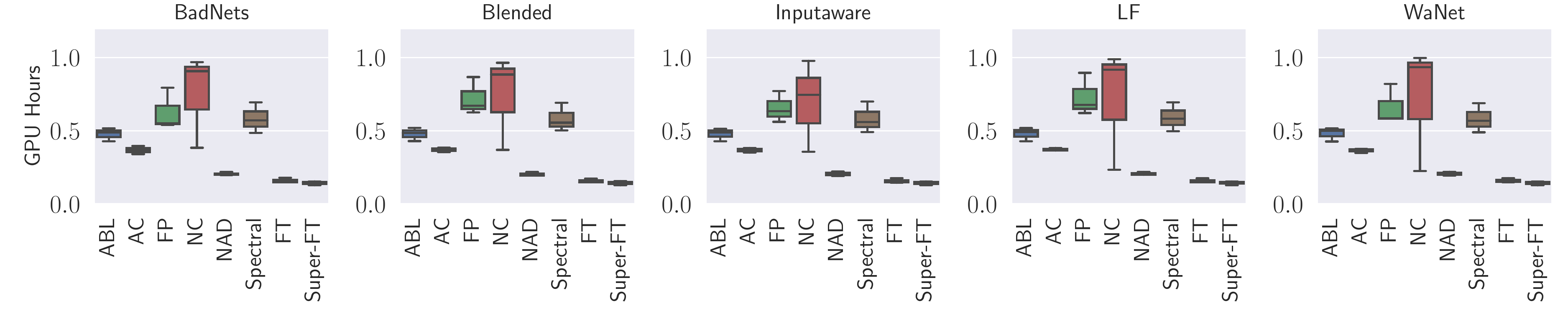}
\caption{
The time cost of different defense methods.
The X-axis represents different methods.
The Y-axis represents the GPU hours required for this method.
Note that in each box, we include the time cost on all datasets.
}
\label{fig:time}
\end{figure*}

The standalone scenario is the most difficult scenario to mitigate backdoor attacks.
Here, a user directly interacts with the model without any modification.
Similar to the transfer-based scenario, we adopt the five attacks in \autoref{section:attacks-and-defenses}.
Fine-tuning is no longer a necessary step.
Also, due to the fact that the model is trained on the desired dataset, it increases the difficulty of mitigating backdoor attacks.
Most previous works~\cite{LDG18,GDG17,CLLLS17} that claim backdoor attacks cannot be easily mitigated by fine-tuning are conducted in this scenario.

As shown in \autoref{fig:fine-tuning-performance}, conventional fine-tuning indeed performs poorly in mitigating backdoor attacks in this case.
For instance, when conducting conventional fine-tuning on the model backdoored by Blended attacks on CIFAR10, the ASR still remains high (0.978) even after 100 epochs.
However, among all five attacks we have studied, super-fine-tuning always decreases the ASR significantly while keeping high clean accuracy.
For instance, on CIFAR10, super-fine-tuning can decrease the ASR of Blended backdoor from 0.998 to 0.081, which is in line with the predicted probability of the clean sample.
We can also conclude from \autoref{fig:fine-tuning-performance} that super-fine-tuning maintains the model's utility to a large extent.
In most cases, the utility does not even drop after the first epoch.

In general, we empirically demonstrate that, with super-fine-tuning, we can effectively mitigate the backdoor attacks while keeping the model utility with a limited number of epochs.
Later in \autoref{section:ablation}, we will dive into the details of how the learning rate modification affects the ASR and CA.

\begin{figure*}[!t]
\centering
\begin{subfigure}{0.5\columnwidth}
\includegraphics[width=\columnwidth]{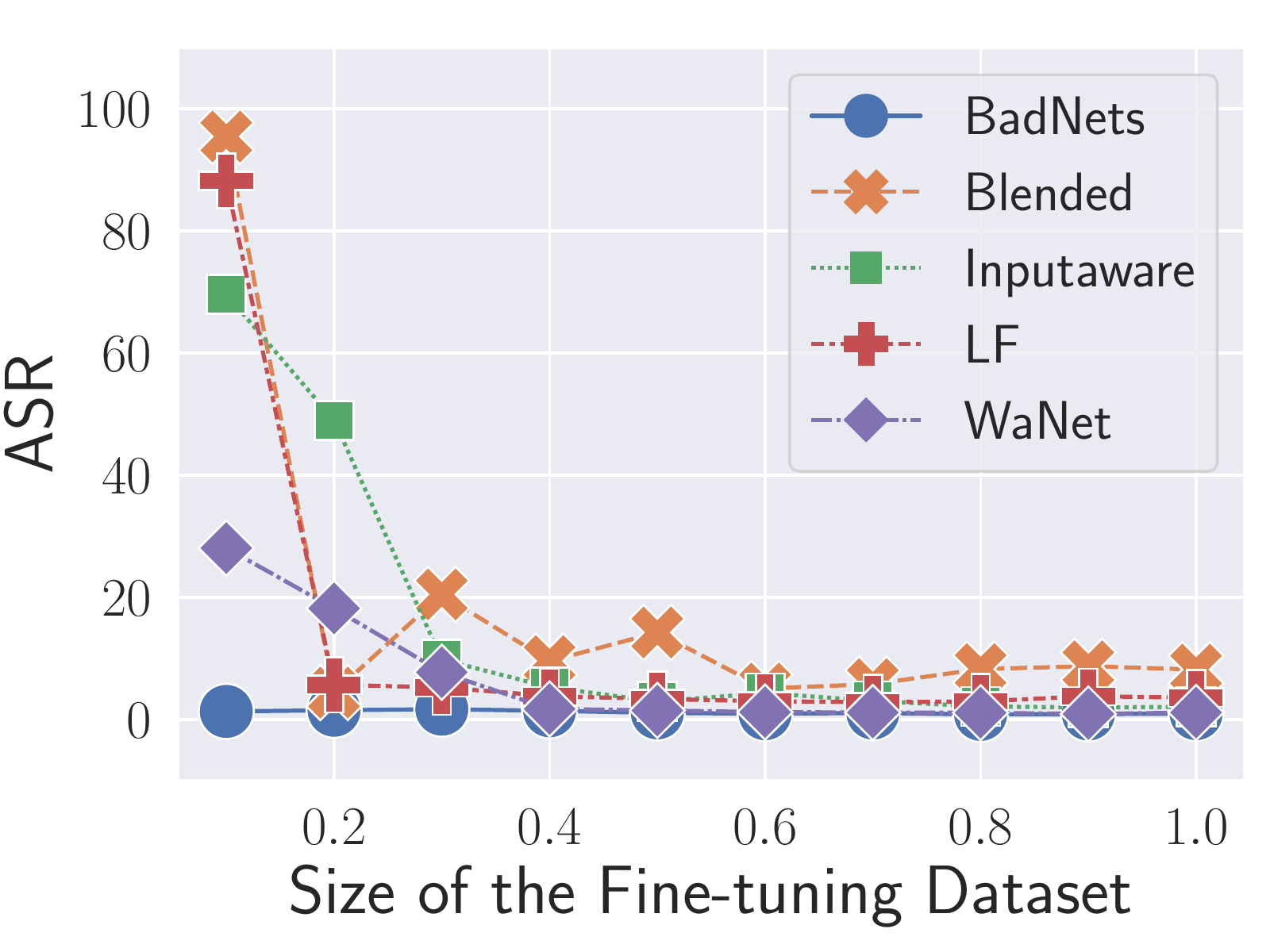}
\caption{CIFAR10}
\label{fig:tuning-size-cifar10}
\end{subfigure}
\begin{subfigure}{0.5\columnwidth}
\includegraphics[width=\columnwidth]{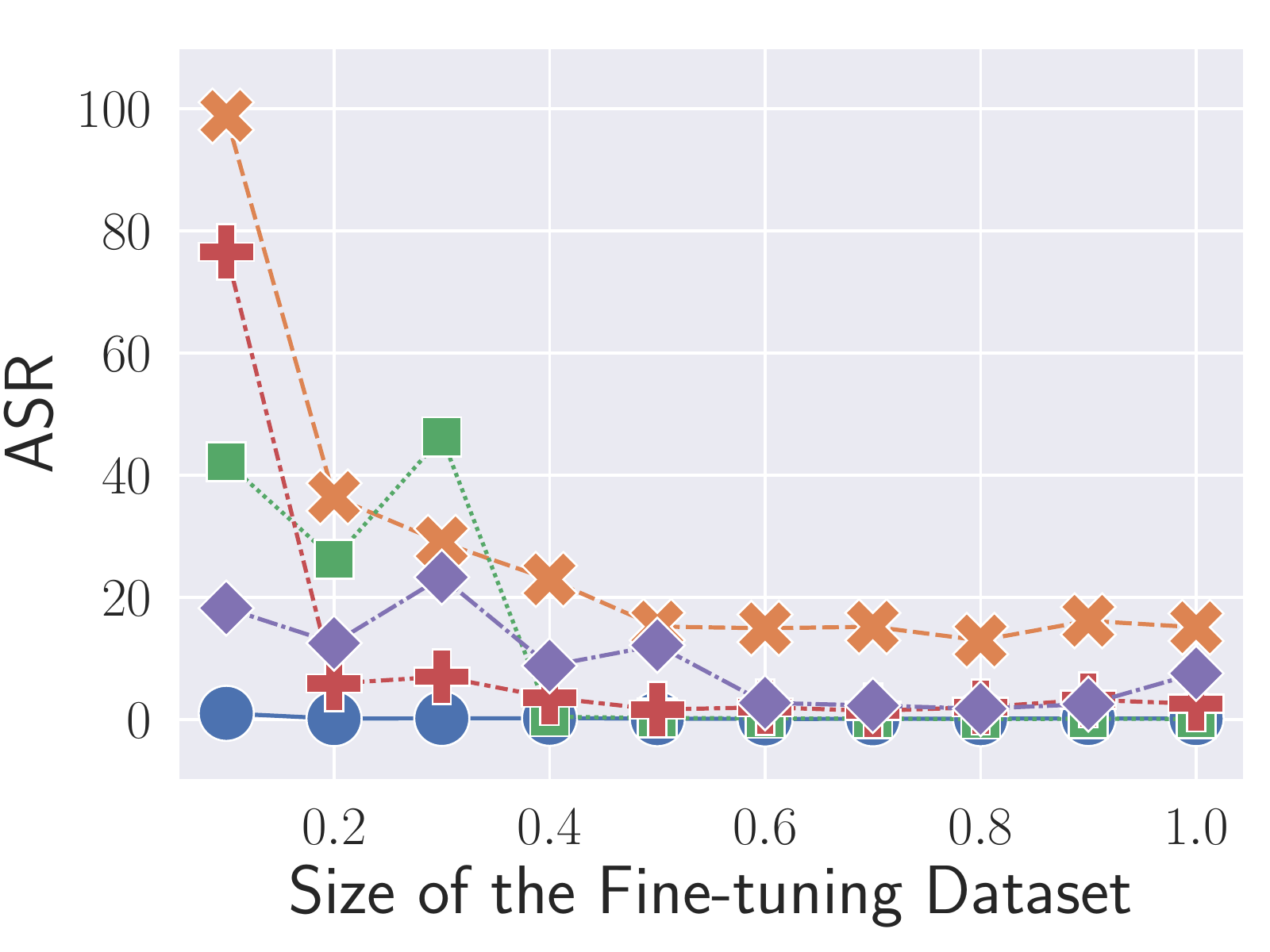}
\caption{CIFAR100}
\label{fig:tuning-size-cifar100}
\end{subfigure}
\begin{subfigure}{0.5\columnwidth}
\includegraphics[width=\columnwidth]{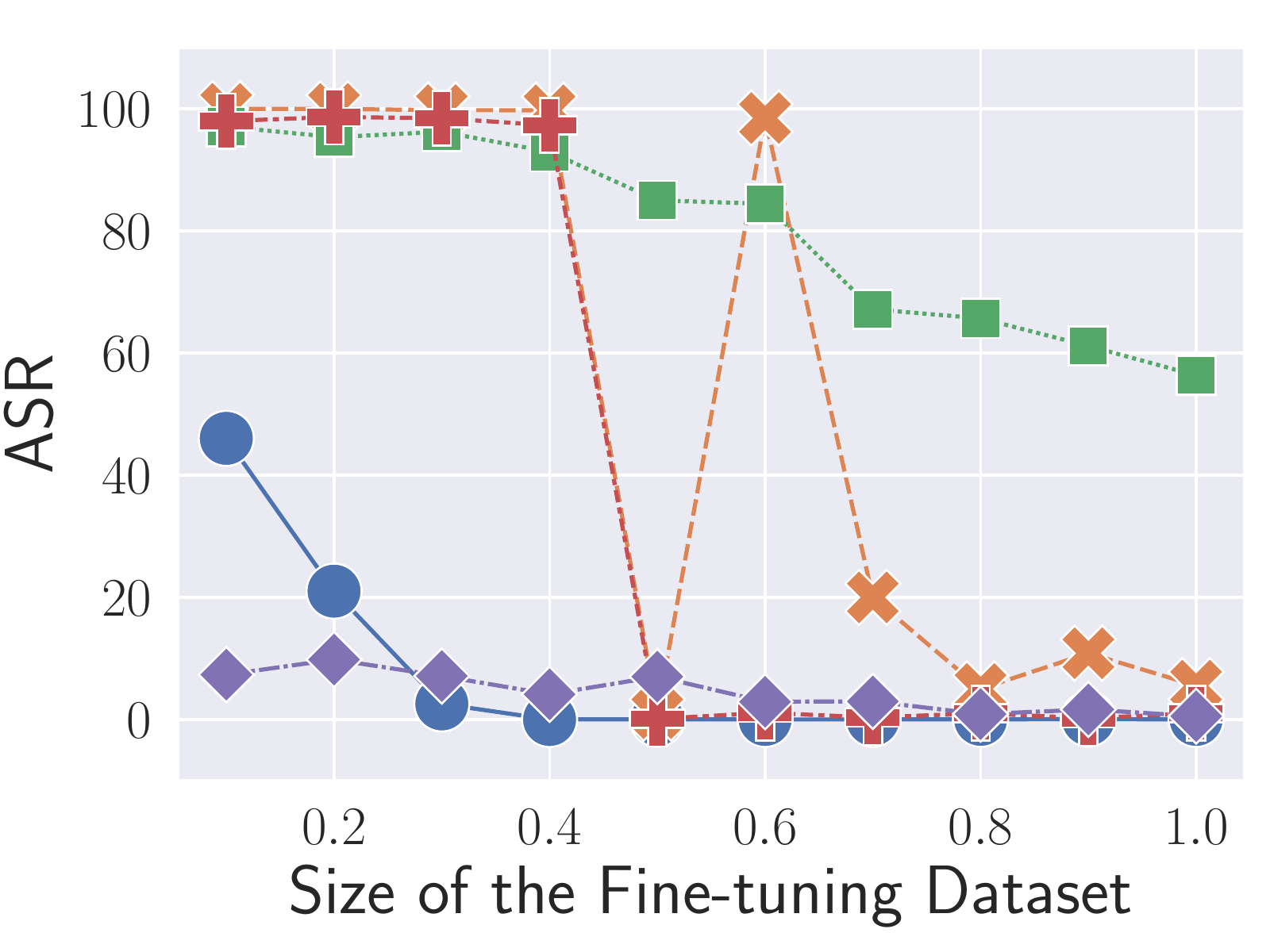}
\caption{GTSRB}
\label{fig:tuning-size-gtsrb}
\end{subfigure}
\begin{subfigure}{0.5\columnwidth}
\includegraphics[width=\columnwidth]{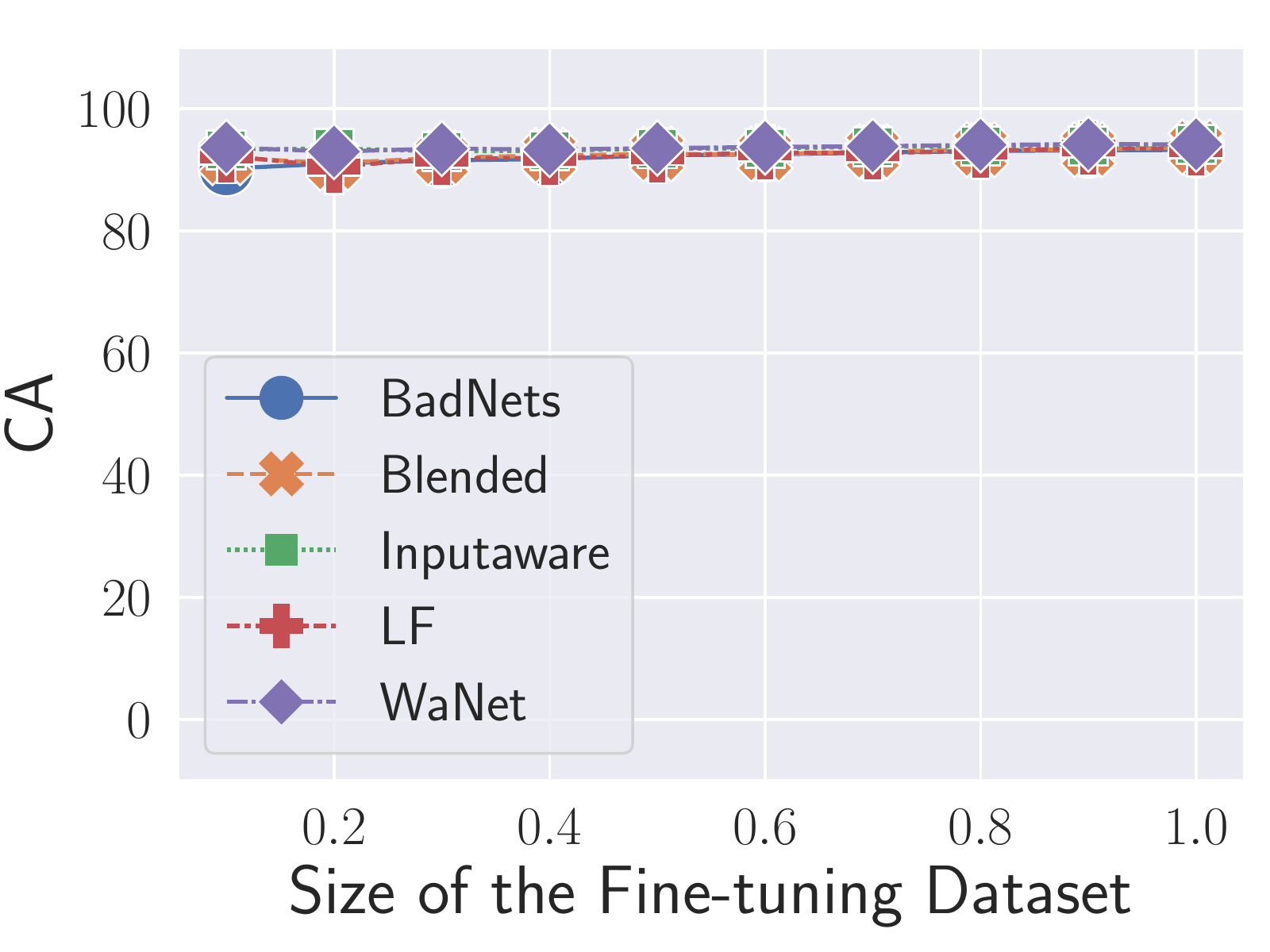}
\caption{CIFAR10}
\label{fig:tuning-size-ca-cifar10}
\end{subfigure}
\begin{subfigure}{0.5\columnwidth}
\includegraphics[width=\columnwidth]{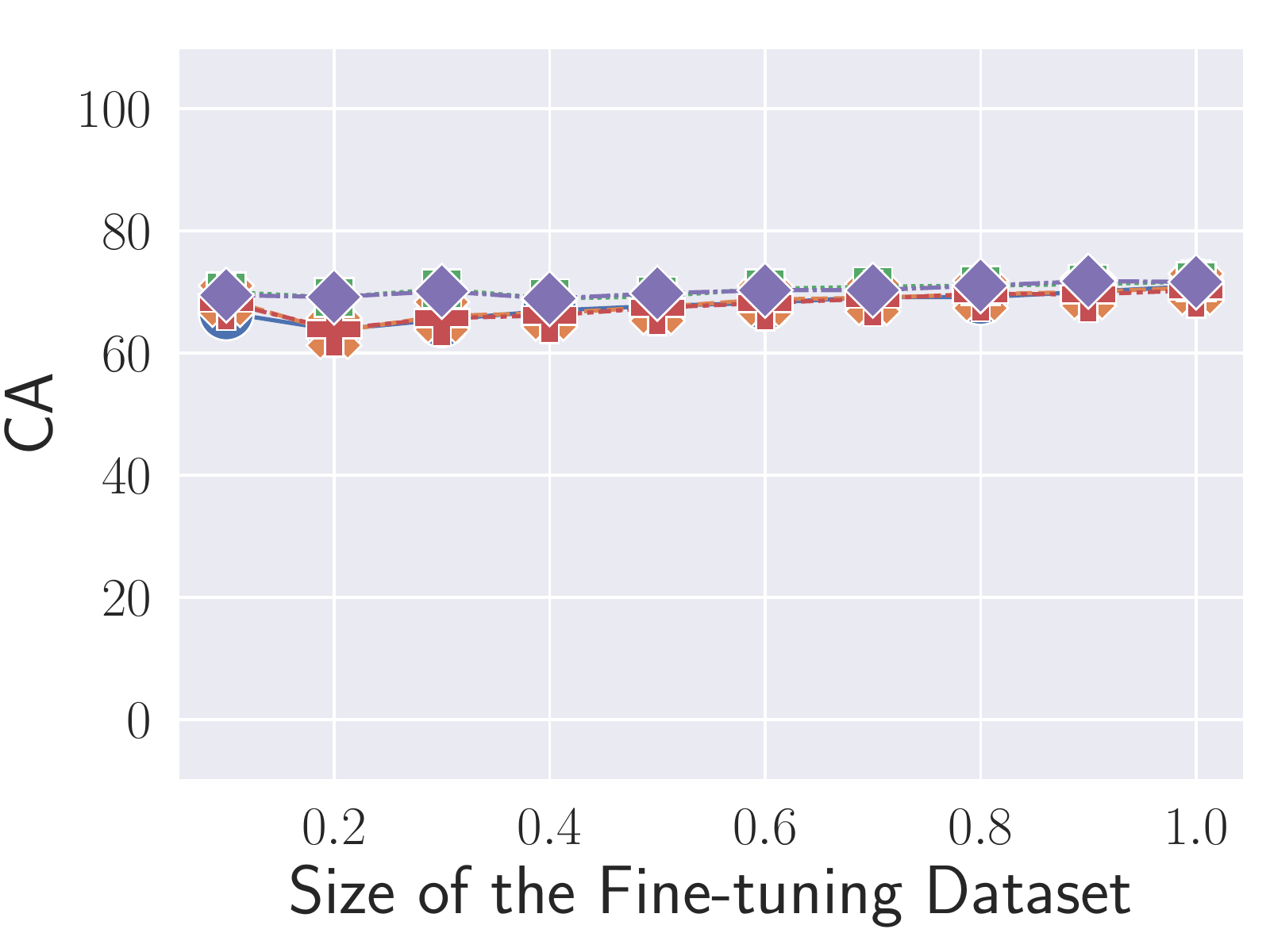}
\caption{CIFAR100}
\label{fig:tuning-size-ca-cifar100}
\end{subfigure}
\begin{subfigure}{0.5\columnwidth}
\includegraphics[width=\columnwidth]{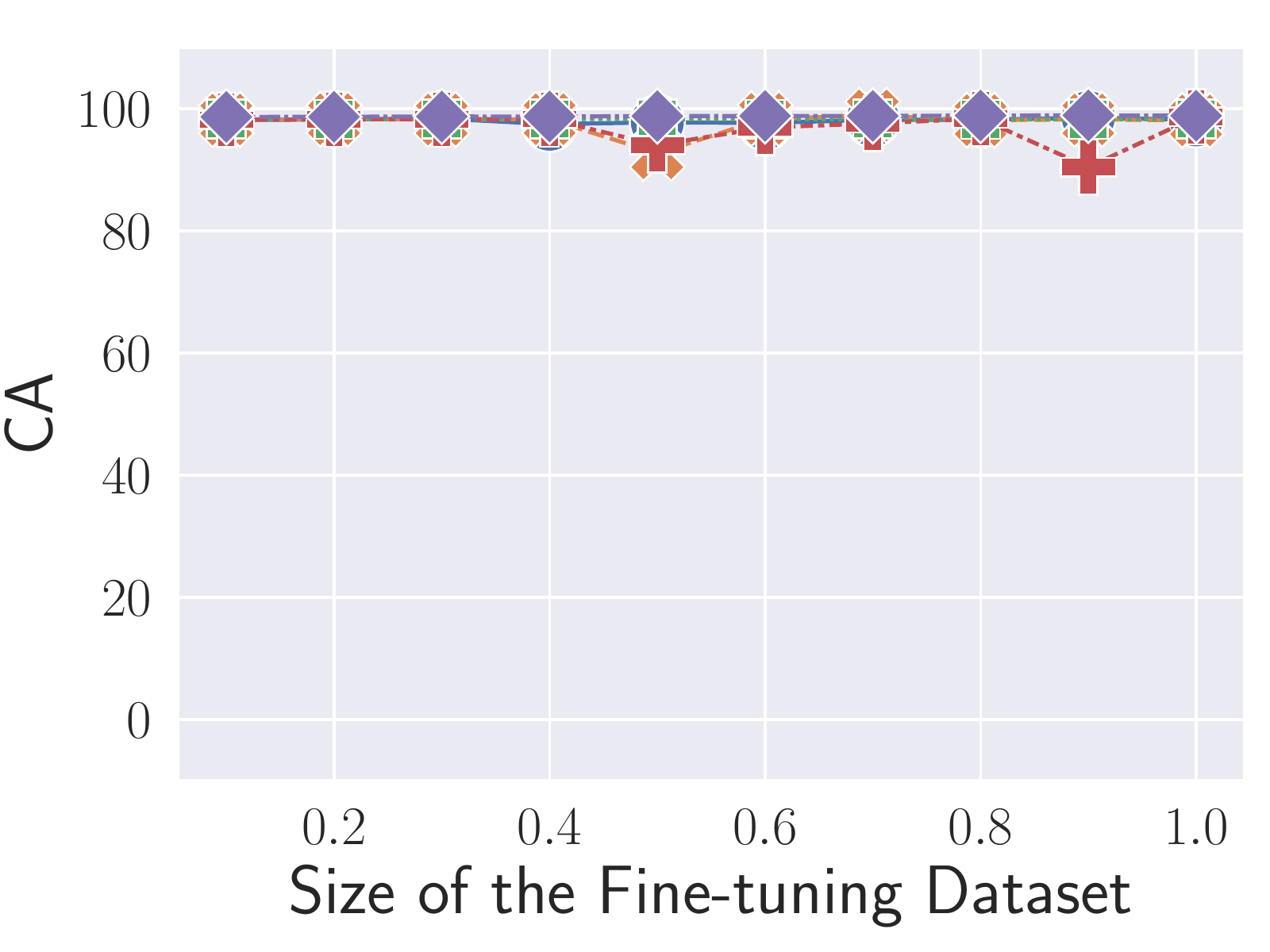}
\caption{GTSRB}
\label{fig:tuning-size-ca-gtsrb}
\end{subfigure}
\caption{
The impact of fine-tuning dataset size on defense performance.
The first row shows fine-tuning dataset size's impacts on attack success rate.
The second row shows fine-tuning dataset size's impacts on clean samples accuracy.
The X-axis represents the ratio of the fine-tuning dataset, which is used to conduct fine-tuning.
}
\label{fig:tuning-size}
\end{figure*}

% ----------------------------------------------------
\subsection{Comparison to Other Defense Methods}
\label{section:compare}
% ----------------------------------------------------

Previously, we have shown that super-fine-tuning can effectively mitigate backdoor attacks with limited computational resources.
In this section, we compare super-fine-tuning with other existing state-of-the-art defense methods to show that super-fine-tuning is the most effective and efficient one.
Note that here we only focus on the standalone scenario since fine-tuning is a necessary step in the other two scenarios, which means fine-tuning as a defense is zero-cost.
Also, fine-tuning or super-fine-tuning can decrease the ASR to a large extent while maintaining the model's utility.

The results on CIFAR10 are shown in \autoref{fig:compare-asr-ca}.
We also show the results on CIFAR100 and GTSRB in \autoref{fig:compare-asr-ca-cifar100} and \autoref{fig:compare-asr-ca-gtsrb} in the appendix.
Note that in \autoref{fig:compare-asr-ca}, the X-axis is CA and the Y-axis is ASR.
Therefore, in each sub-figure, the closer to the lower right corner, the better the defense performance.
Among all defense methods against different attacks, super-fine-tuning, in general, achieves the lowest ASR while maintaining the highest CA.
For instance, to mitigate the BadNets attack on CIFAR10, super-fine-tuning can achieve 0.932 CA with only 0.009 ASR, which constitutes the best performance among all defense methods.
We can also see that other defense methods cannot always guarantee performance in defending against all attacks.
For instance, although only NC and super-fine-tuning can mitigate Blended attacks on CIFAR10, NC cannot detect Inputaware, LF, and WaNet attacks.

We then consider another important aspect, i.e., each defense's computational cost.
The results are shown in \autoref{fig:time}.
We can observe that, among all defenses against different attacks, NC has the largest computational cost, while super-fine-tuning has the lowest computational cost.
For instance, to detect and remove BadNets on CIFAR100, NC takes 0.997 GPU hours, while super-fine-tuning only needs 0.147 GPU hours, which is significantly lower.

In general, we conclude that super-fine-tuning outperforms other defenses in terms of the lowest ASR, highest CA, and lowest computational cost.

% ----------------------------------------------------
\subsection{Ablation Study}
\label{section:ablation}
% ----------------------------------------------------

Here, we conduct some ablation studies to show the impact of fine-tuning dataset size and learning rate on the backdoor removal performance.
Note that we only focus on the standalone scenario here because: (i) in both encoder-based and transfer-based scenarios, fine-tuning is a necessary step, so we do not modify the fine-tuning dataset size and learning rate;
(ii) standalone is the most challenging scenario, as we mentioned before.

\begin{figure}[!t]
\centering
\begin{subfigure}{0.49\columnwidth}
\includegraphics[width=\columnwidth]{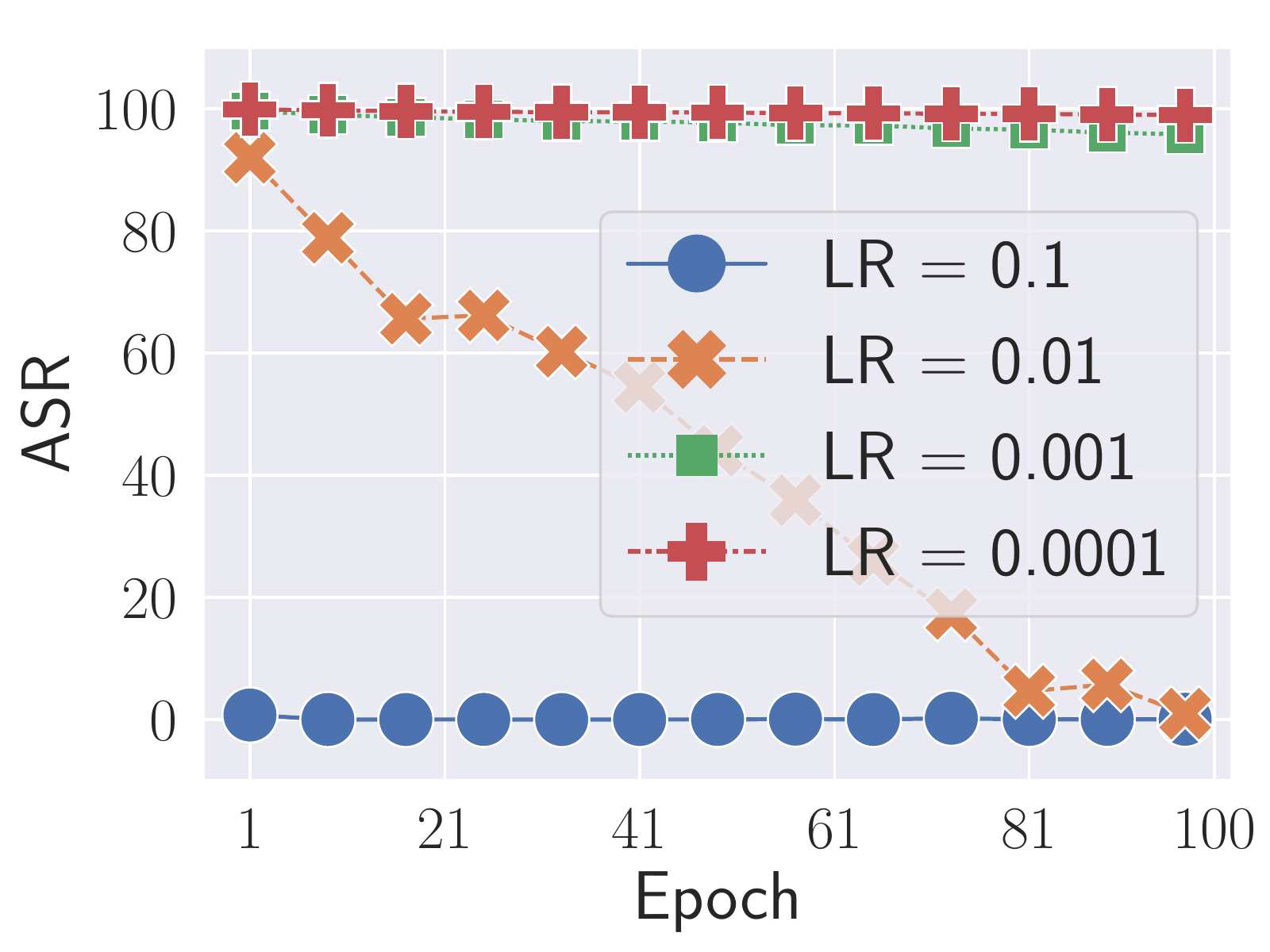}
\caption{ASR}
\label{fig:exp_lr_asr}
\end{subfigure}
\begin{subfigure}{0.49\columnwidth}
\includegraphics[width=\columnwidth]{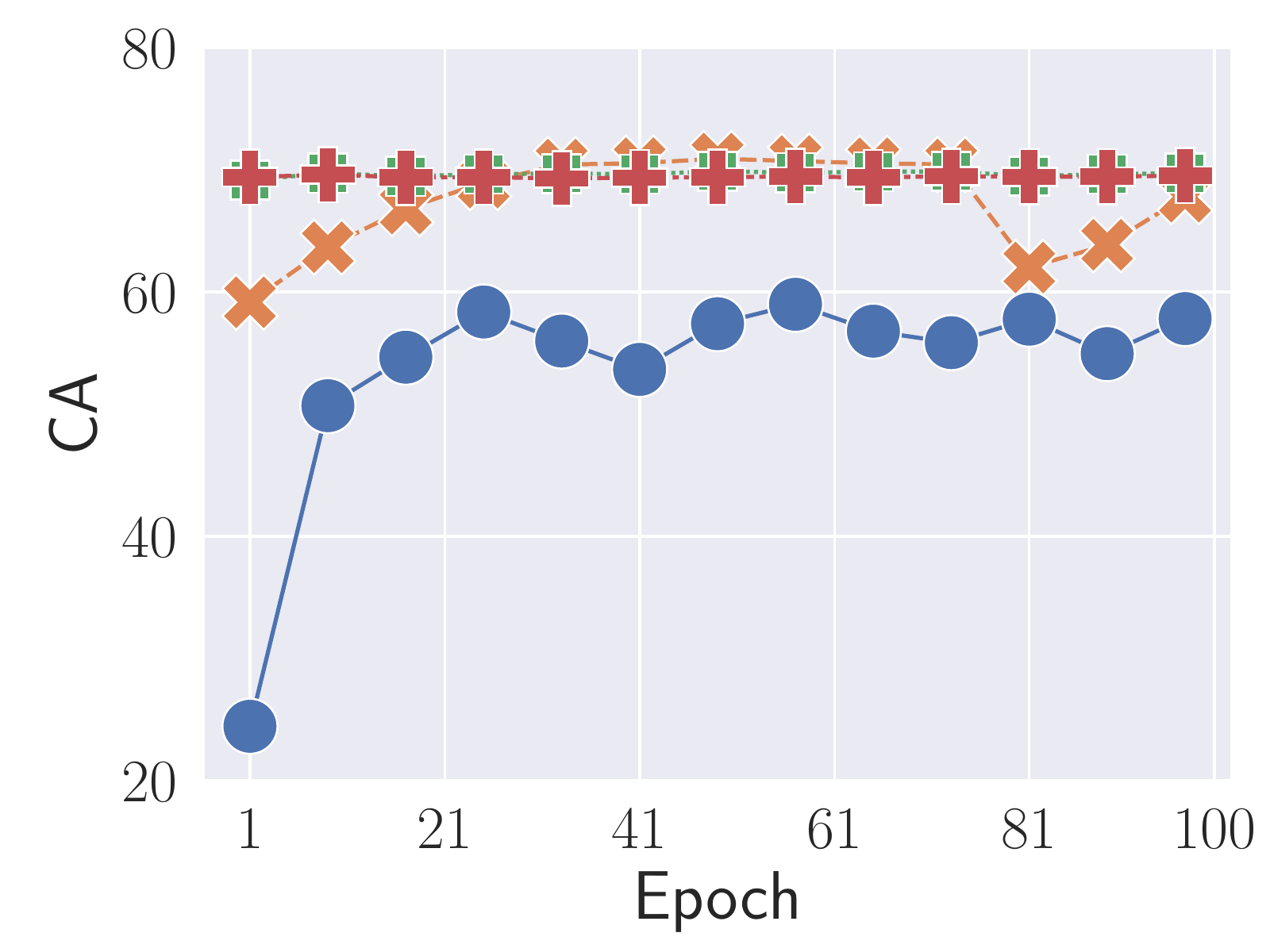}
\caption{CA}
\label{fig:exp_lr_ca}
\end{subfigure}
\caption{
The impact of different learning rates of conventional fine-tuning on removing backdoor attacks.
The X-axis represents training epochs.
The Y-axis represents the accuracy of backdoor samples and clean samples.
}
\label{fig:exp_lr}
\end{figure}

\begin{figure*}[!t]
\centering
\begin{subfigure}{0.5\columnwidth}
\includegraphics[width=\columnwidth]{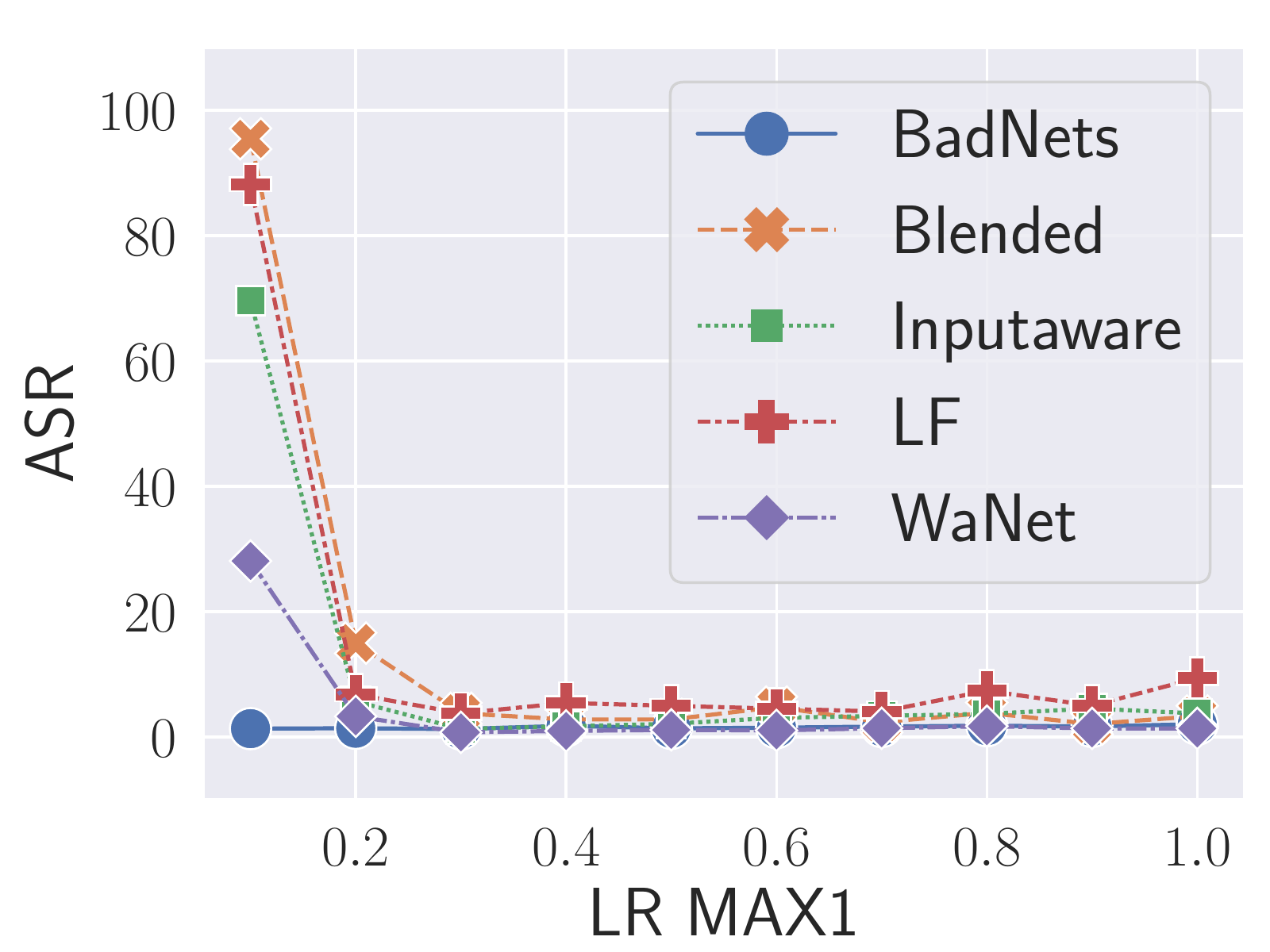}
\caption{CIFAR10}
\label{fig:largest-learning-rate-cifar10}
\end{subfigure}
\begin{subfigure}{0.5\columnwidth}
\includegraphics[width=\columnwidth]{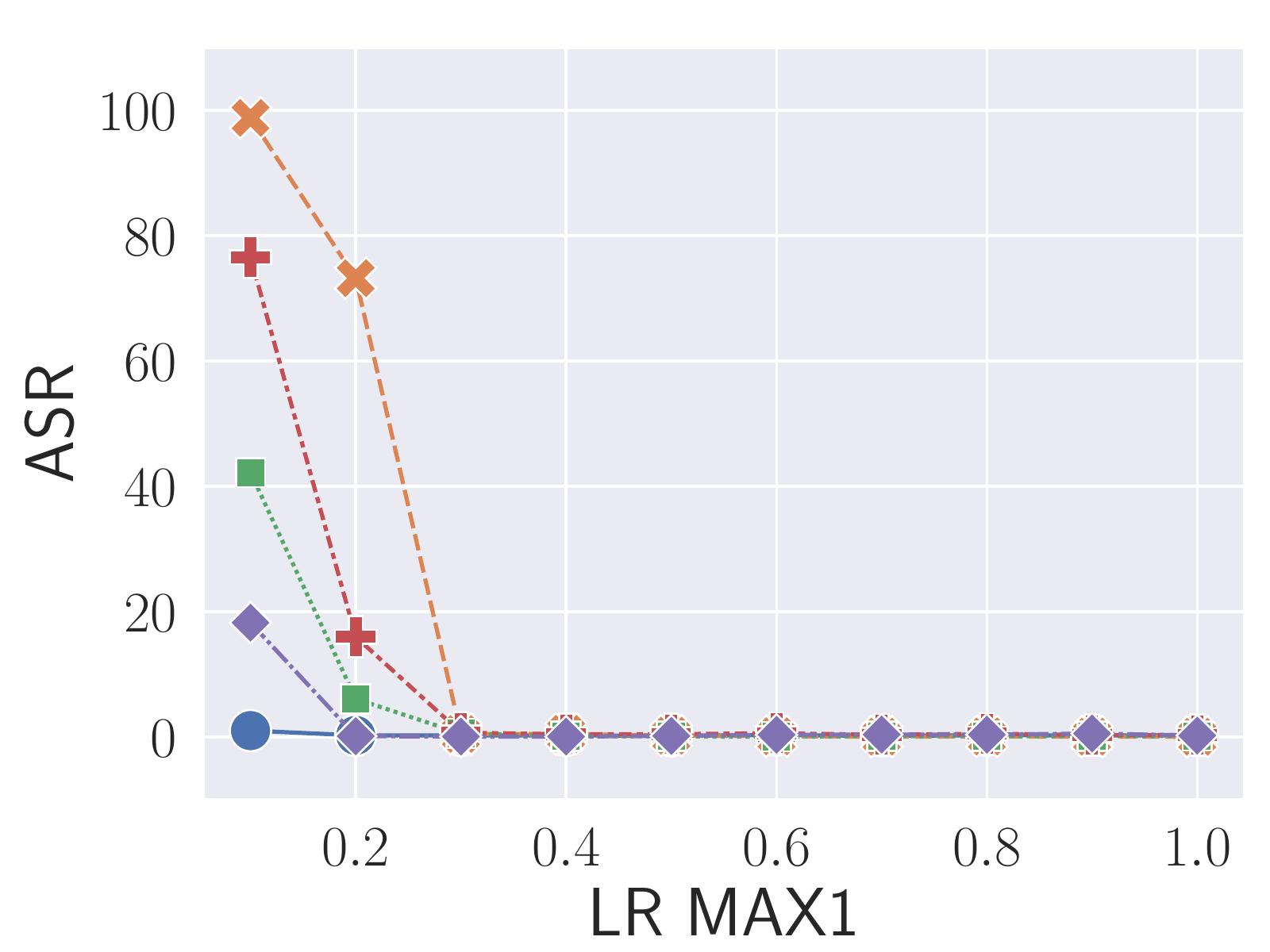}
\caption{CIFAR100}
\label{fig:largest-learning-rate-cifar100}
\end{subfigure}
\begin{subfigure}{0.5\columnwidth}
\includegraphics[width=\columnwidth]{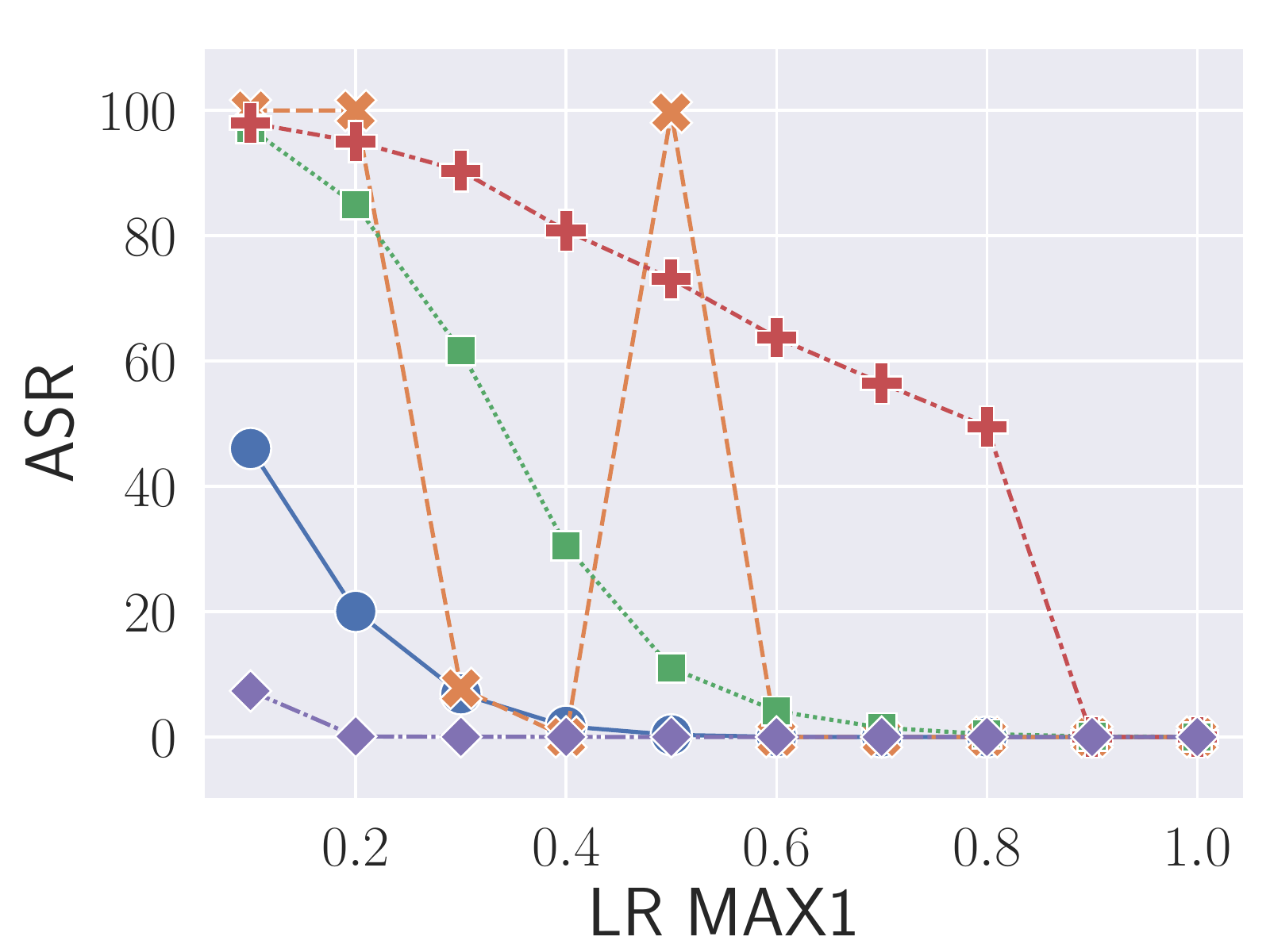}
\caption{GTSRB}
\label{fig:largest-learning-rate-gtsrb}
\end{subfigure}
\begin{subfigure}{0.5\columnwidth}
\includegraphics[width=\columnwidth]{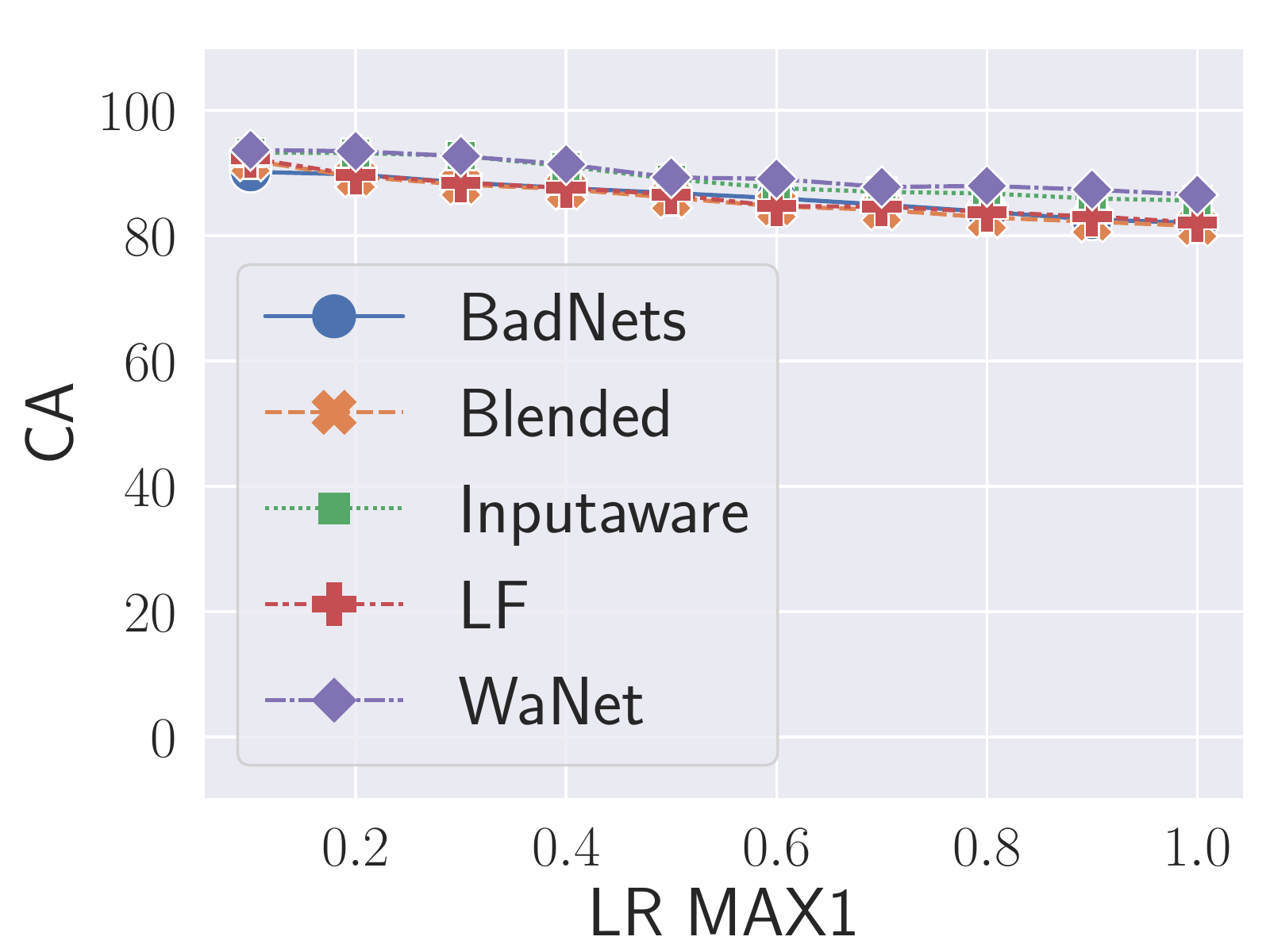}
\caption{CIFAR10}
\label{fig:largest-learning-rate-ca-cifar10}
\end{subfigure}
\begin{subfigure}{0.5\columnwidth}
\includegraphics[width=\columnwidth]{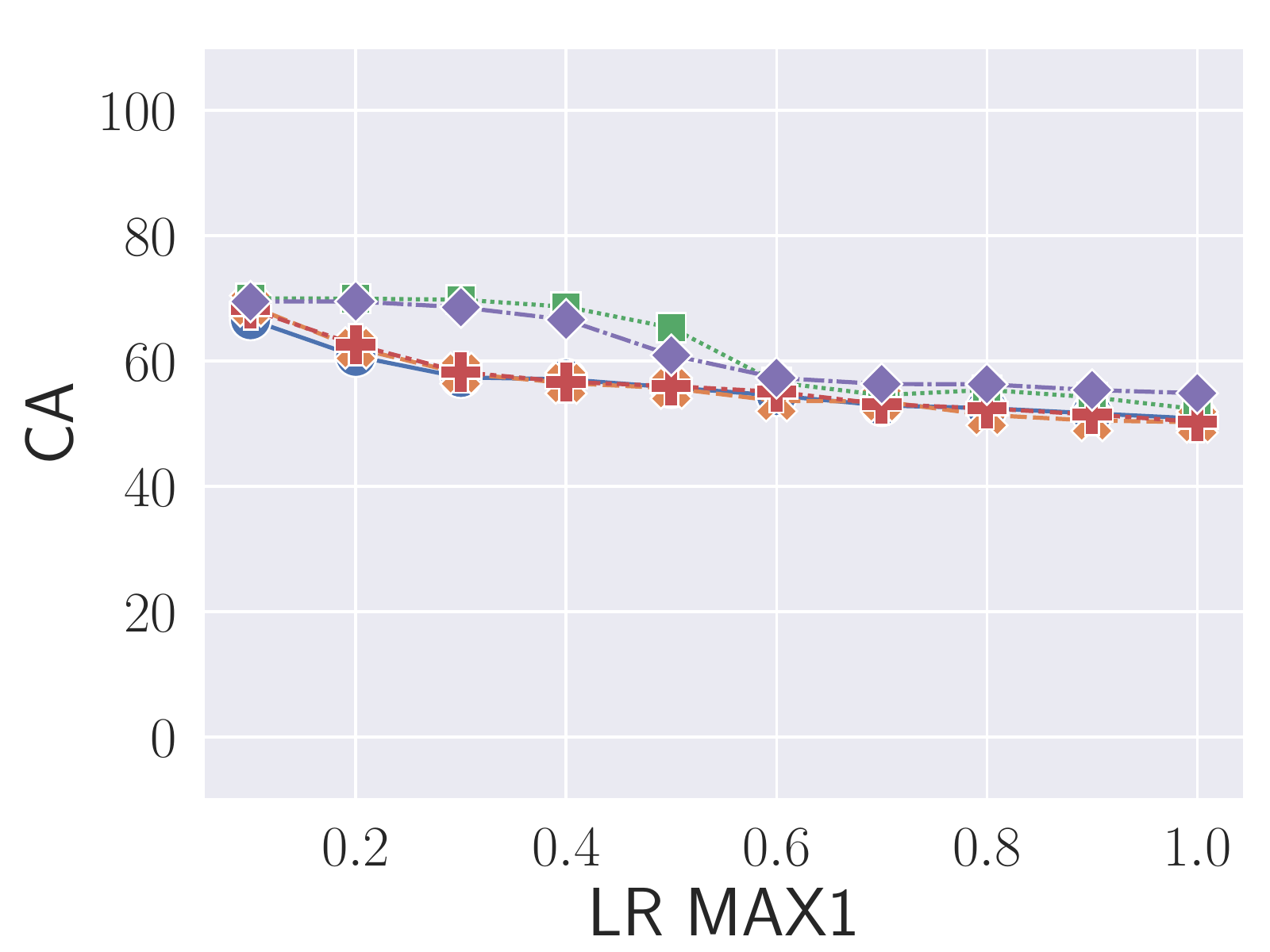}
\caption{CIFAR100}
\label{fig:largest-learning-rate-ca-cifar100}
\end{subfigure}
\begin{subfigure}{0.5\columnwidth}
\includegraphics[width=\columnwidth]{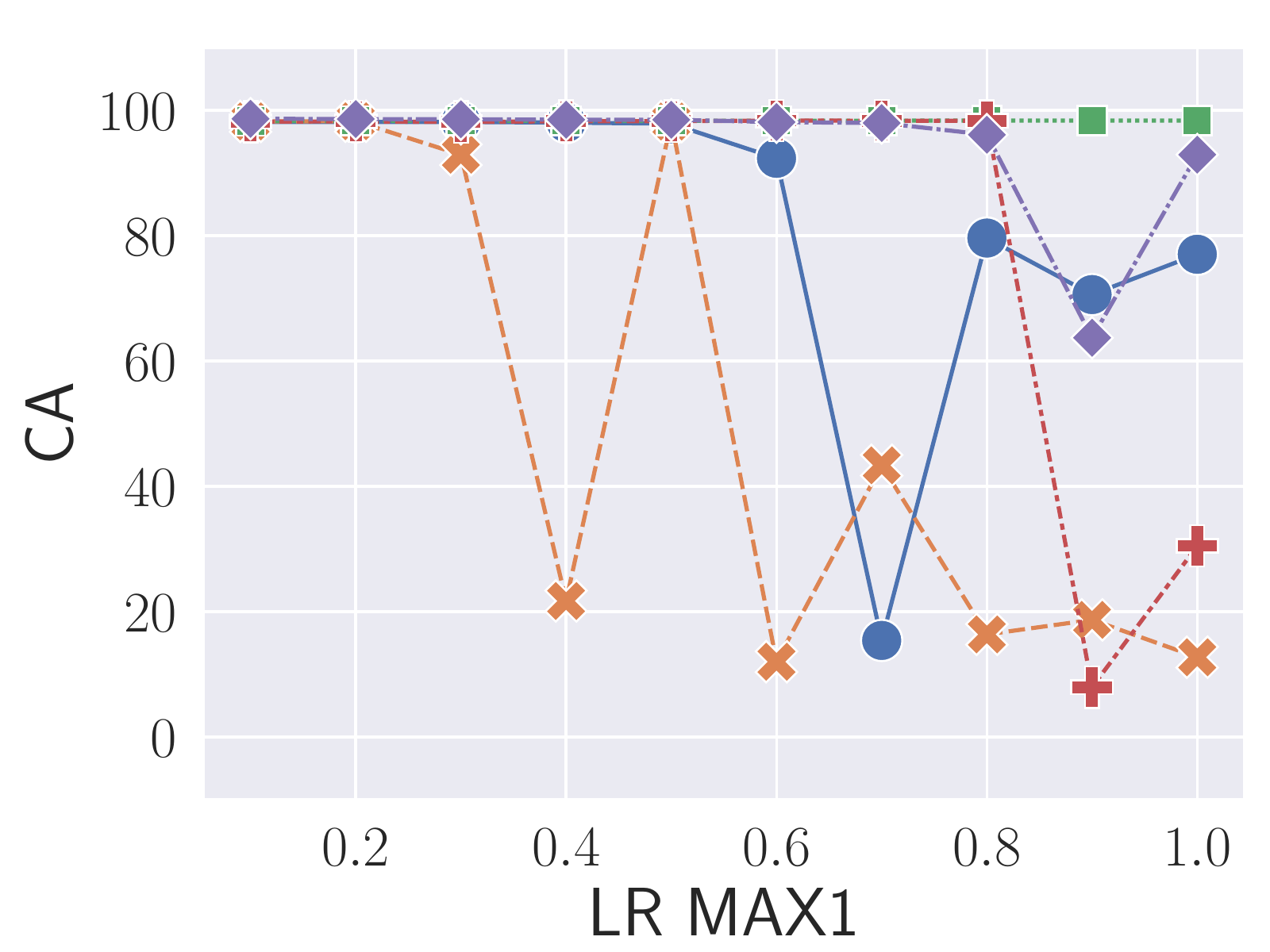}
\caption{GTSRB}
\label{fig:largest-learning-rate-ca-gtsrb}
\end{subfigure}
\caption{
Impact of LR MAX1 of super-fine-tuning on defense performance.
The first row shows LR MAX1's impacts on attack success rate.
The second row shows LR MAX1's impacts on clean sample accuracy.
The X-axis represents how many data samples are used to conduct fine-tuning.
Note that we only use 10\% of the fine-tuning dataset to conduct super-fine-tuning.
}
\label{fig:largest-learning-rate}
\end{figure*}

\mypara{Impact of Fine-Tuning Dataset Size}
We first explore the impact of fine-tuning dataset size.
Previously, we used the whole dataset to conduct super-fine-tuning.
We have shown that, even with the whole dataset, super-fine-tuning consumes limited computational resources compared to other methods.
Then, we further explore how much data is sufficient to conduct a successful super-fine-tuning.
We show our experimental results in \autoref{fig:tuning-size}.
We can see that even with 20\% of the fine-tuning dataset, super-fine-tuning can effectively mitigate the backdoor attacks in most cases.
For instance, with 20\% of the fine-tuning dataset (CIFAR10), super-fine-tuning reduces the ASR of the Blended backdoor attack to 0.044.
Also, from \autoref{fig:tuning-size}, we can see that the size of the fine-tuning dataset has a limited impact on the utility of the model.
The clean accuracy remains high with 10\% to 100\% of the fine-tuning dataset.
Therefore, it can be concluded that super-fine-tuning requires significantly fewer fine-tuning data samples for a stable performance, which further reduces the computational cost.

Note that super-fine-tuning is less effective with 10\% of the fine-tuning dataset.
For instance, when only using a 10\% clean training dataset and 0.1 as LR MAX1, the defender can only achieve 0.954 ASR with the Blended attack on CIFAR10.
We show later that the backdoor attacks can still be effectively mitigated by increasing LR MAX1 if the defender only has 10\% of the fine-tuning dataset.

\mypara{Impact of Learning Rate Change}
During our experiments, we first find that backdoor attacks are very sensitive to different learning rates.
We show different learning rates' results of conventional fine-tuning in \autoref{fig:exp_lr}.
It can be seen that if the defender uses the same small learning rate as in the pre-training phase, the ASR remains high even after 100 epochs.
However, with an increased learning rate (from 0.0001 to 0.001), backdoor triggers are forgotten gradually in 100 epochs.
Moreover, when the learning rate increases to 0.1, the backdoor can be immediately removed within one epoch.
We can conclude that learning rates have a significant impact on backdoor removal.
In particular, larger learning rates tend to mitigate backdoor attacks faster.

From \autoref{fig:exp_lr}, we can also find that although increasing learning rates can effectively mitigate backdoor attacks, it also causes utility drops.
From the utility perspective, small learning rates lead to higher clean accuracy.
Therefore, combining large and small learning rates becomes a promising idea to achieve both goals.
This is also the general intuition for super-fine-tuning.

In our previous super-fine-tuning experiments, we set LR MAX1 to 0.1 as we find that 0.1 is enough for removing backdoors with sufficient fine-tuning datasets.
Here, we also explore the impact of LR MAX1 on super-fine-tuning.
To better show the learning rate's impact, we only use 10\% of the fine-tuning dataset.
In the previous section, when the defender only has 10\% of the fine-tuning dataset, super-fine-tuning does not achieve good performance, especially in Blended and LF attacks.

Our experimental results are shown in \autoref{fig:largest-learning-rate}.
When increasing the LR MAX1 from 0.1 to 0.3, even when using 10\% of the fine-tuning dataset, super-fine-tuning can still successfully remove backdoors.
For instance, when the LR MAX1 is 0.1, Blended attacks on CIFAR100 still achieve 0.988 ASR under super-fine-tuning.
However, when LR MAX1 increases to 0.3, the ASR drops to 0.004.
Although increasing LR MAX1 can more effectively remove the backdoors, it also leads to a small drop in the model's utility.
We show these results in \autoref{fig:largest-learning-rate}.
For instance, when the learning rate increases from 0.1 to 0.3, the utility of the fine-tuned model from Blended on CIFAR10 drops from 0.919 to 0.881.
Therefore, if users have enough clean data, we recommend using 0.1 as the largest learning rate.
However, when users only have limited data, increasing the largest learning rate (e.g., from 0.1 to 0.3) also helps mitigate almost all attacks without suffering a large utility drop.

% ----------------------------------------------------
\subsection{Summary}
% ----------------------------------------------------

In this section, we have shown that backdoor attacks can be easily defended by fine-tuning or super-fine-tuning.
We show that in the encoder-based and transfer-based scenarios, fine-tuning as the necessary step can naturally remove the existing backdoors.
Also, our proposed super-fine-tuning method can better mitigate the backdoor attacks in the transfer-based scenario.
In the standalone scenario, super-fine-tuning can effectively prevent backdoor attacks with a limited size of the training dataset and limited computational resources compared to other existing defenses.
Our ablation study on the fine-tuning dataset size and learning rate setting further demonstrates the effectiveness and efficiency of super-fine-tuning.

\begin{figure*}[!t]
\centering
\begin{subfigure}{0.65\columnwidth}
\includegraphics[width=\columnwidth]{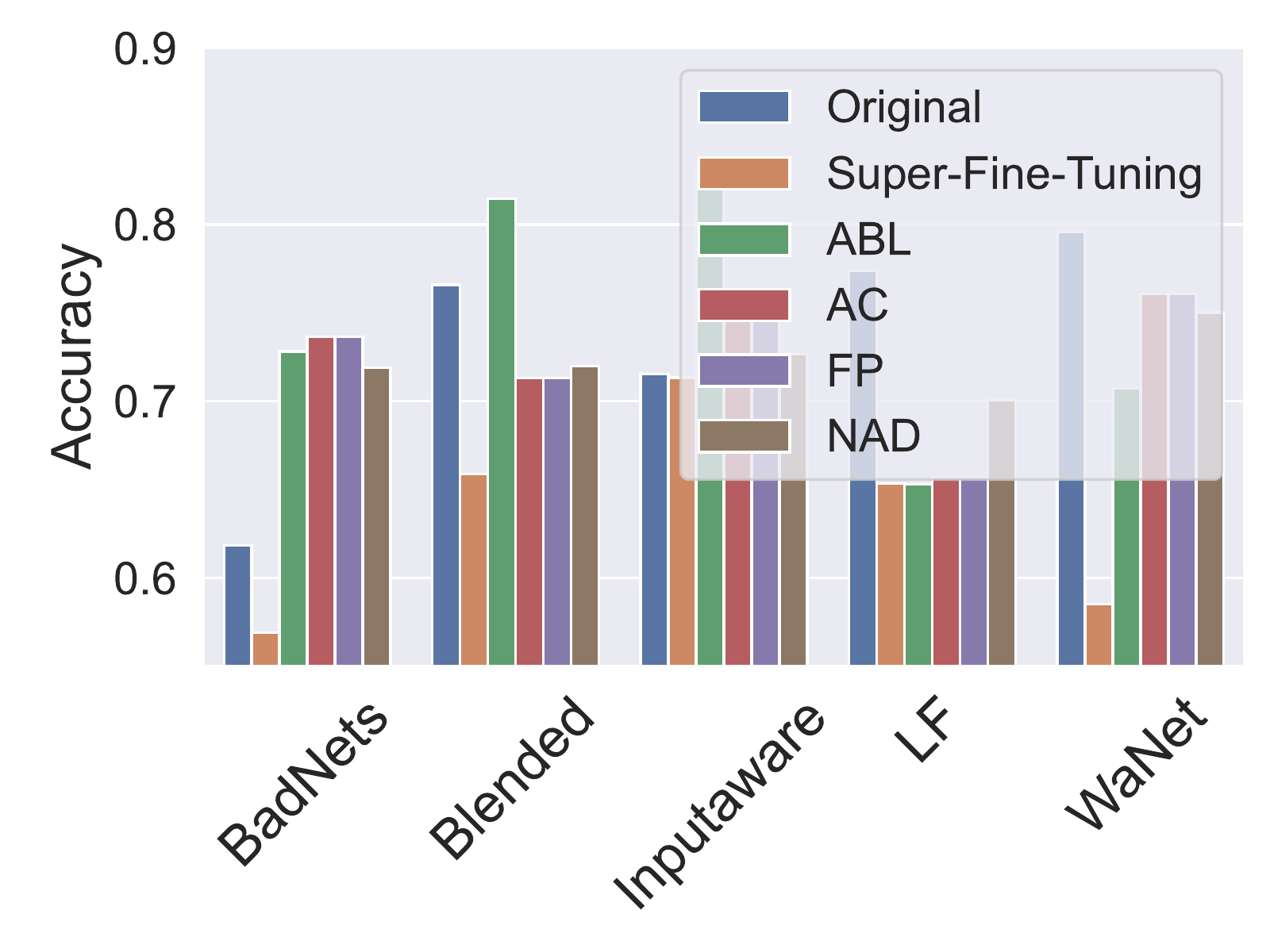}
\caption{CIFAR10}
\label{fig:mem_cifar10}
\end{subfigure}
\begin{subfigure}{0.65\columnwidth}
\includegraphics[width=\columnwidth]{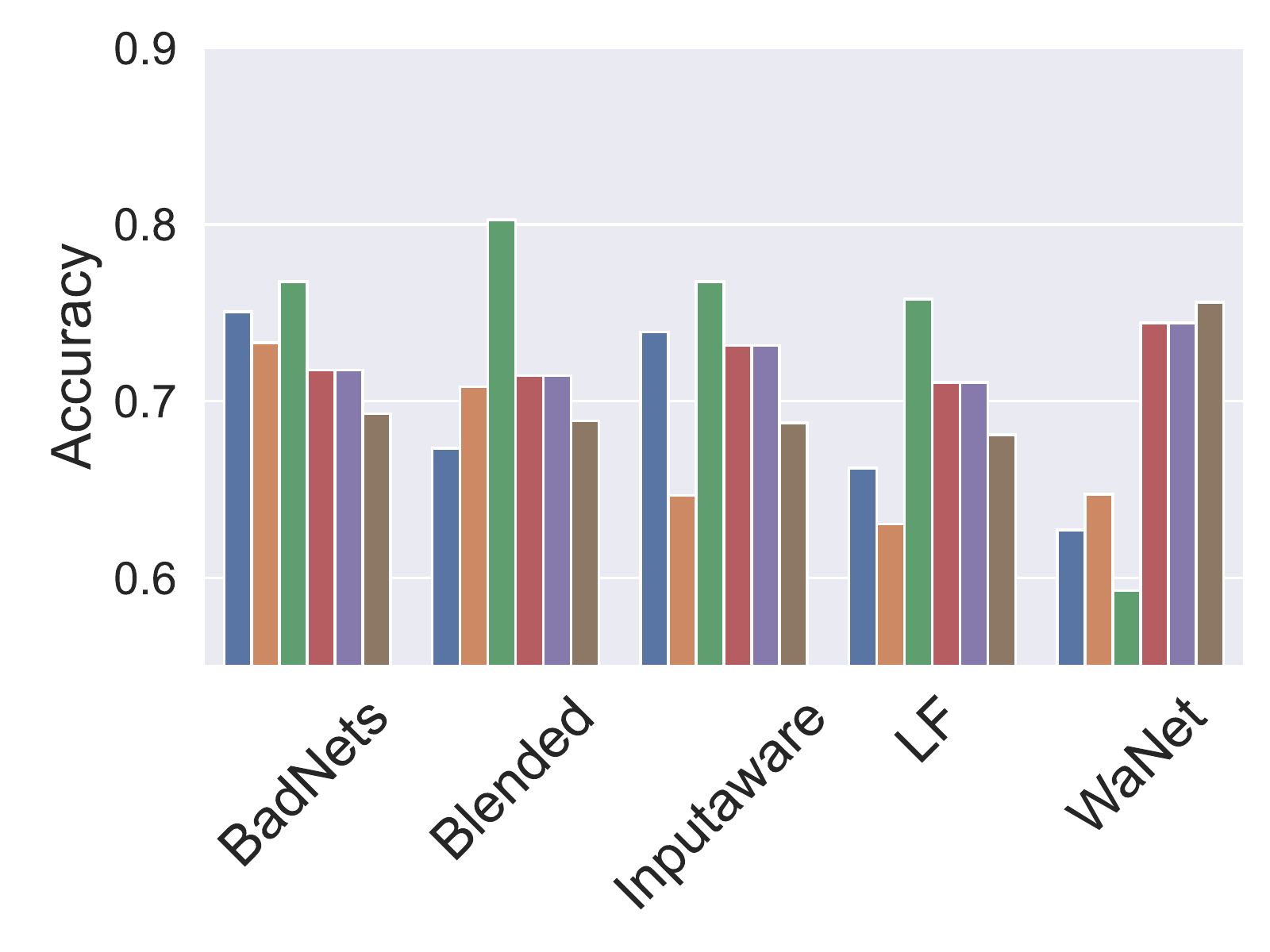}
\caption{CIFAR100}
\label{fig:mem_cifar100}
\end{subfigure}
\begin{subfigure}{0.65\columnwidth}
\includegraphics[width=\columnwidth]{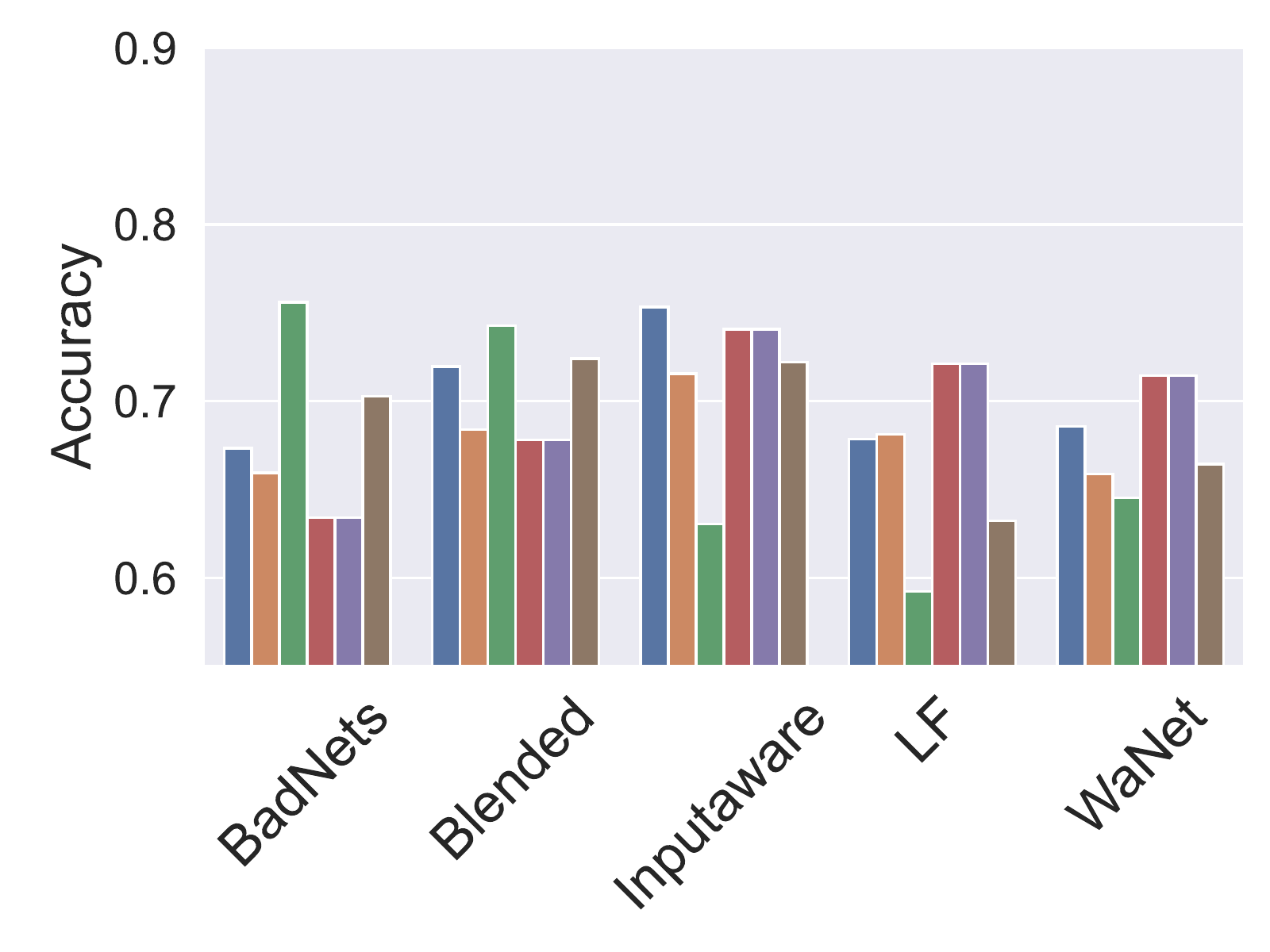}
\caption{GTSRB}
\label{fig:mem_gtsrb}
\end{subfigure}
\caption{
The performance of membership inference attacks on different defended models and backdoored models.
The X-axis represents different attack methods.
The Y-axis represents membership inference attack accuracy.
}
\label{fig:mem}
\end{figure*}

% ----------------------------------------------------
\section{Backdoor Sequela}
% ----------------------------------------------------

We have previously shown that super-fine-tuning outperforms other defenses against backdoor attacks.
However, since the defense modifies the model's parameters, it is worthwhile to explore whether the model will be more or less vulnerable to certain attacks after removing the backdoor attacks.
To this end, we coin the term and investigate backdoor sequela.

Due to the fact that super-fine-tuning needs to use a clean dataset to make the model forget the backdoor, it is natural to wonder whether the process will lead the model to better remember the clean dataset.
To verify this, we conduct membership inference attacks against the backdoored models and the models defended by super-fine-tuning (and other defenses) to see whether the membership leakage becomes larger after applying the defense.
Also, since the backdoors are easily removed after a few epochs, we are also curious if it is easier to re-inject the backdoor into the model.
Note that, following the same reasons as in \autoref{section:compare}, we only consider the standalone scenario in this section.

% ----------------------------------------------------
\subsection{Membership Inference Attack}
% ----------------------------------------------------

We first explore whether the model is more vulnerable to membership inference attacks~\cite{SSSS17} or not after fine-tuning.
Membership inference attacks aim to infer whether a given sample is in the training set of a target model or not.
A successful membership inference attack can cause severe privacy leakage.
Normally, there are three different ways to conduct membership inference attacks: neural network-based attacks~\cite{NSH18,SSSS17}, metric-based attacks~\cite{LF20,SM21,SSM19,YGFJ18}, and query-based attacks~\cite{CTCP21,LZ21}.
In this work, we use the neural network-based attack due to its popularity.

\mypara{Threat Model}
We first assume that the adversary only has black-box access to the target model, which means they can only query the model and obtain the output.
Then, following previous works~\cite{NSH19,LWHSZBCFZ22}, we further assume that the adversary has part of the target model's training data (treated as members) and testing data (non-members).
The adversary can use them for training an attack model and inferring the membership status for other data samples.
Note that we adopt the strongest attacker assumption defined in~\cite{LWHSZBCFZ22} to estimate the worst-case scenario for membership leakage.

\mypara{Methodology}
Our method can be described in two steps:

\begin{enumerate}
    \item The adversary first queries the target model with both the target model's (partial) training and testing samples, and they label the corresponding outputs as members and non-members.
    \item Second, the adversary uses the outputs and the corresponding labels to train their attack model, which is a three-layer neural network model.
\end{enumerate}

The evaluations are conducted on both the backdoored model and the super-fine-tuned model to see whether fine-tuning will increase or decrease membership inference risks.

\mypara{Experimental Settings}
We evaluate the membership inference attack in the standalone scenario, which means that the fine-tuning dataset is the same as the pre-training dataset.
For each dataset, we randomly sample half of its testing samples and the same number of training samples as the attack training dataset.
Then, we select the other half of its testing samples (serving as non-members) and the same number of training samples (serving as members with no overlap on the attack training dataset) to evaluate the attack performance.
Note that we use the datasets and backdoor attacks/defenses introduced in \autoref{section:attacks-and-defenses}.

\begin{figure*}[!t]
\centering
\begin{subfigure}{0.5\columnwidth}
\includegraphics[width=\columnwidth]{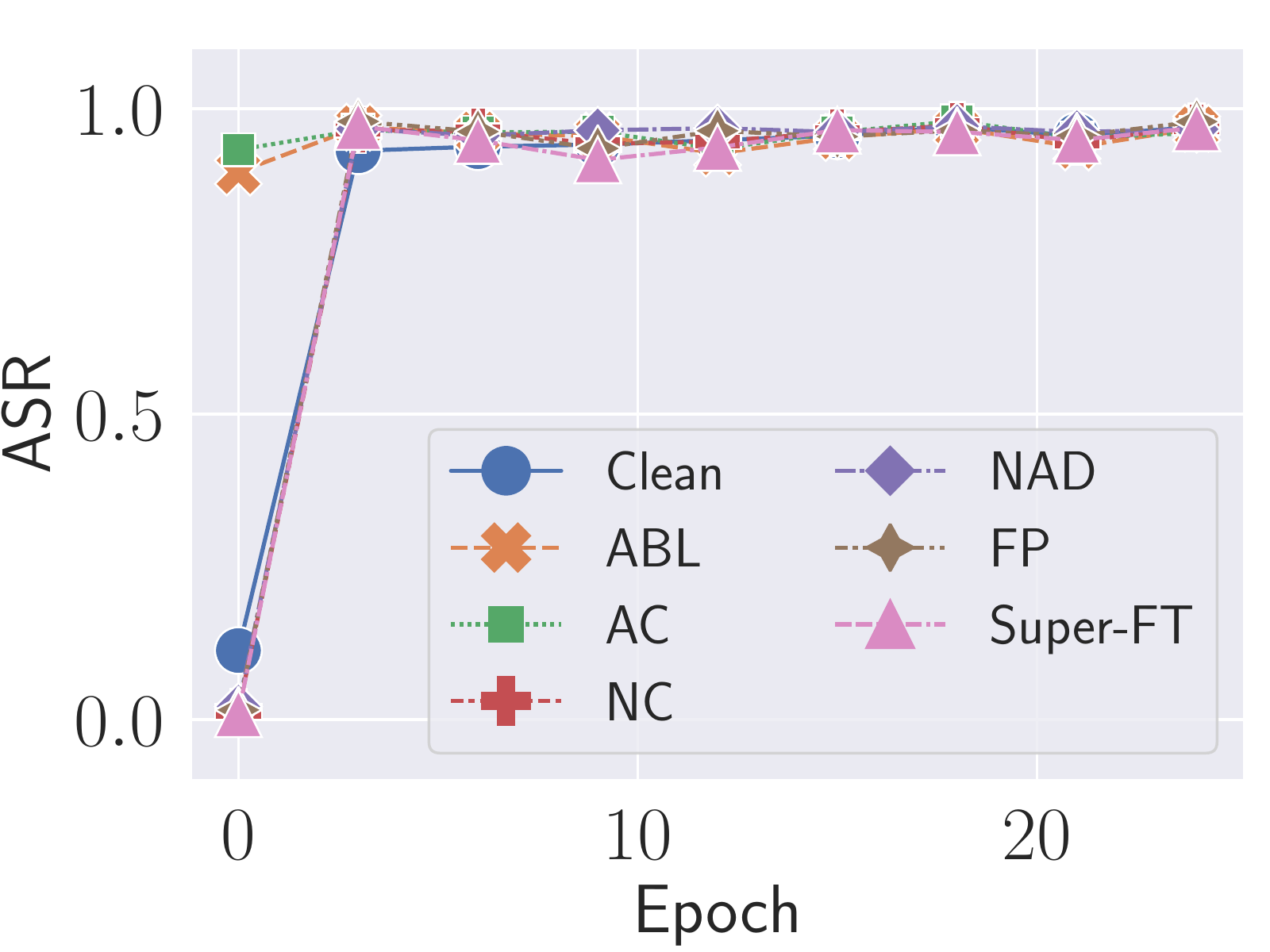}
\caption{Poison Ratio: 0.1}
\label{fig:rebadnet-0.1}
\end{subfigure}
\begin{subfigure}{0.5\columnwidth}
\includegraphics[width=\columnwidth]{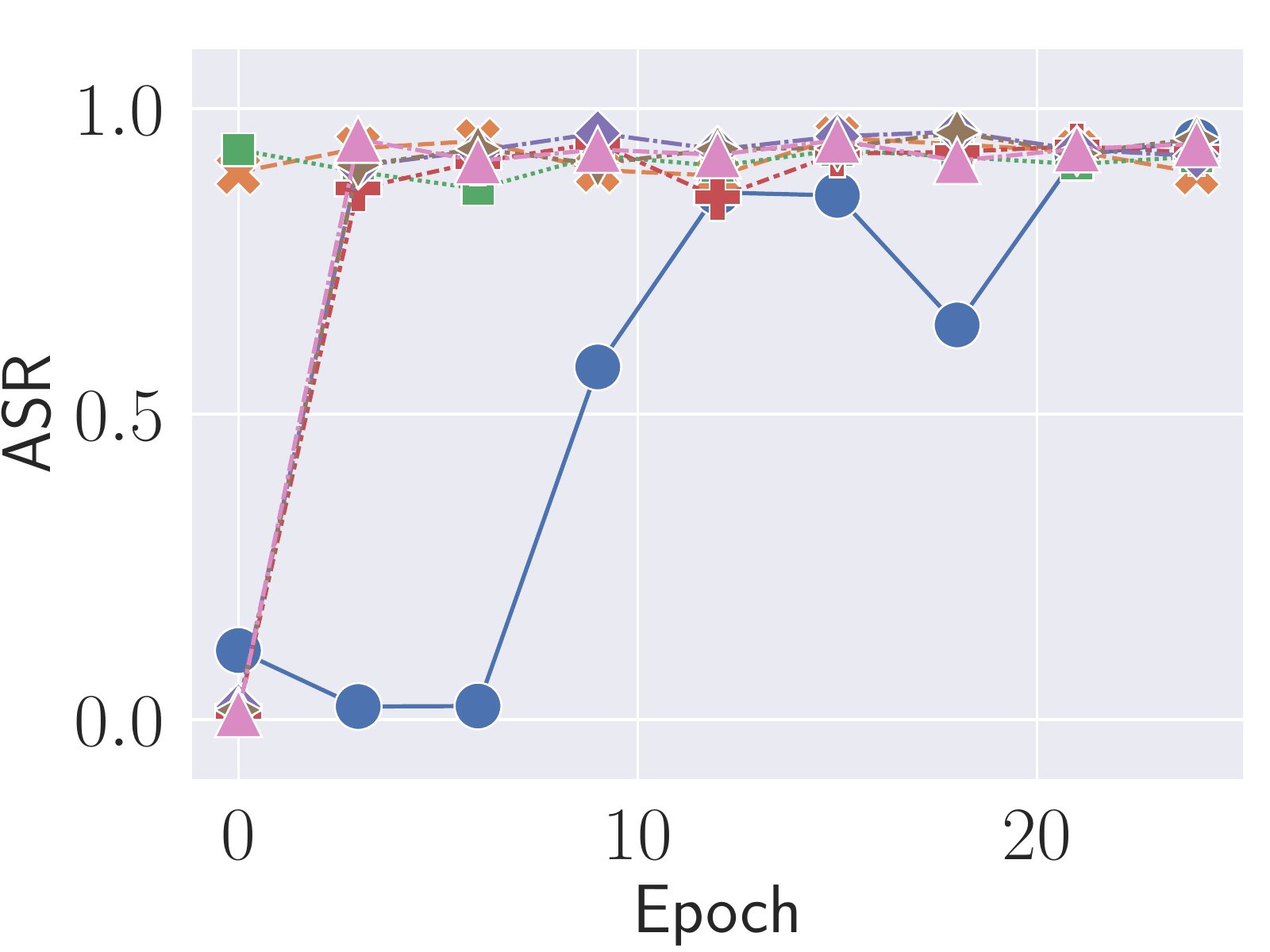}
\caption{Poison Ratio: 0.01}
\label{fig:rebadnet-0.01}
\end{subfigure}
\begin{subfigure}{0.5\columnwidth}
\includegraphics[width=\columnwidth]{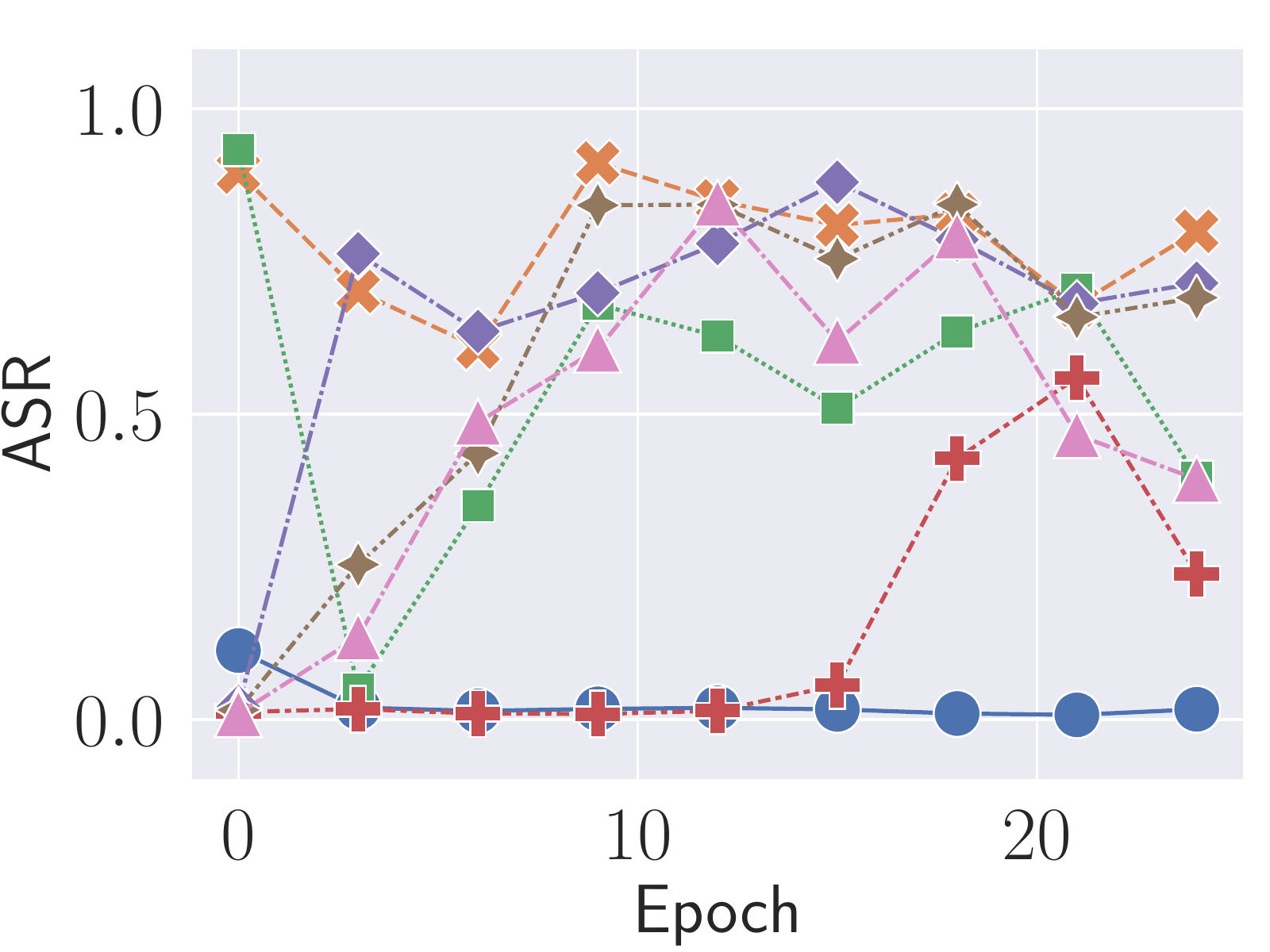}
\caption{Poison Ratio: 0.001}
\label{fig:rebadnet-0.001}
\end{subfigure}
\begin{subfigure}{0.5\columnwidth}
\includegraphics[width=\columnwidth]{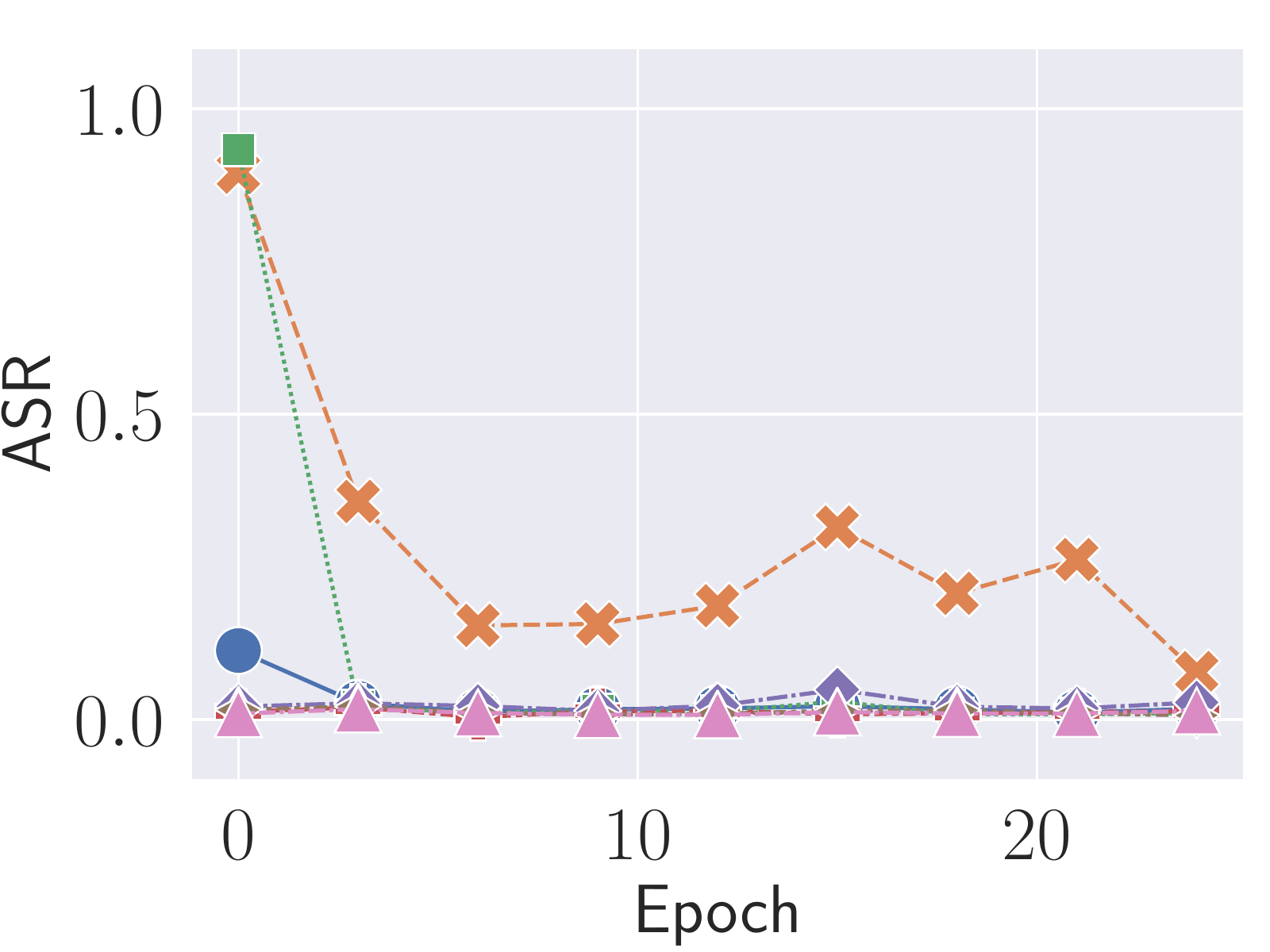}
\caption{Poison Ratio: 0.0001}
\label{fig:rebadnet-0.0001}
\end{subfigure}
\caption{
Performance of BadNets backdoor re-injection attacks on different defense methods.
The X-axis represents training epochs in the re-injection phase.
The Y-axis represents the accuracy of poison samples.
Note that epoch 0 represents different defense methods' results before re-injection.
}
\label{fig:reinject}
\end{figure*}

\mypara{Results}
We show our membership inference results in \autoref{fig:mem}.
Surprisingly, we observe that instead of increasing the privacy risks, super-fine-tuning mitigates the performance of membership inference.
In almost all cases, fine-tuned models have lower attack performance than the original models.
For instance, membership inference attacks on the original model (backdoored by BadNets) on CIFAR10 can achieve 0.618 accuracy, while the performance drops to 0.569 after conducting super-fine-tuning on the original model.
This is contradictory to previous work~\cite{SZHBFB19}, where more training epochs led to higher attack performance due to the increasing overfitting level.
We suspect the reason is that super-fine-tuning actually increases the generalization ability of the model, which leads to a lower overfitting level.

Though super-fine-tuning does not cause backdoor sequela with respect to membership inference, we do observe that some of the other defense methods make membership inference more unstable.
For instance, in most cases, ABL makes the model more vulnerable to membership inference attacks (accuracy increases from 0.765 to 0.815 in Blended CIFAR10).
However, when ABL is used to mitigate LF attacks on CIFAR10, the membership inference risk drops from 0.726 to 0.653.

From \autoref{fig:mem}, we can also see that different kinds of backdoor attacks make a huge difference in membership inference performance.
For example, the adversary can achieve a 0.765 accuracy when conducting membership inference attacks using CIFAR10 on Blended attacks.
However, they can only achieve a 0.618 accuracy when conducting the same attacks on the BadNets model.
We leave it to future work to further explore the relationship between backdoor attacks and membership inference attacks.

% ----------------------------------------------------
\subsection{Backdoor Re-injection Attack}
% ----------------------------------------------------

Another backdoor sequela we study is the backdoor re-injection attack.
Since super-fine-tuning can easily remove backdoors within a few epochs, it is interesting to see whether the fine-tuned models are more vulnerable to injecting (the same) backdoor attacks again.
To our knowledge, there is no prior work measuring whether existing backdoor defense methods will make the re-injection process easier.

\mypara{Threat Model}
We first assume that the adversary has white-box access to the fine-tuned model.
Also, the adversary has knowledge of the previous backdoor attack on the model.
The goal of the adversary is to re-inject the same backdoor into the model while keeping model utility.

\mypara{Methodology}
To measure the vulnerability of the fine-tuned models on backdoor re-injection attacks, we take the following steps:

\begin{enumerate}
    \item The adversary first generates new triggered samples based on the knowledge of the previous attack.
    \item Then, the adversary re-trains the fine-tuned model with the poison dataset and the corresponding training process to re-inject the backdoor into the fine-tuned model.
\end{enumerate}

\mypara{Results}
To measure how easily the adversary can re-inject a backdoor into the model, we consider training epochs, poison ratio, and ASR as the evaluation metrics.
If the model is more vulnerable to backdoor re-injection attacks, the adversary only needs fewer epochs and a smaller poison ratio to achieve a similar (or even better) ASR compared to injecting the backdoor on a clean model.
Note that in this setting, we assume a very powerful adversary who has knowledge of the previous attack.
Thus, the adversary can follow the same procedure to conduct the attack, i.e., they use the exact same injection process and strategies, including the same attack method, the same learning rate, etc.

\begin{figure*}[!t]
\centering
\begin{subfigure}{0.5\columnwidth}
\includegraphics[width=\columnwidth]{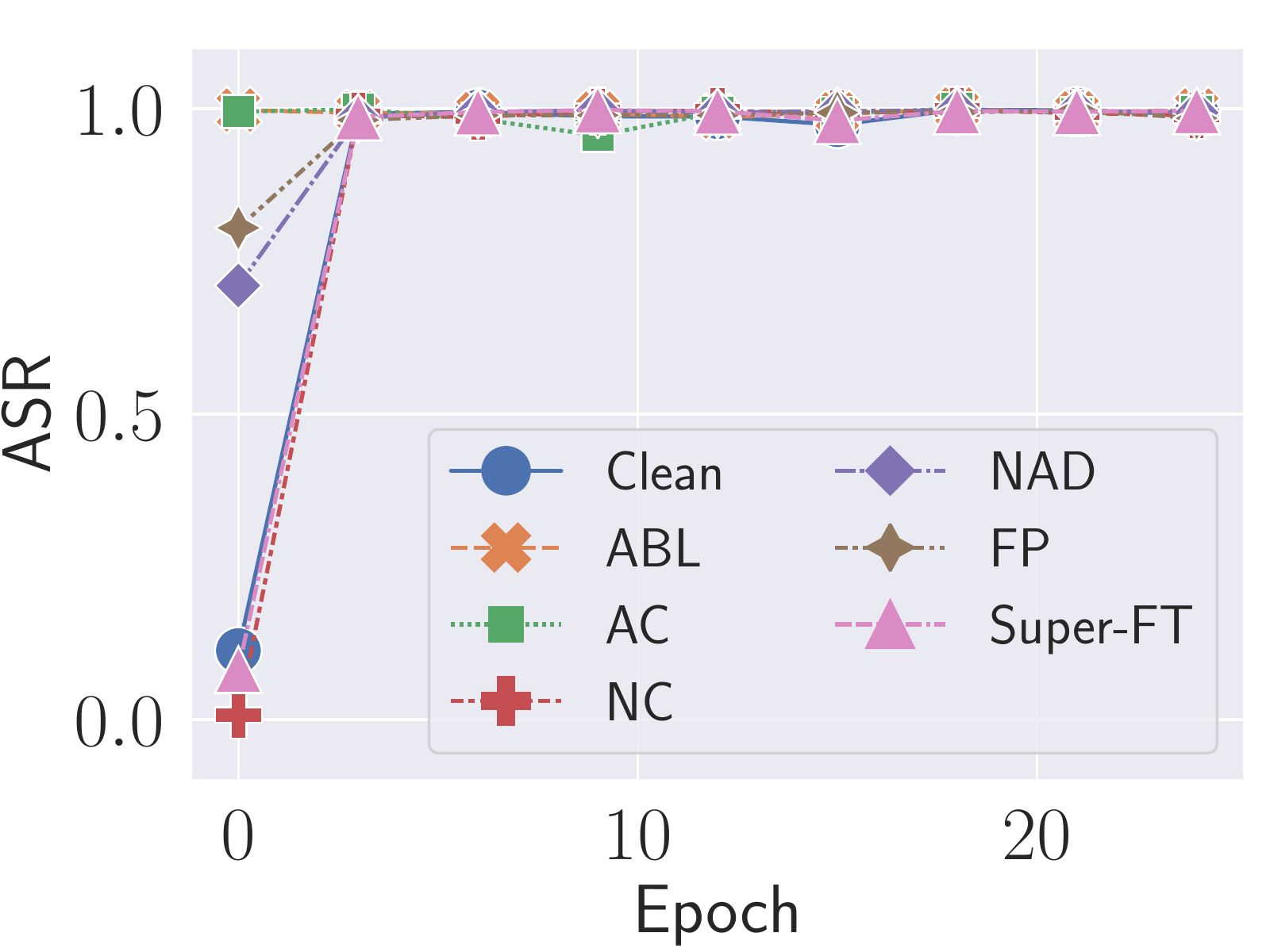}
\caption{Poison Ratio: 0.1}
\label{fig:rebl-0.1}
\end{subfigure}
\begin{subfigure}{0.5\columnwidth}
\includegraphics[width=\columnwidth]{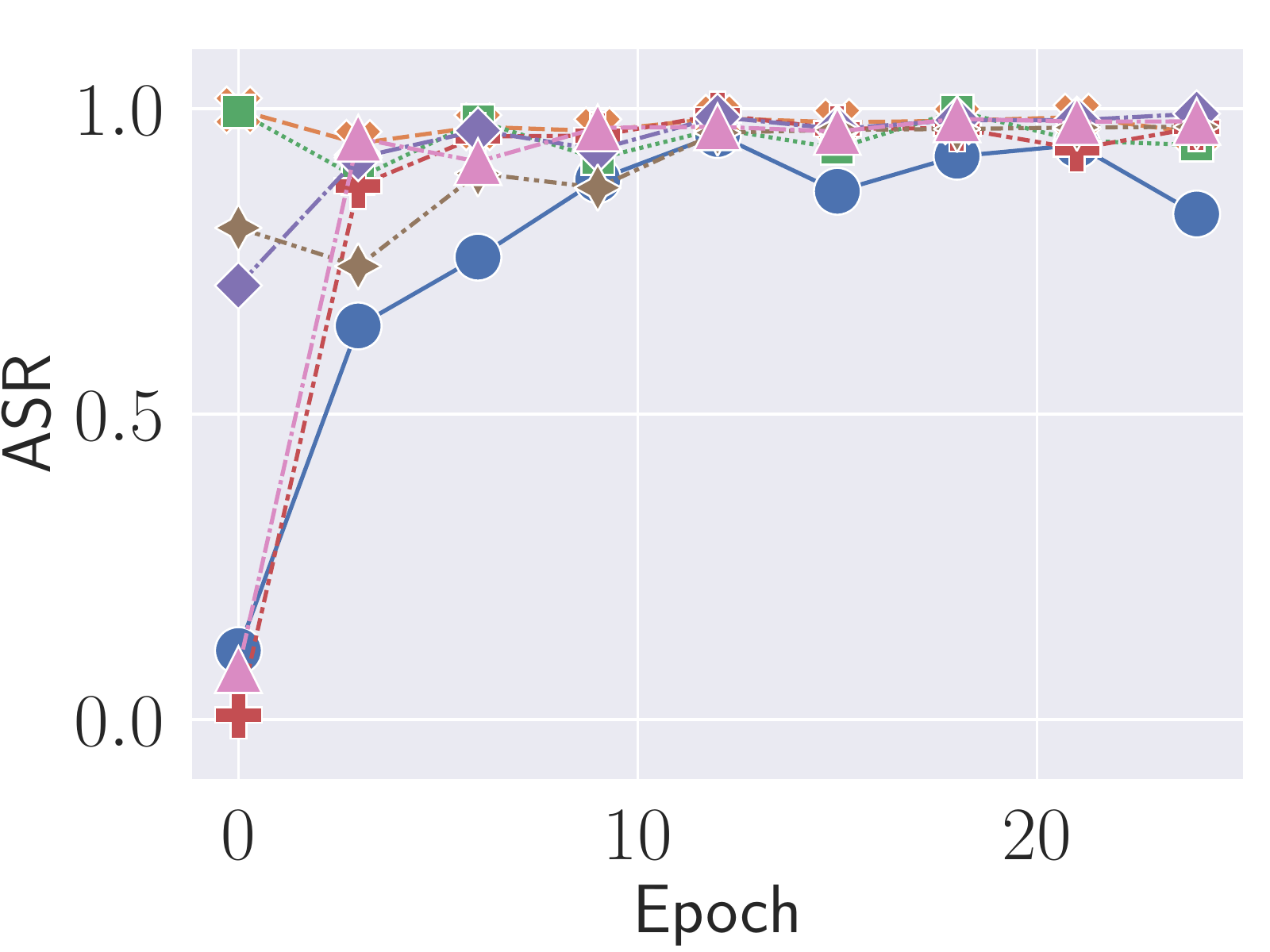}
\caption{Poison Ratio: 0.01}
\label{fig:rebl-0.01}
\end{subfigure}
\begin{subfigure}{0.5\columnwidth}
\includegraphics[width=\columnwidth]{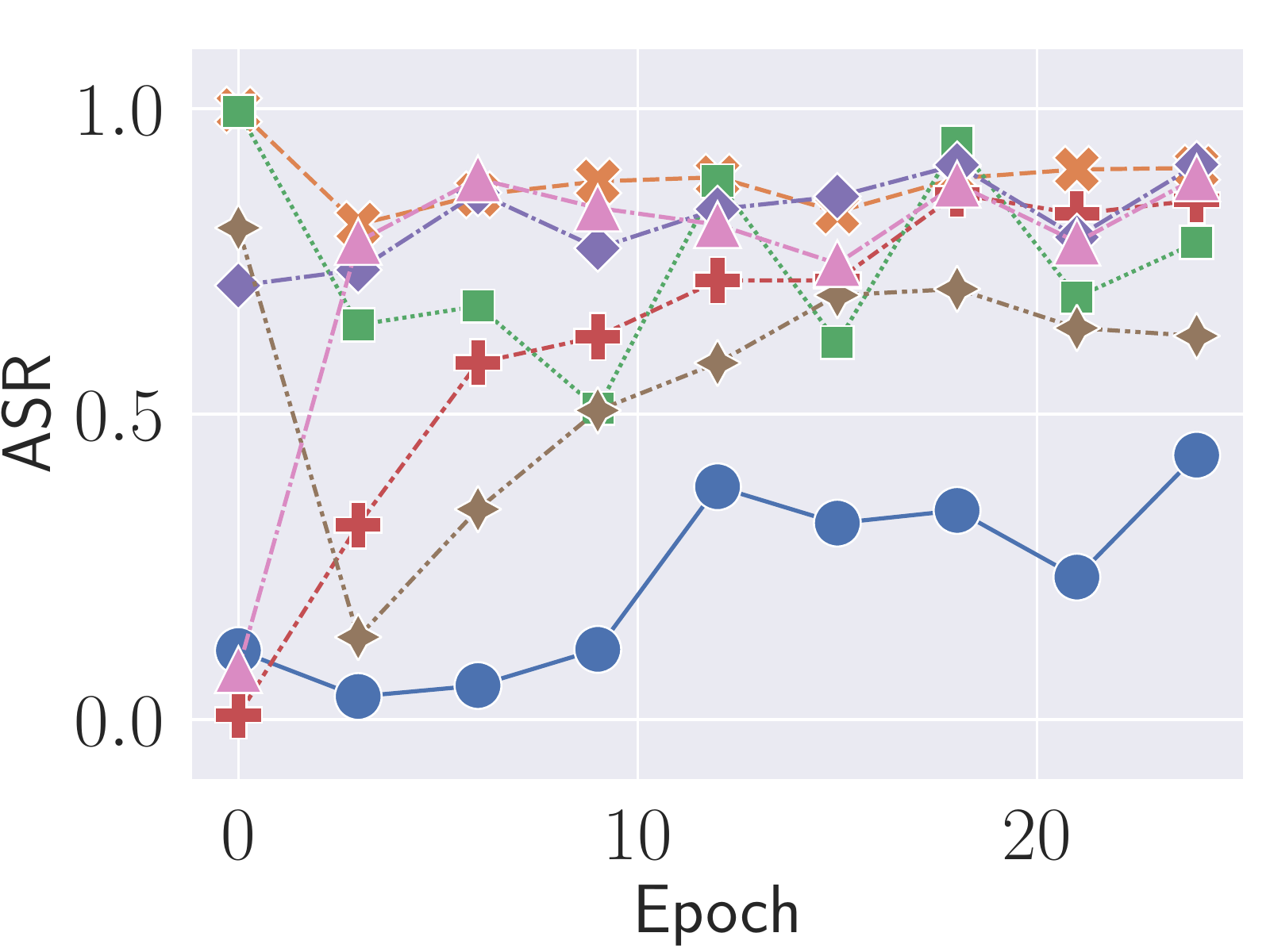}
\caption{Poison Ratio: 0.001}
\label{fig:rebl-0.001}
\end{subfigure}
\begin{subfigure}{0.5\columnwidth}
\includegraphics[width=\columnwidth]{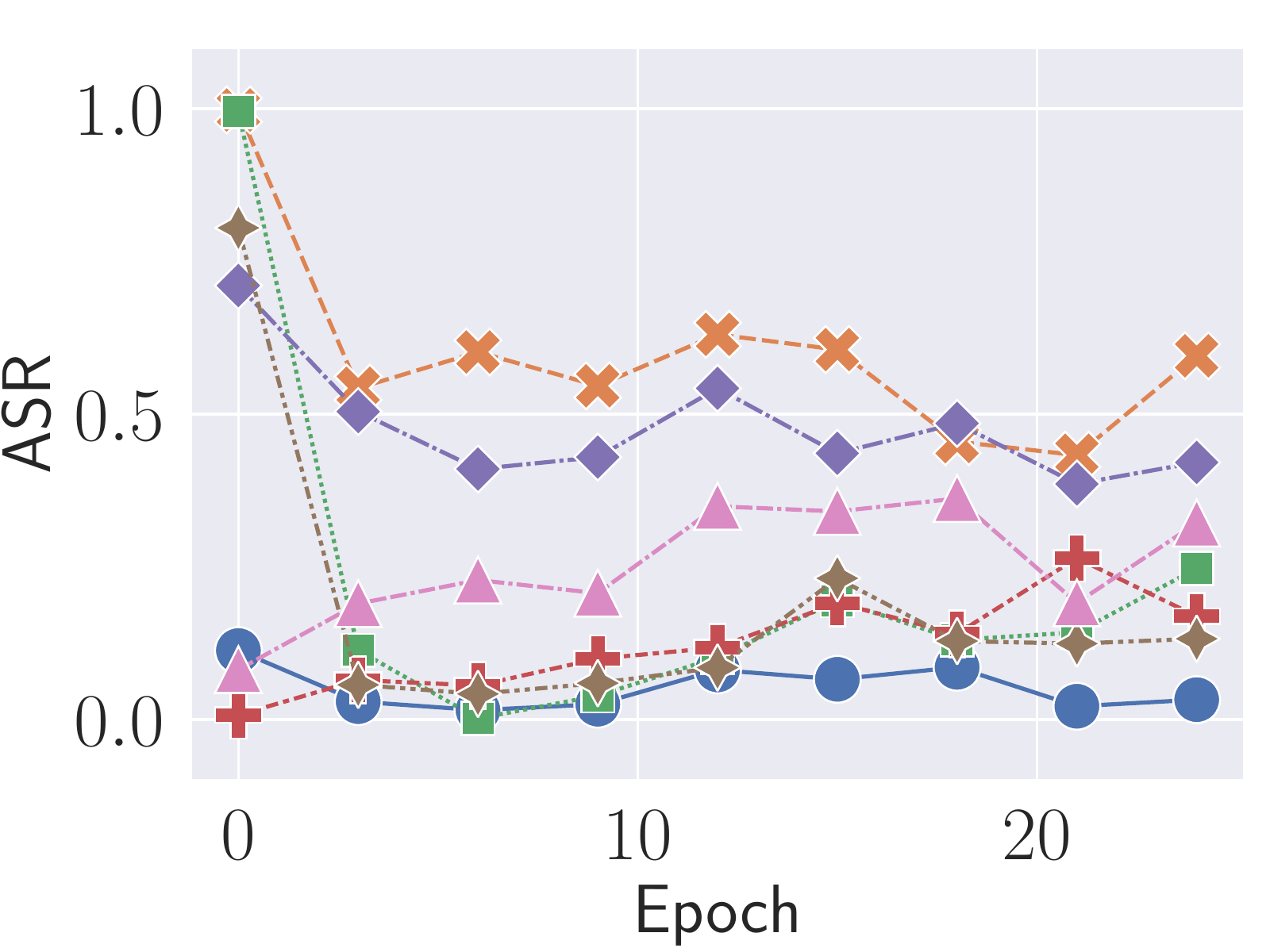}
\caption{Poison Ratio: 0.0001}
\label{fig:rebl-0.0001}
\end{subfigure}
\caption{
Performance of Blended backdoor re-injection attacks on different defense methods.
The X-axis represents training epochs in the re-injection phase.
The Y-axis represents the accuracy of poison samples.
}
\label{fig:reinject-blended}
\end{figure*}

\begin{figure*}[!t]
\centering
\begin{subfigure}{0.5\columnwidth}
\includegraphics[width=\columnwidth]{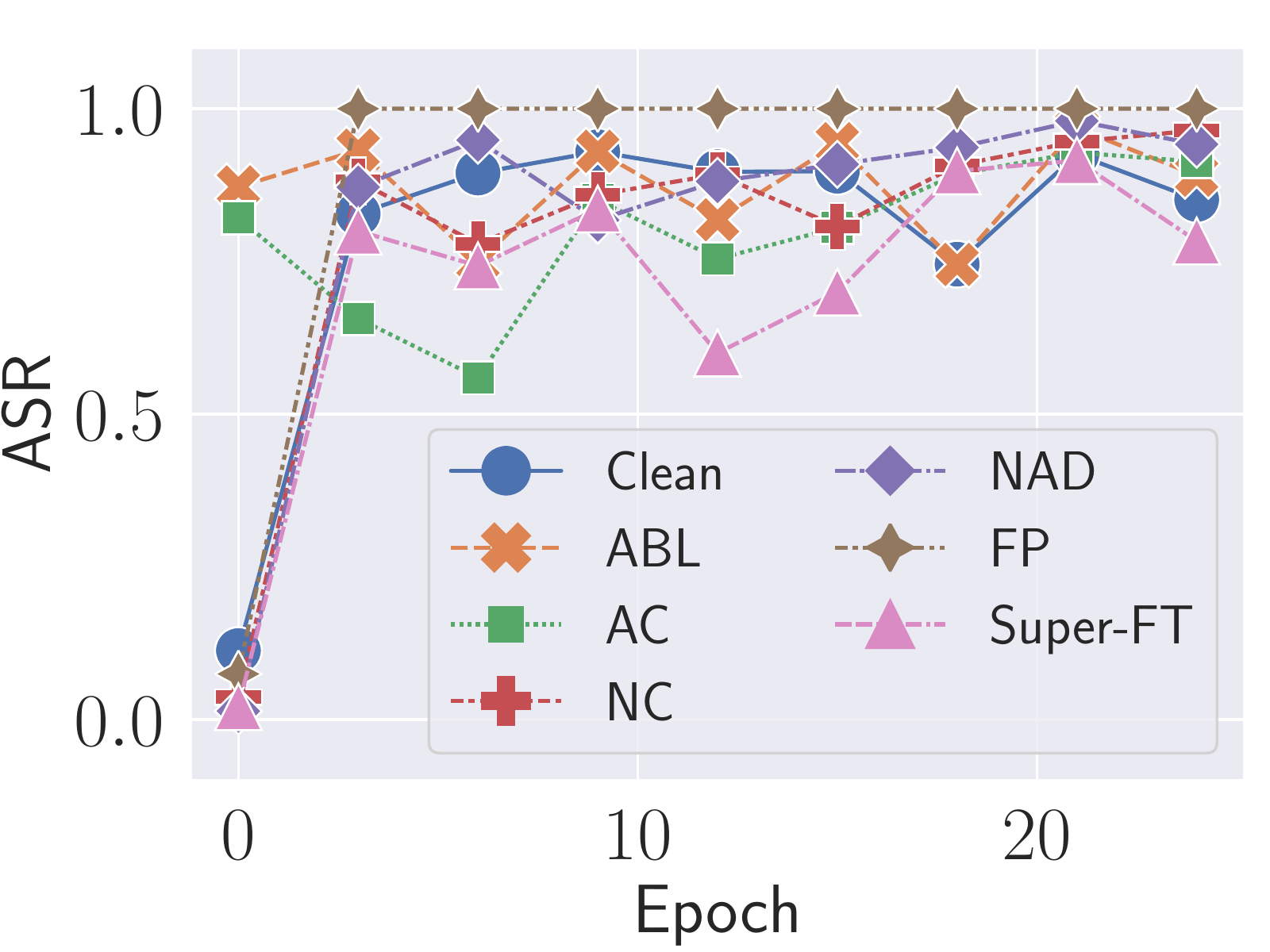}
\caption{Poison Ratio: 0.1}
\label{fig:reip-0.1}
\end{subfigure}
\begin{subfigure}{0.5\columnwidth}
\includegraphics[width=\columnwidth]{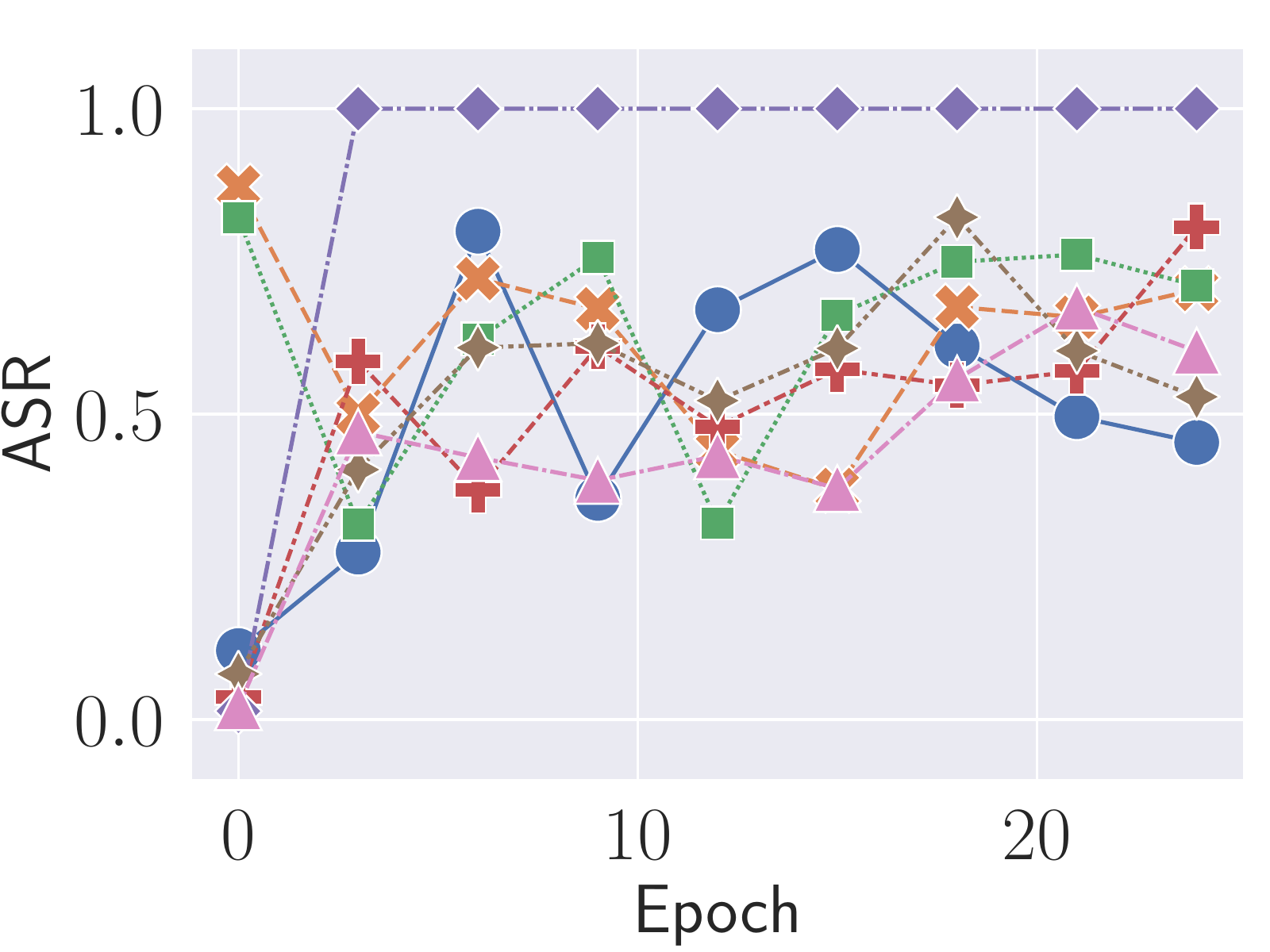}
\caption{Poison Ratio: 0.01}
\label{fig:reip-0.01}
\end{subfigure}
\begin{subfigure}{0.5\columnwidth}
\includegraphics[width=\columnwidth]{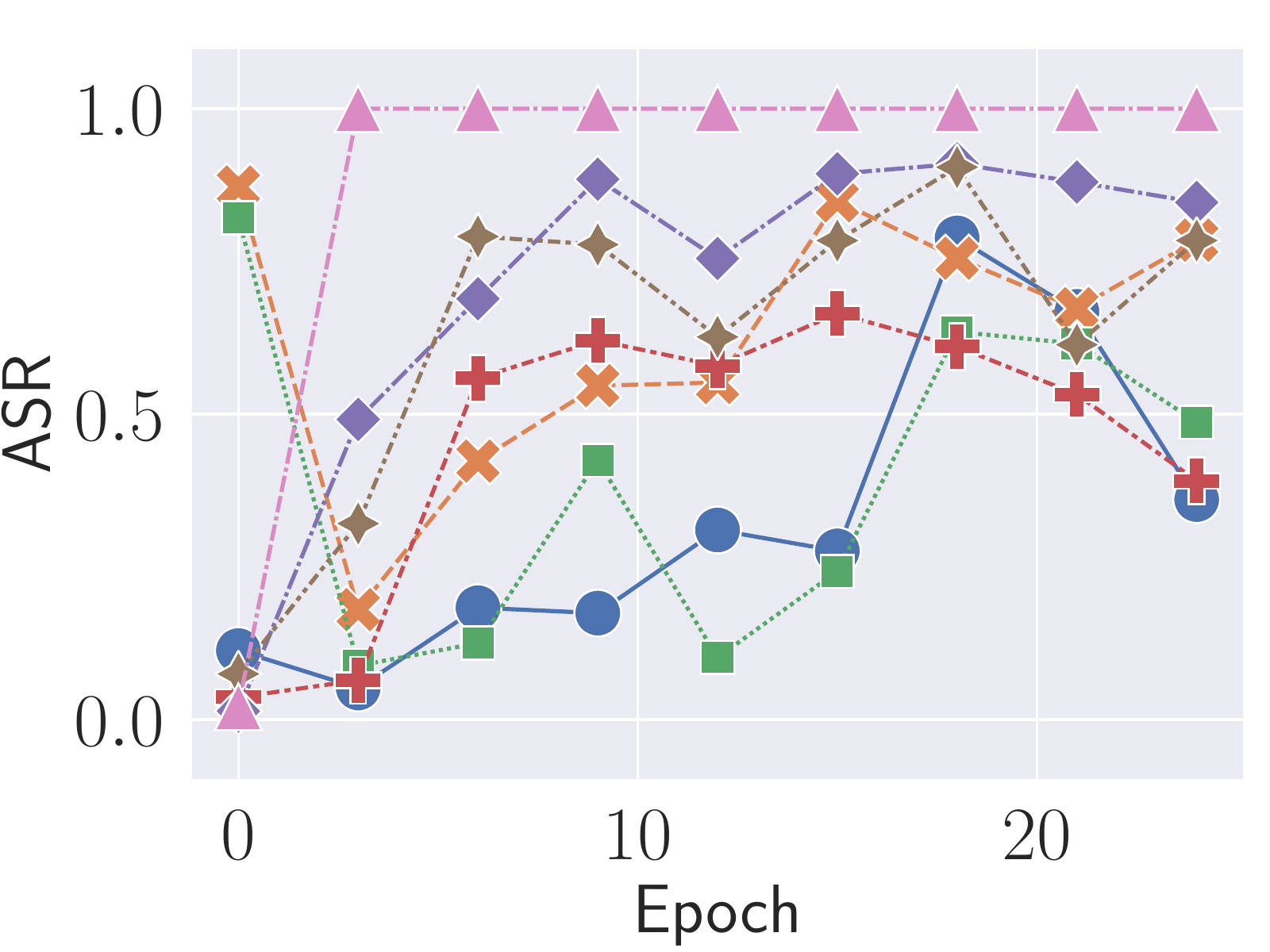}
\caption{Poison Ratio: 0.001}
\label{fig:reip-0.001}
\end{subfigure}
\begin{subfigure}{0.5\columnwidth}
\includegraphics[width=\columnwidth]{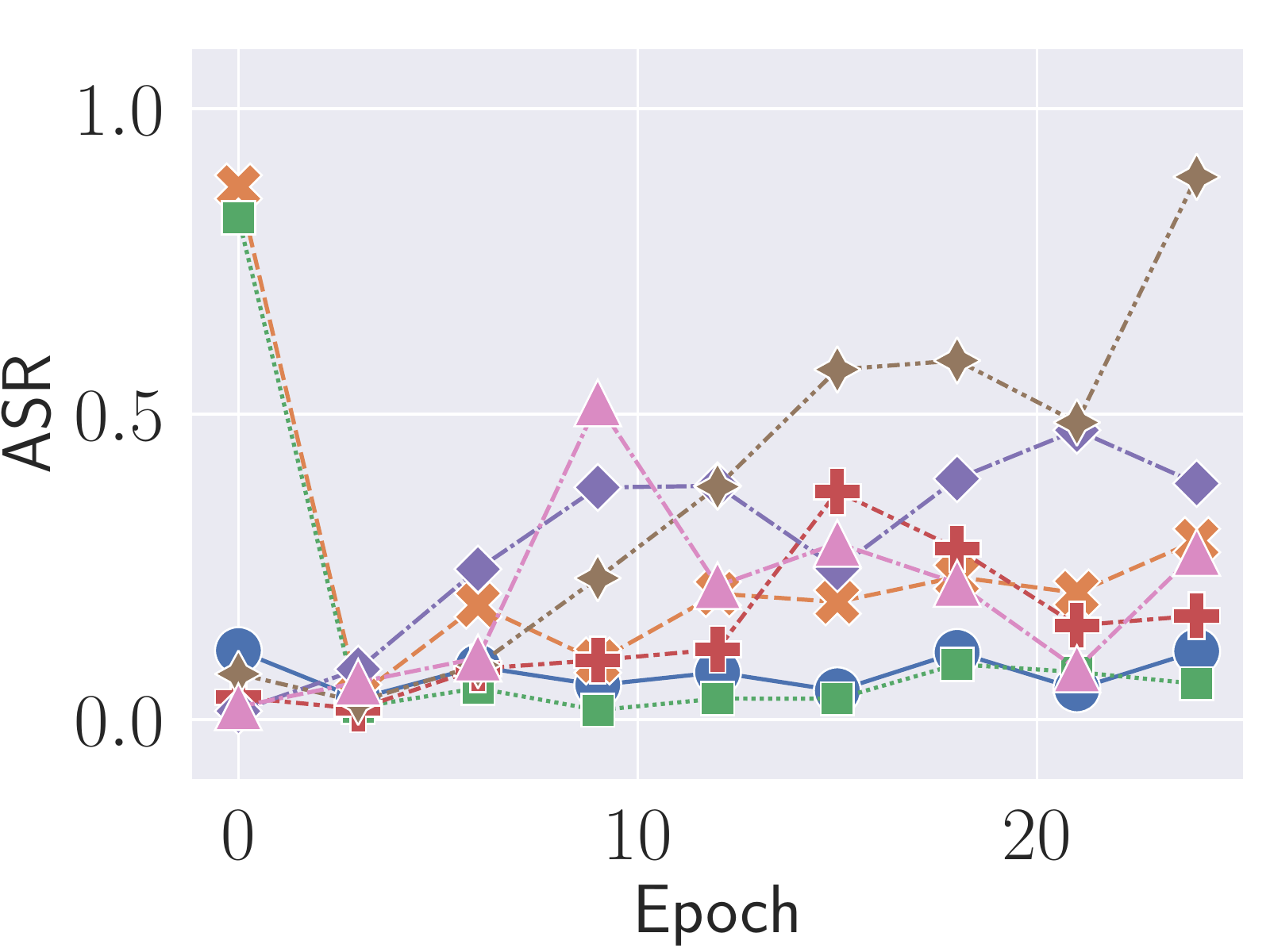}
\caption{Poison Ratio: 0.0001}
\label{fig:reip-0.0001}
\end{subfigure}
\caption{
Performance of Inputaware backdoor re-injection attacks on different defense methods.
The X-axis represents training epochs in the re-injection phase.
The Y-axis represents the accuracy of poison samples.
}
\label{fig:reinject-inputaware}
\end{figure*}

\autoref{fig:reinject} shows our results of BadNets backdoor re-injection attacks.
Other attacks' results, including Blended (\autoref{fig:reinject-blended}), Inputaware (\autoref{fig:reinject-inputaware}), LF (\autoref{fig:reinject-lf}), and WaNet (\autoref{fig:reinject-wanet}) are shown in the appendix.
For the poison ratio, we adjust it from 0.1 to 0.0001 for completeness.
Note that 0.1 is the poison ratio we used in previous backdoor attacks.
For the training process, we only show the first 25 epochs to see whether the backdoor can be injected within a few epochs.

From \autoref{fig:reinject}, it can first be observed that almost all defense methods we have tried can make the defended model more vulnerable to backdoor re-injection attacks.
For instance, when the attack method is BadNets with a poison ratio of 0.01 (\autoref{fig:rebadnet-0.01}), the adversary needs 15 epochs to insert the backdoor into the clean model (ASR=0.819) while the backdoor re-injection attacks against defended models only need 3-6 epochs to achieve even higher ASR.
We also see that, in all cases, fine-tuned models are not the most vulnerable models to backdoor re-injection attacks.
For instance, when the poison ratio is 0.001, the adversary can achieve 0.023 ASR in three epochs on models defended by super-fine-tuning, while other defense methods like ABL increases the ASR to 0.763.

% ----------------------------------------------------
\subsection{Summary}
% ----------------------------------------------------

In this section, we propose the new term, \emph{backdoor sequela}, to measure how backdoor defense methods affect a model's vulnerability to other attacks.
Specifically, we consider membership inference attacks and backdoor re-injection attacks.
Our evaluation results show that super-fine-tuning can even make the model more robust to membership inference attacks.
We also find that, in general, existing defense methods considered in our experiments make the defended models more vulnerable to backdoor re-injection attacks compared to the attacks on the clean model.
To our knowledge, we are the first to study backdoor sequela, and we argue that backdoor sequela should be considered as an important metric to evaluate a backdoor defense's efficacy.
We plan to investigate more backdoor sequela, i.e., attacks, in the future.

% ----------------------------------------------------
\section{Related Work}
\label{section:relatedwork}
% ----------------------------------------------------

% ----------------------------------------------------
\subsection{Backdoor Attacks}
% ----------------------------------------------------

Backdoor attacks, as one of the major threats to ML systems, have been widely studied.
BadNets~\cite{GDG17} is the first work to show that the adversary can insert a backdoor into the machine learning model via poisoning training datasets.
Then, targeted backdoor~\cite{CLLLS17} was proposed to show that with undetectable and random-position triggers, the adversary can still successfully launch backdoor attacks.
After that, more works focus on how to design better trigger patterns for backdoor attacks~\cite{NT20,SWBMZ22,LMBL20}.
Zheng et al.~\cite{ZPMJ21} propose backdoor attacks that are undetectable from the frequency perspective.
Nguyen et al.~\cite{NT20} argue that different images should have different triggers.
Therefore, they propose the generator to produce triggers for different images.
Besides better trigger patterns, there are also many works focused on designing better injection processes~\cite{LMALZWZ18,YLZZ19,ZMZBCJ20,LMBL20,BKT19}.
In these works, they find that the adversary can even inject a backdoor into the model without changing the label of the poison samples.
Other studies on backdoor attacks in various scenarios include~\cite{JLG22,SHLSBZ22}.

% ----------------------------------------------------
\subsection{Backdoor Defenses}
% ----------------------------------------------------

Following the increasing popularity of backdoor attacks, various defense methods have been proposed.
Current defense methods can be divided into three categories.
The most popular defense methods are based on reverse engineering, where the defender aims to reverse the possible backdoor triggers to judge whether the given model is the backdoored model or not~\cite{WYSLVZZ19,CFZK19,LLTMAZ19,HAS19,GWXDS19}.
Besides reverse engineering, another line of works~\cite{TLM18,CCBLELMS18,GXWCRN19,UPWLRC22} focuses on mitigating the backdoor by detecting the poison samples.
Due to the fact that backdoor attacks may involve more carefully designed triggers to bypass potential defenses, such detection methods look like the arms race with evolving attacks.
The last type of backdoor defense method is based on meta-learning to learn the difference between the backdoored models and the clean models by training with a large number of backdoored shadow models~\cite{XWLBGL21}.
Besides different detection methods, fine-tuning has also been proposed to mitigate backdoor attacks.
Previous fine-tuning-based methods either adapt pruning~\cite{LDG18} or distillation~\cite{LLKLLM212}.
Liu et al.~\cite{LDG18} argue that fine-tuning itself cannot effectively mitigate backdoor attacks.
However, pruning, distillation, or other methods based on fine-tuning will cost much more computational resources and sacrifice the models' utility.
In this work, we show for the first time that carefully designed fine-tuning is sufficient to remove the backdoor and maintain the model's utility with limited cost (e.g., with limited epochs).

% ----------------------------------------------------
\section{Conclusion}
% ----------------------------------------------------

In this paper, we have demonstrated that fine-tuning is a very effective backdoor removal method.
We consider three scenarios, namely encoder-based, transfer-based, and standalone.
Our experimental results show that in the encoder-based scenario, whole model conventional fine-tuning can effectively remove backdoors within a few epochs.
As fine-tuning is a necessary step for users to train downstream classifiers, it can be argued that fine-tuning as a defense method incurs \emph{zero-cost}.
In the transfer-based scenario, fine-tuning is still a necessary step.
However, we find that conventional fine-tuning cannot always effectively remove backdoor attacks.
Therefore, we propose super-fine-tuning, a newly designed fine-tuning method for backdoor removal tasks.
Our experimental results show that super-fine-tuning can effectively mitigate backdoor attacks in this scenario.
In the most difficult standalone scenario, we show that super-fine-tuning is still effective.
We also compare super-fine-tuning with state-of-the-art defense methods and demonstrate that super-fine-tuning outperforms them.

Furthermore, we propose a new term, \emph{backdoor sequela}, to measure the defended model's vulnerability to other attacks.
Experiments show that super-fine-tuning does not have a strong impact on the defended models with respect to membership inference and backdoor re-injection attacks.
We hope that, in the future, backdoor sequela will be considered an important aspect for judging a backdoor defense's efficacy.

Our results demonstrate that backdoor defenses can be performed in an easier way than previously considered.
Fine-tuning or super-fine-tuning is sufficient in most cases.
We hope our methods can help ML model owners better shield their models from backdoor attacks.
Also, it further calls for the design of more advanced attacks in order to comprehensively assess machine learning models' vulnerabilities to backdoor attacks.

% ----------------------------------------------------
\begin{small}
\bibliographystyle{plain}
\bibliography{normal_generated_py3}

\begin{thebibliography}{10}

\bibitem{CIFAR}
\url{https://www.cs.toronto.edu/~kriz/cifar.html}.

\bibitem{STL10}
\url{https://cs.stanford.edu/\%7Eacoates/stl10/}.

\bibitem{GTSRB}
\url{http://benchmark.ini.rub.de/?section=gtsrb}.

\bibitem{SVHN}
\url{http://ufldl.stanford.edu/housenumbers/}.

\bibitem{BKT19}
Mauro Barni, Kassem Kallas, and Benedetta Tondi.
\newblock {A New Backdoor Attack in {CNNS} by Training Set Corruption Without
  Label Poisoning}.
\newblock In {\em {IEEE International Conference on Image Processing (ICIP)}},
  pages 101--105. IEEE, 2019.

\bibitem{CCBLELMS18}
Bryant Chen, Wilka Carvalho, Nathalie Baracaldo, Heiko Ludwig, Benjamin
  Edwards, Taesung Lee, Ian~M. Molloy, and Biplav Srivastava.
\newblock {Detecting Backdoor Attacks on Deep Neural Networks by Activation
  Clustering}.
\newblock {\em {CoRR abs/1811.03728}}, 2018.

\bibitem{CFZK19}
Huili Chen, Cheng Fu, Jishen Zhao, and Farinaz Koushanfar.
\newblock {DeepInspect: {A} Black-box Trojan Detection and Mitigation Framework
  for Deep Neural Networks}.
\newblock In {\em {International Joint Conferences on Artifical Intelligence
  (IJCAI)}}, pages 4658--4664. IJCAI, 2019.

\bibitem{CKNH20}
Ting Chen, Simon Kornblith, Mohammad Norouzi, and Geoffrey~E. Hinton.
\newblock {A Simple Framework for Contrastive Learning of Visual
  Representations}.
\newblock In {\em {International Conference on Machine Learning (ICML)}}, pages
  1597--1607. PMLR, 2020.

\bibitem{CH21}
Xinlei Chen and Kaiming He.
\newblock {Exploring Simple Siamese Representation Learning}.
\newblock In {\em {IEEE Conference on Computer Vision and Pattern Recognition
  (CVPR)}}, pages 15750--15758. IEEE, 2021.

\bibitem{CLLLS17}
Xinyun Chen, Chang Liu, Bo~Li, Kimberly Lu, and Dawn Song.
\newblock {Targeted Backdoor Attacks on Deep Learning Systems Using Data
  Poisoning}.
\newblock {\em {CoRR abs/1712.05526}}, 2017.

\bibitem{CTCP21}
Christopher A.~Choquette Choo, Florian Tram{\`e}r, Nicholas Carlini, and
  Nicolas Papernot.
\newblock {Label-Only Membership Inference Attacks}.
\newblock In {\em {International Conference on Machine Learning (ICML)}}, pages
  1964--1974. PMLR, 2021.

\bibitem{DDSLLF09}
Jia Deng, Wei Dong, Richard Socher, Li{-}Jia Li, Kai Li, and Li~Fei{-}Fei.
\newblock {ImageNet: {A} large-scale hierarchical image database}.
\newblock In {\em {IEEE Conference on Computer Vision and Pattern Recognition
  (CVPR)}}, pages 248--255. IEEE, 2009.

\bibitem{GXWCRN19}
Yansong Gao, Change Xu, Derui Wang, Shiping Chen, Damith~C Ranasinghe, and
  Surya Nepal.
\newblock {STRIP: A Defence Against Trojan Attacks on Deep Neural Networks}.
\newblock In {\em {Annual Computer Security Applications Conference (ACSAC)}},
  pages 113--125. ACM, 2019.

\bibitem{GSATRBDPGAPKMV20}
Jean{-}Bastien Grill, Florian Strub, Florent Altch{\'{e}}, Corentin Tallec,
  Pierre~H. Richemond, Elena Buchatskaya, Carl Doersch, Bernardo~{\'{A}}vila
  Pires, Zhaohan Guo, Mohammad~Gheshlaghi Azar, Bilal Piot, Koray Kavukcuoglu,
  R{\'{e}}mi Munos, and Michal Valko.
\newblock {Bootstrap Your Own Latent - {A} New Approach to Self-Supervised
  Learning}.
\newblock In {\em {Annual Conference on Neural Information Processing Systems
  (NeurIPS)}}. NeurIPS, 2020.

\bibitem{GDG17}
Tianyu Gu, Brendan Dolan-Gavitt, and Siddharth Grag.
\newblock {Badnets: Identifying Vulnerabilities in the Machine Learning Model
  Supply Chain}.
\newblock {\em {CoRR abs/1708.06733}}, 2017.

\bibitem{GWXDS19}
Wenbo Guo, Lun Wang, Xinyu Xing, Min Du, and Dawn Song.
\newblock {{TABOR:} {A} Highly Accurate Approach to Inspecting and Restoring
  Trojan Backdoors in {AI} Systems}.
\newblock {\em {CoRR abs/1908.01763}}, 2019.

\bibitem{HCXLDG21}
Kaiming He, Xinlei Chen, Saining Xie, Yanghao Li, Piotr Doll{\'{a}}r, and
  Ross~B. Girshick.
\newblock {Masked Autoencoders Are Scalable Vision Learners}.
\newblock {\em {CoRR abs/2111.06377}}, 2021.

\bibitem{HFWXG20}
Kaiming He, Haoqi Fan, Yuxin Wu, Saining Xie, and Ross~B. Girshick.
\newblock {Momentum Contrast for Unsupervised Visual Representation Learning}.
\newblock In {\em {IEEE Conference on Computer Vision and Pattern Recognition
  (CVPR)}}, pages 9726--9735. IEEE, 2020.

\bibitem{HZRS16}
Kaiming He, Xiangyu Zhang, Shaoqing Ren, and Jian Sun.
\newblock {Deep Residual Learning for Image Recognition}.
\newblock In {\em {IEEE Conference on Computer Vision and Pattern Recognition
  (CVPR)}}, pages 770--778. IEEE, 2016.

\bibitem{HZRS162}
Kaiming He, Xiangyu Zhang, Shaoqing Ren, and Jian Sun.
\newblock {Identity Mappings in Deep Residual Networks}.
\newblock In {\em {European Conference on Computer Vision (ECCV)}}, pages
  630--645. Springer, 2016.

\bibitem{HLXCZ22}
Xinlei He, Zheng Li, Weilin Xu, Cory Cornelius, and Yang Zhang.
\newblock {Membership-Doctor: Comprehensive Assessment of Membership Inference
  Against Machine Learning Models}.
\newblock {\em {CoRR abs/2208.10445}}, 2022.

\bibitem{HAS19}
Xijie Huang, Moustafa Alzantot, and Mani~B. Srivastava.
\newblock {NeuronInspect: Detecting Backdoors in Neural Networks via Output
  Explanations}.
\newblock {\em {CoRR abs/1911.07399}}, 2019.

\bibitem{JLG22}
Jinyuan Jia, Yupei Liu, and Neil~Zhenqiang Gong.
\newblock {BadEncoder: Backdoor Attacks to Pre-trained Encoders in
  Self-Supervised Learning}.
\newblock In {\em {IEEE Symposium on Security and Privacy (S\&P)}}. IEEE, 2022.

\bibitem{KTWSTIMLK20}
Prannay Khosla, Piotr Teterwak, Chen Wang, Aaron Sarna, Yonglong Tian, Phillip
  Isola, Aaron Maschinot, Ce~Liu, and Dilip Krishnan.
\newblock {Supervised Contrastive Learning}.
\newblock In {\em {Annual Conference on Neural Information Processing Systems
  (NeurIPS)}}. NeurIPS, 2020.

\bibitem{KB15}
Diederik~P. Kingma and Jimmy Ba.
\newblock {Adam: A Method for Stochastic Optimization}.
\newblock In {\em {International Conference on Learning Representations
  (ICLR)}}, 2015.

\bibitem{KSL19}
Simon Kornblith, Jonathon Shlens, and Quoc~V. Le.
\newblock {Do Better ImageNet Models Transfer Better?}
\newblock In {\em {IEEE Conference on Computer Vision and Pattern Recognition
  (CVPR)}}, pages 2661--2671. IEEE, 2019.

\bibitem{LF20}
Klas Leino and Matt Fredrikson.
\newblock {Stolen Memories: Leveraging Model Memorization for Calibrated
  White-Box Membership Inference}.
\newblock In {\em {USENIX Security Symposium (USENIX Security)}}, pages
  1605--1622. USENIX, 2020.

\bibitem{LCYLRBS20}
Hao Li, Pratik Chaudhari, Hao Yang, Michael Lam, Avinash Ravichandran, Rahul
  Bhotika, and Stefano Soatto.
\newblock {Rethinking the Hyperparameters for Fine-tuning}.
\newblock In {\em {International Conference on Learning Representations
  (ICLR)}}, 2020.

\bibitem{LLKLLM21}
Yige Li, Xixiang Lyu, Nodens Koren, Lingjuan Lyu, Bo~Li, and Xingjun Ma.
\newblock {Anti-Backdoor Learning: Training Clean Models on Poisoned Data}.
\newblock In {\em {Annual Conference on Neural Information Processing Systems
  (NeurIPS)}}, pages 14900--14912. NeurIPS, 2021.

\bibitem{LLKLLM212}
Yige Li, Xixiang Lyu, Nodens Koren, Lingjuan Lyu, Bo~Li, and Xingjun Ma.
\newblock {Neural Attention Distillation: Erasing Backdoor Triggers from Deep
  Neural Networks}.
\newblock In {\em {International Conference on Learning Representations
  (ICLR)}}, 2021.

\bibitem{LLWLHL21}
Yuezun Li, Yiming Li, Baoyuan Wu, Longkang Li, Ran He, and Siwei Lyu.
\newblock {Invisible Backdoor Attack with Sample-Specific Triggers}.
\newblock In {\em {IEEE International Conference on Computer Vision (ICCV)}},
  pages 16443--16452. IEEE, 2021.

\bibitem{LZ21}
Zheng Li and Yang Zhang.
\newblock {Membership Leakage in Label-Only Exposures}.
\newblock In {\em {ACM SIGSAC Conference on Computer and Communications
  Security (CCS)}}, pages 880--895. ACM, 2021.

\bibitem{LDG18}
Kang Liu, Brendan Dolan{-}Gavitt, and Siddharth Garg.
\newblock {Fine-Pruning: Defending Against Backdooring Attacks on Deep Neural
  Networks}.
\newblock In {\em {Research in Attacks, Intrusions, and Defenses (RAID)}},
  pages 273--294. Springer, 2018.

\bibitem{LLTMAZ19}
Yingqi Liu, Wen-Chuan Lee, Guanhong Tao, Shiqing Ma, Yousra Aafer, and Xiangyu
  Zhang.
\newblock {ABS: Scanning Neural Networks for Back-Doors by Artificial Brain
  Stimulation}.
\newblock In {\em {ACM SIGSAC Conference on Computer and Communications
  Security (CCS)}}, pages 1265--1282. ACM, 2019.

\bibitem{LMALZWZ18}
Yingqi Liu, Shiqing Ma, Yousra Aafer, Wen-Chuan Lee, Juan Zhai, Weihang Wang,
  and Xiangyu Zhang.
\newblock {Trojaning Attack on Neural Networks}.
\newblock In {\em {Network and Distributed System Security Symposium (NDSS)}}.
  Internet Society, 2018.

\bibitem{LWHSZBCFZ22}
Yugeng Liu, Rui Wen, Xinlei He, Ahmed Salem, Zhikun Zhang, Michael Backes,
  Emiliano~De Cristofaro, Mario Fritz, and Yang Zhang.
\newblock {ML-Doctor: Holistic Risk Assessment of Inference Attacks Against
  Machine Learning Models}.
\newblock In {\em {USENIX Security Symposium (USENIX Security)}}, pages
  4525--4542. USENIX, 2022.

\bibitem{LMBL20}
Yunfei Liu, Xingjun Ma, James Bailey, and Feng Lu.
\newblock {Reflection Backdoor: A Natural Backdoor Attack on Deep Neural
  Networks}.
\newblock In {\em {European Conference on Computer Vision (ECCV)}}, pages
  182--199. Springer, 2020.

\bibitem{NSH18}
Milad Nasr, Reza Shokri, and Amir Houmansadr.
\newblock {Machine Learning with Membership Privacy using Adversarial
  Regularization}.
\newblock In {\em {ACM SIGSAC Conference on Computer and Communications
  Security (CCS)}}, pages 634--646. ACM, 2018.

\bibitem{NSH19}
Milad Nasr, Reza Shokri, and Amir Houmansadr.
\newblock {Comprehensive Privacy Analysis of Deep Learning: Passive and Active
  White-box Inference Attacks against Centralized and Federated Learning}.
\newblock In {\em {IEEE Symposium on Security and Privacy (S\&P)}}, pages
  1021--1035. IEEE, 2019.

\bibitem{NT20}
Tuan~Anh Nguyen and Anh Tran.
\newblock {Input-Aware Dynamic Backdoor Attack}.
\newblock In {\em {Annual Conference on Neural Information Processing Systems
  (NeurIPS)}}. NeurIPS, 2020.

\bibitem{NT21}
Tuan~Anh Nguyen and Anh~Tuan Tran.
\newblock {WaNet - Imperceptible Warping-based Backdoor Attack}.
\newblock In {\em {International Conference on Learning Representations
  (ICLR)}}, 2021.

\bibitem{PSZJVLLW20}
Ren Pang, Hua Shen, Xinyang Zhang, Shouling Ji, Yevgeniy Vorobeychik, Xiapu
  Luo, Alex~X. Liu, and Ting Wang.
\newblock {A Tale of Evil Twins: Adversarial Inputs versus Poisoned Models}.
\newblock In {\em {ACM SIGSAC Conference on Computer and Communications
  Security (CCS)}}, pages 85--99. ACM, 2020.

\bibitem{PZGXJCW20}
Ren Pang, Zheng Zhang, Xiangshan Gao, Zhaohan Xi, Shouling Ji, Peng Cheng, and
  Ting Wang.
\newblock {{TROJANZOO:} Everything You Ever Wanted to Know about Neural
  Backdoors (But Were Afraid to Ask)}.
\newblock {\em {CoRR abs/2012.09302}}, 2020.

\bibitem{SWBMZ22}
Ahmed Salem, Rui Wen, Michael Backes, Shiqing Ma, and Yang Zhang.
\newblock {Dynamic Backdoor Attacks Against Machine Learning Models}.
\newblock In {\em {IEEE European Symposium on Security and Privacy (Euro
  S\&P)}}, pages 703--718. IEEE, 2022.

\bibitem{SZHBFB19}
Ahmed Salem, Yang Zhang, Mathias Humbert, Pascal Berrang, Mario Fritz, and
  Michael Backes.
\newblock {ML-Leaks: Model and Data Independent Membership Inference Attacks
  and Defenses on Machine Learning Models}.
\newblock In {\em {Network and Distributed System Security Symposium (NDSS)}}.
  Internet Society, 2019.

\bibitem{SHLSBZ22}
Xinyue Shen, Xinlei He, Zheng Li, Yun Shen, Michael Backes, and Yang Zhang.
\newblock {Backdoor Attacks in the Supply Chain of Masked Image Modeling}.
\newblock {\em {CoRR abs/2210.01632}}, 2022.

\bibitem{SSSS17}
Reza Shokri, Marco Stronati, Congzheng Song, and Vitaly Shmatikov.
\newblock {Membership Inference Attacks Against Machine Learning Models}.
\newblock In {\em {IEEE Symposium on Security and Privacy (S\&P)}}, pages
  3--18. IEEE, 2017.

\bibitem{ST18}
Leslie~N. Smith and Nicholay Topin.
\newblock {Super-Convergence: Very Fast Training of Neural Networks Using Large
  Learning Rates}.
\newblock {\em {CoRR abs/1708.07120}}, 2018.

\bibitem{SM21}
Liwei Song and Prateek Mittal.
\newblock {Systematic Evaluation of Privacy Risks of Machine Learning Models}.
\newblock In {\em {USENIX Security Symposium (USENIX Security)}}. USENIX, 2021.

\bibitem{SSM19}
Liwei Song, Reza Shokri, and Prateek Mittal.
\newblock {Privacy Risks of Securing Machine Learning Models against
  Adversarial Examples}.
\newblock In {\em {ACM SIGSAC Conference on Computer and Communications
  Security (CCS)}}, pages 241--257. ACM, 2019.

\bibitem{TKI20}
Yonglong Tian, Dilip Krishnan, and Phillip Isola.
\newblock {Contrastive Representation Distillation}.
\newblock In {\em {International Conference on Learning Representations
  (ICLR)}}, 2020.

\bibitem{TSPKSI20}
Yonglong Tian, Chen Sun, Ben Poole, Dilip Krishnan, Cordelia Schmid, and
  Phillip Isola.
\newblock {What Makes for Good Views for Contrastive Learning?}
\newblock In {\em {Annual Conference on Neural Information Processing Systems
  (NeurIPS)}}. NeurIPS, 2020.

\bibitem{TLM18}
Brandon Tran, Jerry Li, and Aleksander Madry.
\newblock {Spectral Signatures in Backdoor Attacks}.
\newblock In {\em {Annual Conference on Neural Information Processing Systems
  (NeurIPS)}}, pages 8011--8021. NeurIPS, 2018.

\bibitem{UPWLRC22}
Sakshi Udeshi, Shanshan Peng, Gerald Woo, Lionell Loh, Louth Rawshan, and
  Sudipta Chattopadhyay.
\newblock {Model Agnostic Defence Against Backdoor Attacks in Machine
  Learning}.
\newblock {\em {IEEE Transactions on Reliability}}, 2022.

\bibitem{WYSLVZZ19}
Bolun Wang, Yuanshun Yao, Shawn Shan, Huiying Li, Bimal Viswanath, Haitao
  Zheng, and Ben~Y. Zhao.
\newblock {Neural Cleanse: Identifying and Mitigating Backdoor Attacks in
  Neural Networks}.
\newblock In {\em {IEEE Symposium on Security and Privacy (S\&P)}}, pages
  707--723. IEEE, 2019.

\bibitem{WCZZWYSZ22}
Baoyuan Wu, Hongrui Chen, Mingda Zhang, Zihao Zhu, Shaokui Wei, Danni Yuan,
  Chao Shen, and Hongyuan Zha.
\newblock {BackdoorBench: {A} Comprehensive Benchmark of Backdoor Learning}.
\newblock {\em {CoRR abs/2206.12654}}, 2022.

\bibitem{XWLBGL21}
Xiaojun Xu, Qi~Wang, Huichen Li, Nikita Borisov, Carl~A. Gunter, and Bo~Li.
\newblock {Detecting AI Trojans Using Meta Neural Analysis}.
\newblock In {\em {IEEE Symposium on Security and Privacy (S\&P)}}. IEEE, 2021.

\bibitem{YLZZ19}
Yuanshun Yao, Huiying Li, Haitao Zheng, and Ben~Y. Zhao.
\newblock {Latent Backdoor Attacks on Deep Neural Networks}.
\newblock In {\em {ACM SIGSAC Conference on Computer and Communications
  Security (CCS)}}, pages 2041--2055. ACM, 2019.

\bibitem{YGFJ18}
Samuel Yeom, Irene Giacomelli, Matt Fredrikson, and Somesh Jha.
\newblock {Privacy Risk in Machine Learning: Analyzing the Connection to
  Overfitting}.
\newblock In {\em {IEEE Computer Security Foundations Symposium (CSF)}}, pages
  268--282. IEEE, 2018.

\bibitem{ZPMJ21}
Yi~Zeng, Won Park, Z.~Morley Mao, and Ruoxi Jia.
\newblock {Rethinking the Backdoor Attacks' Triggers: {A} Frequency
  Perspective}.
\newblock In {\em {IEEE International Conference on Computer Vision (ICCV)}},
  pages 16453--16461. IEEE, 2021.

\bibitem{ZMZBCJ20}
Shihao Zhao, Xingjun Ma, Xiang Zheng, James Bailey, Jingjing Chen, and Yu-Gang
  Jiang.
\newblock {Clean-Label Backdoor Attacks on Video Recognition Models}.
\newblock In {\em {IEEE Conference on Computer Vision and Pattern Recognition
  (CVPR)}}, pages 14443--144528. IEEE, 2020.

\bibitem{ZQDXZZXH19}
Fuzhen Zhuang, Zhiyuan Qi, Keyu Duan, Dongbo Xi, Yongchun Zhu, Hengshu Zhu, Hui
  Xiong, and Qing He.
\newblock {A Comprehensive Survey on Transfer Learning}.
\newblock {\em {CoRR abs/1911.02685}}, 2019.

\end{thebibliography}
\end{small}
% ----------------------------------------------------

% ----------------------------------------------------
\section{Appendix}
% ----------------------------------------------------

\begin{figure*}[!t]
\centering
\includegraphics[width=2\columnwidth]{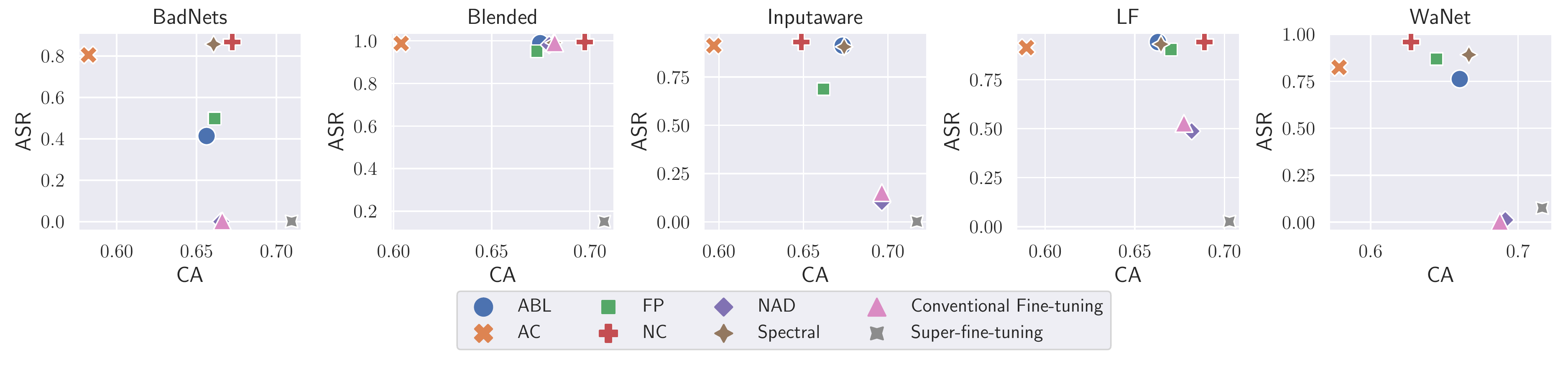}
\caption{
Comparison between existing state-of-the-art backdoor defense methods and super-fine-tuning on CIFAR100.
The X-axis represents accuracy on clean samples.
The Y-axis represents the attack success rate.
Points closer to the lower right corner are better points.
}
\label{fig:compare-asr-ca-cifar100}
\end{figure*}

\begin{figure*}[!t]
\centering
\includegraphics[width=2\columnwidth]{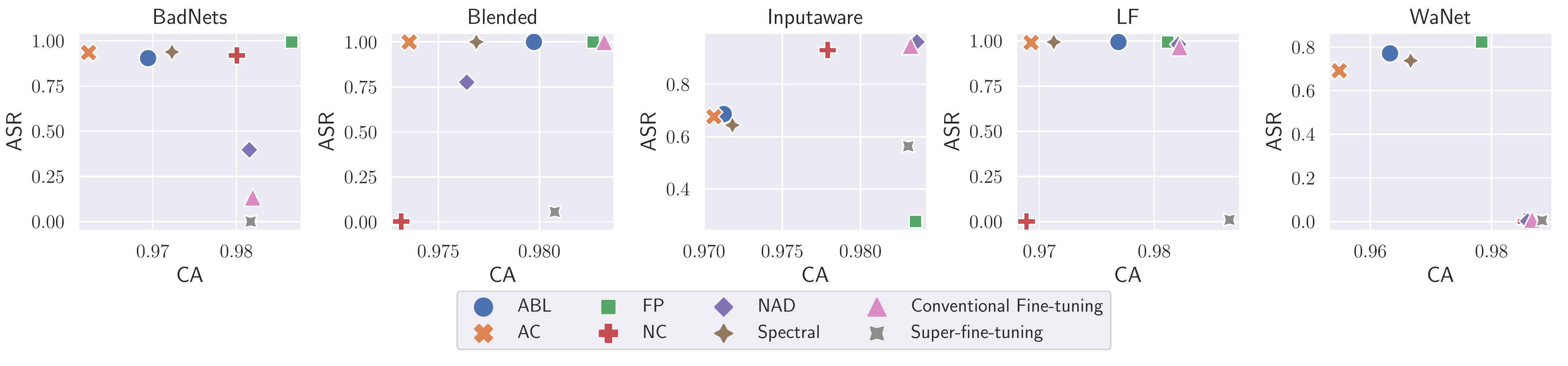}
\caption{
Comparison between existing state-of-the-art backdoor defense methods and super-fine-tuning on GTSRB.
The X-axis represents accuracy on clean samples.
The Y-axis represents the attack success rate.
Points closer to the lower right corner are better points.
}
\label{fig:compare-asr-ca-gtsrb}
\end{figure*}

\begin{figure*}[!t]
\centering
\begin{subfigure}{0.5\columnwidth}
\includegraphics[width=\columnwidth]{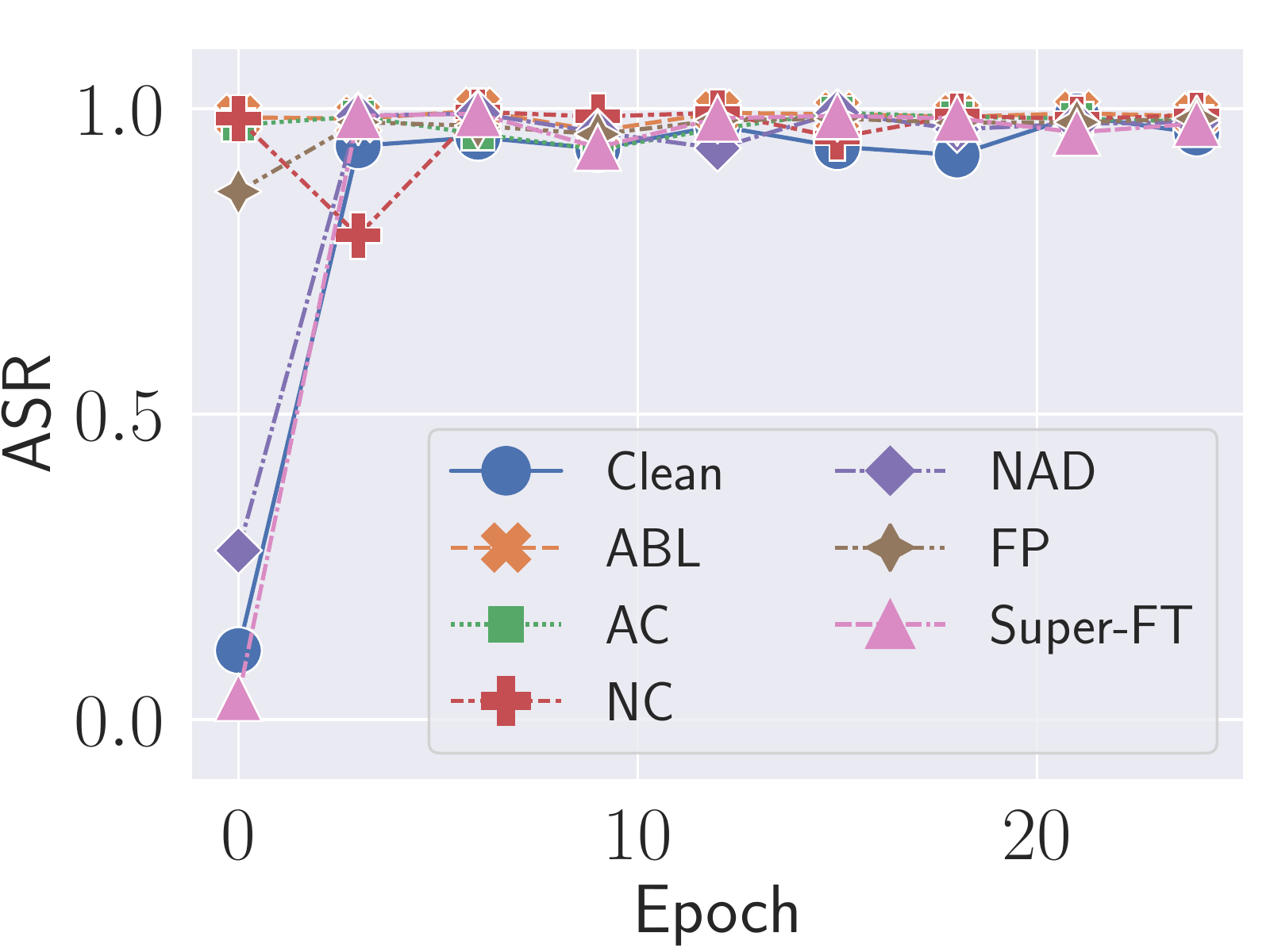}
\caption{Poison Ratio: 0.1}
\label{fig:relf-0.1}
\end{subfigure}
\begin{subfigure}{0.5\columnwidth}
\includegraphics[width=\columnwidth]{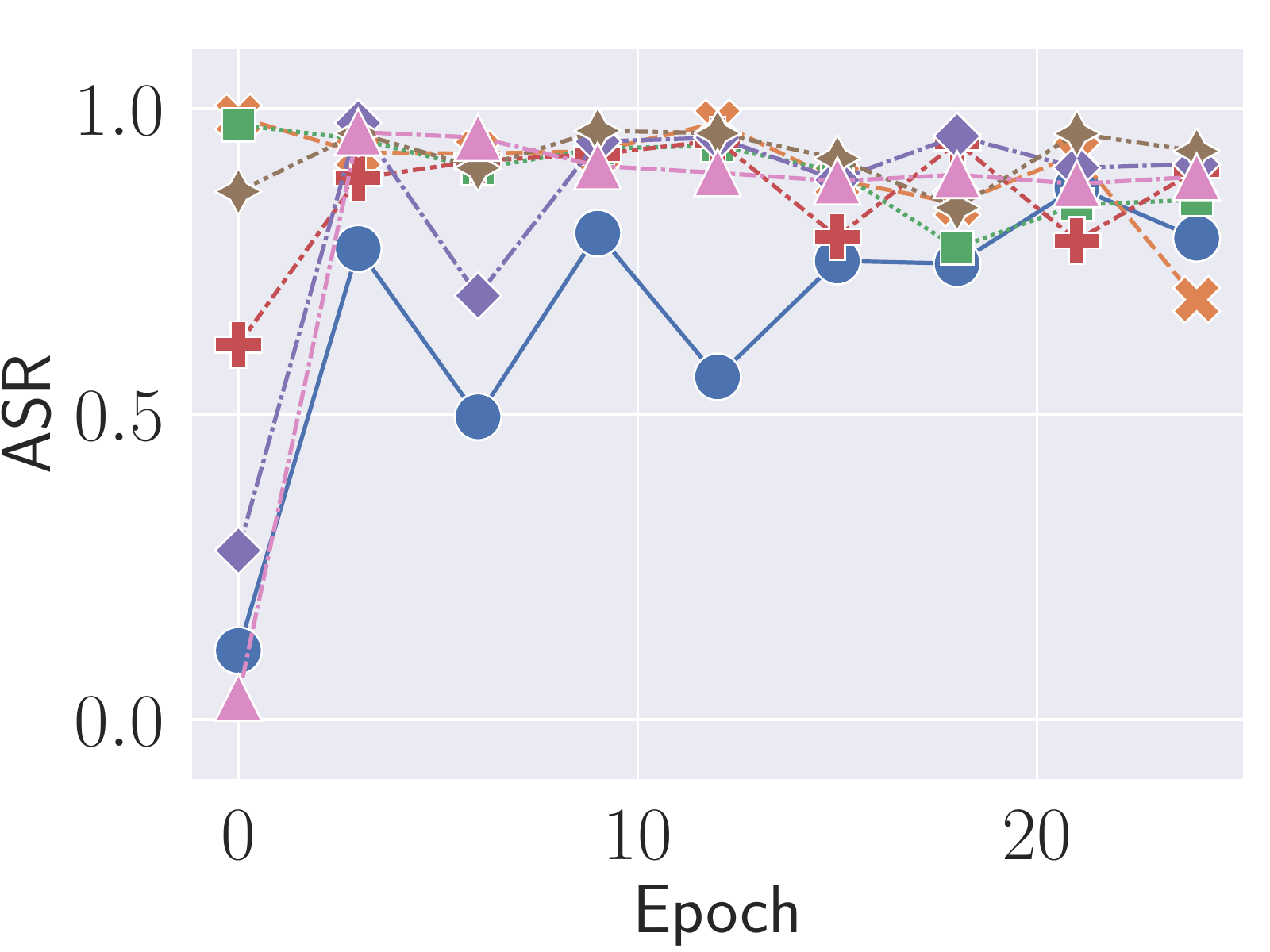}
\caption{Poison Ratio: 0.01}
\label{fig:relf-0.01}
\end{subfigure}
\begin{subfigure}{0.5\columnwidth}
\includegraphics[width=\columnwidth]{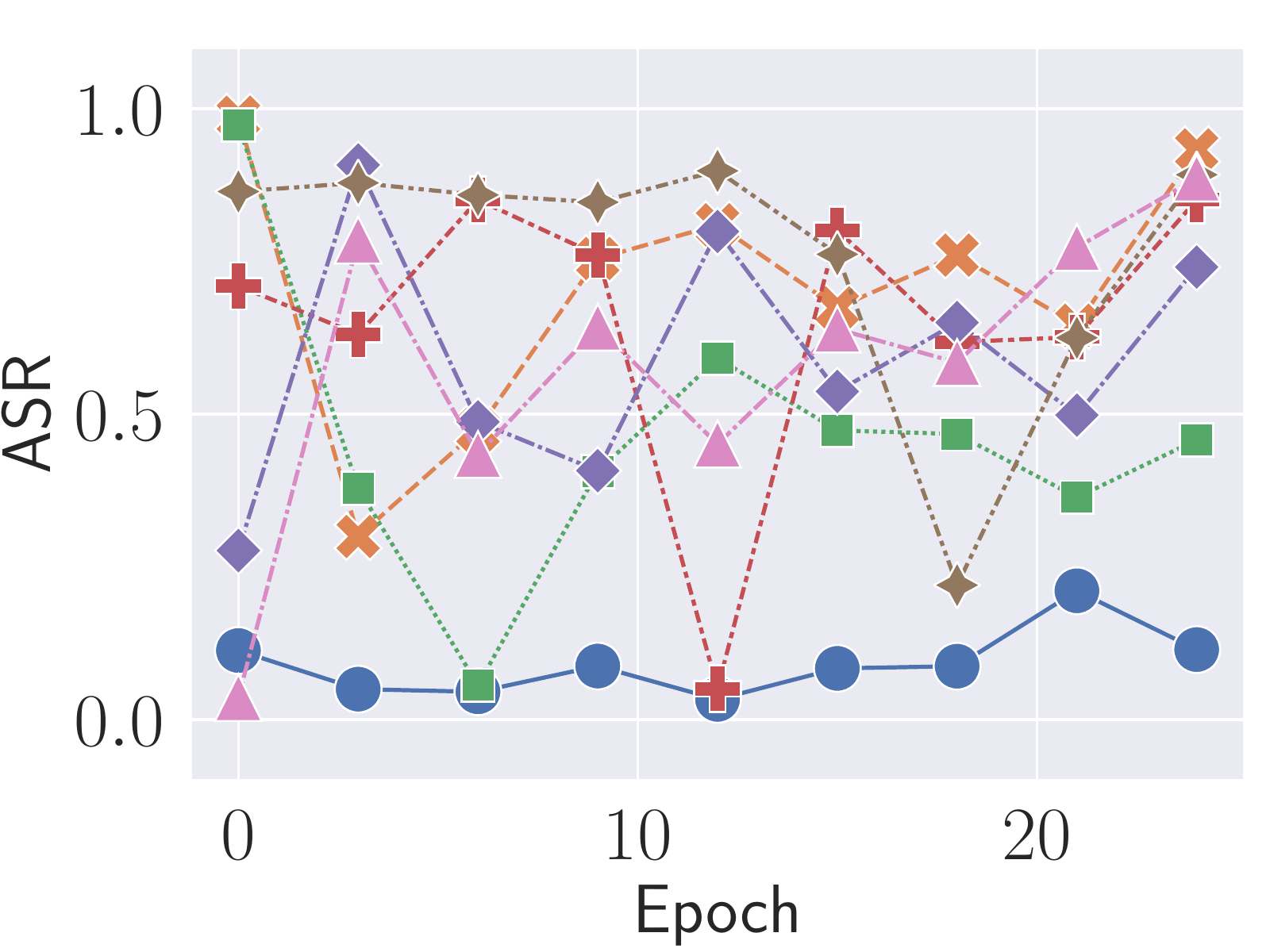}
\caption{Poison Ratio: 0.001}
\label{fig:relf-0.001}
\end{subfigure}
\begin{subfigure}{0.5\columnwidth}
\includegraphics[width=\columnwidth]{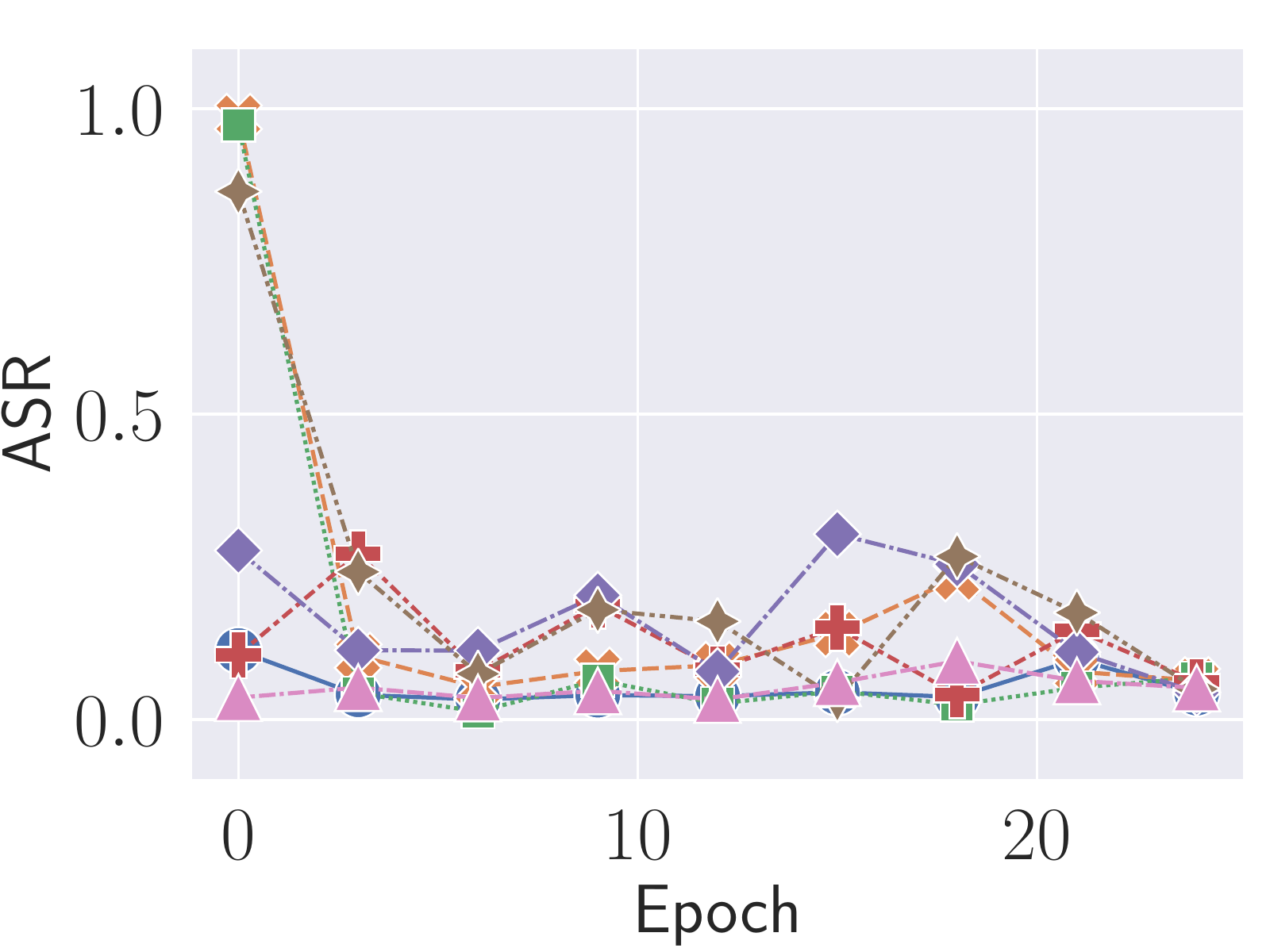}
\caption{Poison Ratio: 0.0001}
\label{fig:relf-0.0001}
\end{subfigure}
\caption{
Performance of LF backdoor re-injection attacks on different defense methods.
The X-axis represents training epochs in the re-injection phase.
The Y-axis represents the accuracy of poison samples.
}
\label{fig:reinject-lf}
\end{figure*}

\begin{figure*}[!t]
\centering
\begin{subfigure}{0.5\columnwidth}
\includegraphics[width=\columnwidth]{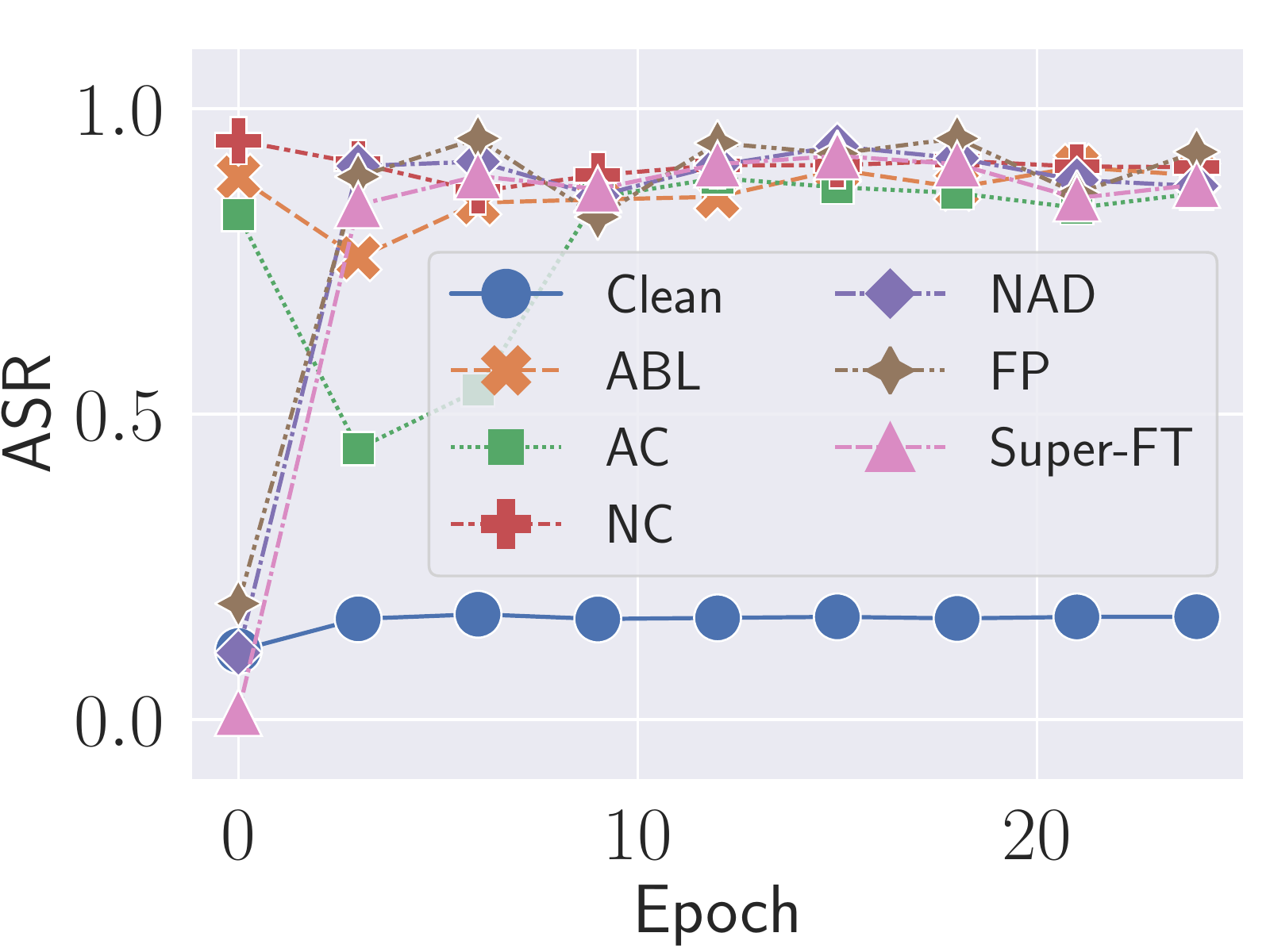}
\caption{Poison Ratio: 0.1}
\label{fig:rewanet-0.1}
\end{subfigure}
\begin{subfigure}{0.5\columnwidth}
\includegraphics[width=\columnwidth]{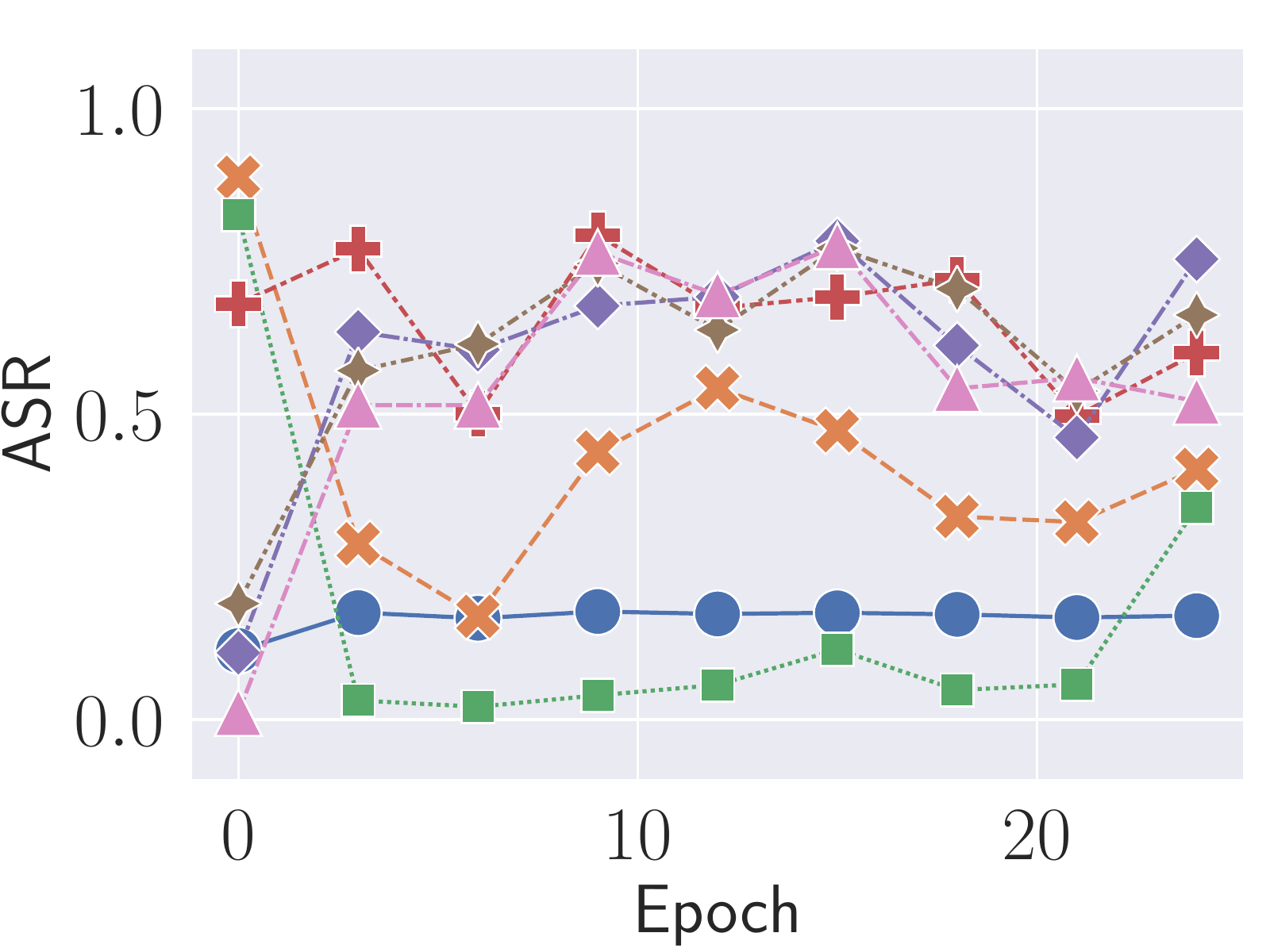}
\caption{Poison Ratio: 0.01}
\label{fig:rewanet-0.01}
\end{subfigure}
\begin{subfigure}{0.5\columnwidth}
\includegraphics[width=\columnwidth]{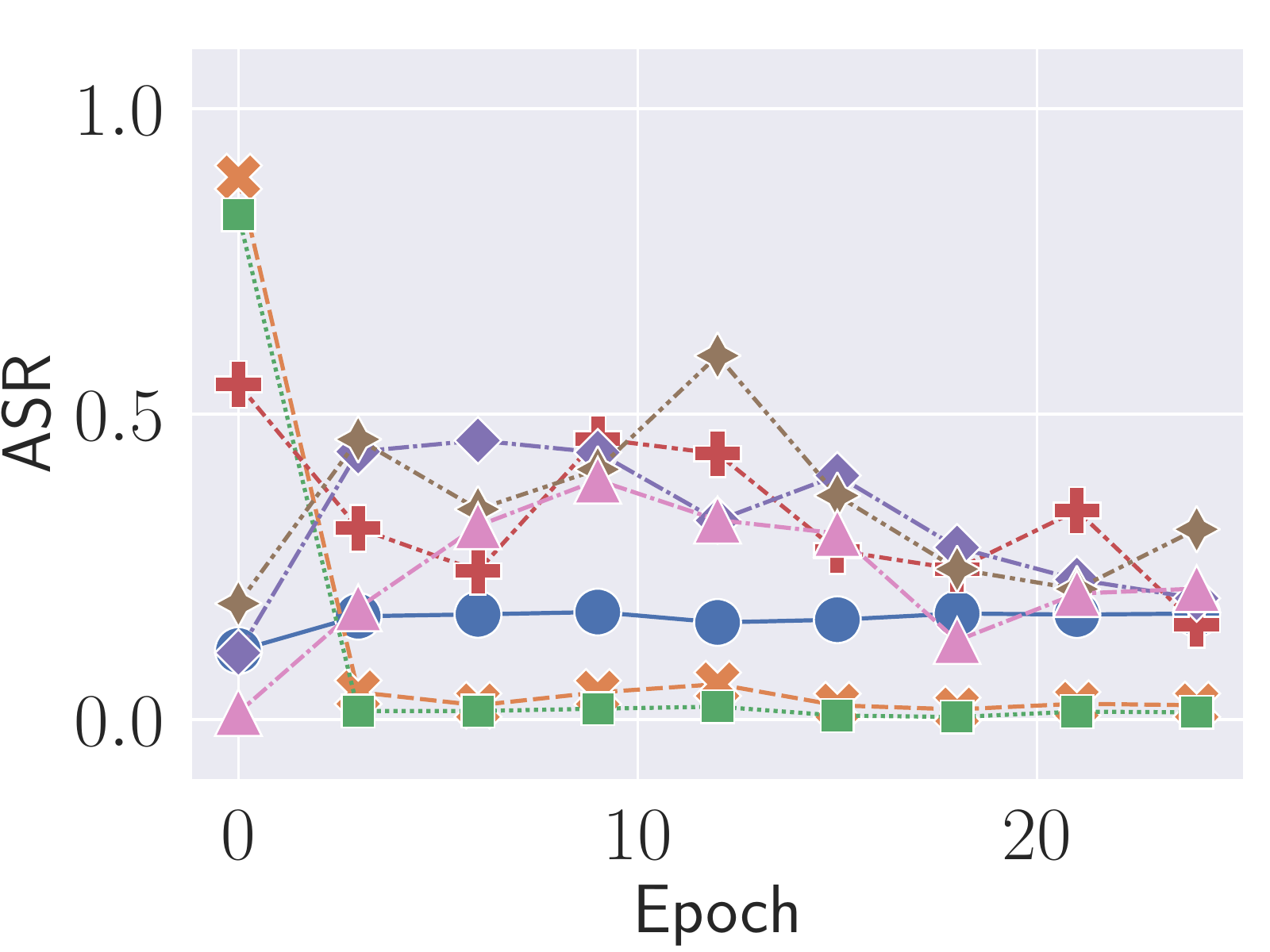}
\caption{Poison Ratio: 0.001}
\label{fig:rewanet-0.001}
\end{subfigure}
\begin{subfigure}{0.5\columnwidth}
\includegraphics[width=\columnwidth]{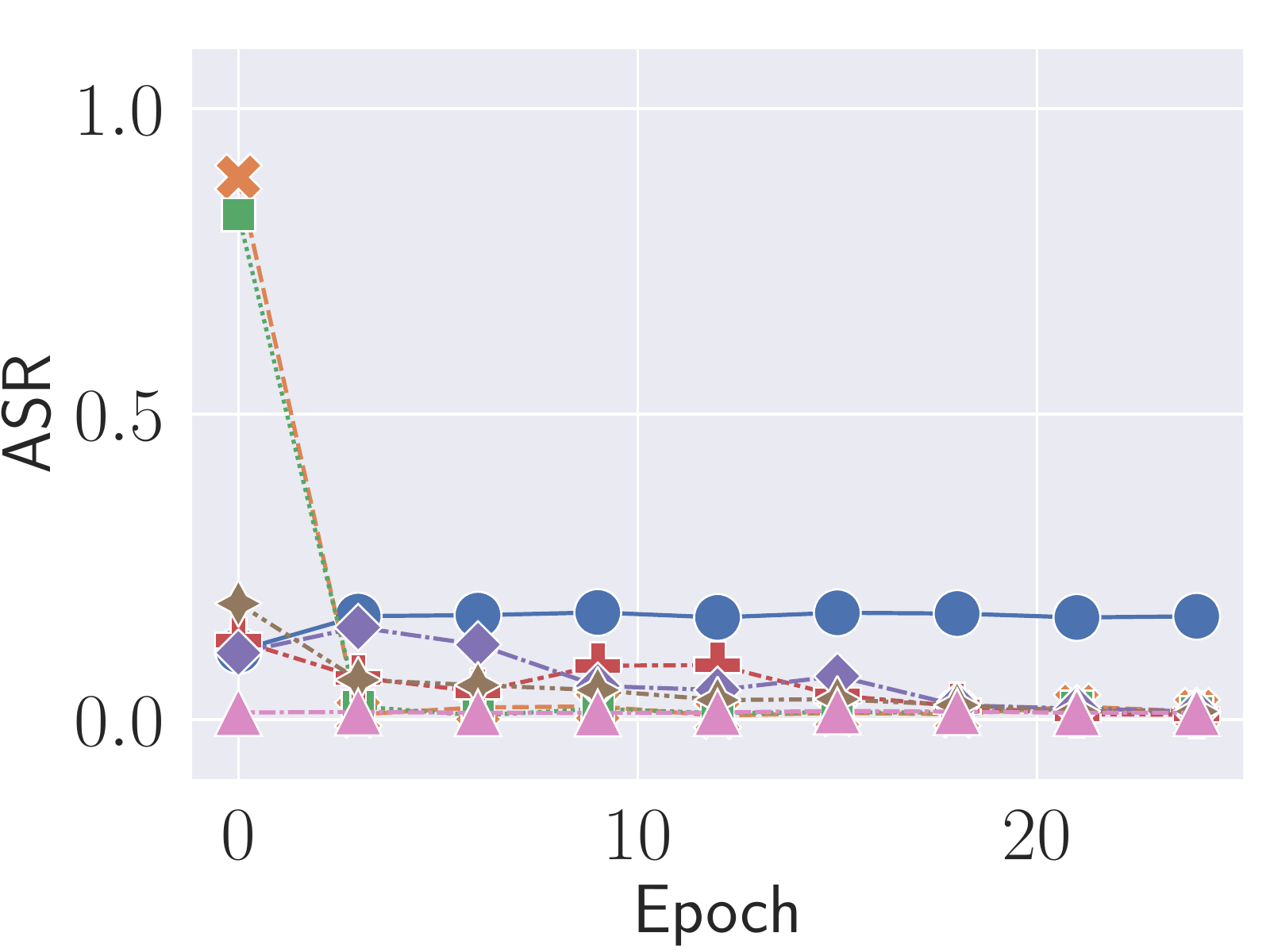}
\caption{Poison Ratio: 0.0001}
\label{fig:rewanet-0.0001}
\end{subfigure}
\caption{
Performance of WaNet backdoor re-injection attacks on different defense methods.
The X-axis represents training epochs in the re-injection phase.
The Y-axis represents the accuracy of poison samples.
}
\label{fig:reinject-wanet}
\end{figure*}

% ----------------------------------------------------
\end{document}